\documentclass[twocolumn,tighten,onecolappendix]{aastex61}

\input epsf.tex

\begin{document}
\title{The Gemini/HST Galaxy Cluster Project: Stellar Populations in the Low Redshift Reference Cluster Galaxies}
\author[0000-0003-3002-1446]{Inger J{\o}rgensen}
\affil{Gemini Observatory, 670 N.\ A`ohoku Pl., Hilo, HI 96720, USA}
\author{Kristin Chiboucas}
\affil{Gemini Observatory, 670 N.\ A`ohoku Pl., Hilo, HI 96720, USA}
\author[0000-0002-8610-0672]{Kristi Webb}
\affil{Department of Physics and Astronomy, University of Victoria, 3800 Finnerty Road, Victoria, British Columbia, V8P 5C2, Canada}
\affil{Department of Physics and Astronomy, University of Waterloo, 200 University Ave.\ W. Waterloo, Ontario, ON N2L 3G1, Canada}
\author[0000-0001-5962-7260]{Charity Woodrum}
\affil{Department of Physics, 1274 University of Oregon, Eugene, OR 97403, USA}
\affil{Steward Observatory, University of Arizona, 933 North Cherry Avenue, Tucson, AZ 85721, USA}

\email{ijorgensen@gemini.edu} 
\email{kchiboucas@gemini.edu}
\email{kristiannewebb@gmail.com}
\email{cwoodrum@email.arizona.edu}

\correspondingauthor{Inger J{\o}rgensen}

\submitjournal{Astronomical Journal}
\date{Accepted for publlication September 26, 2018}

\begin{abstract}
In order to study stellar populations and galaxy structures at intermediate and high redshift ($z=0.2-2.0$) and link these properties
to those of low redshift galaxies, there is a need for well-defined local reference samples.
Especially for galaxies in massive clusters, such samples are often limited to the Coma cluster galaxies.
We present consistently calibrated velocity dispersions and absorption line indices for galaxies in the central 
$2\,R_{\rm 500} \times 2\,R_{\rm 500}$ of four massive clusters at $z<0.1$: Abell 426/Perseus, Abell 1656/Coma, Abell 2029, and Abell 2142.
The measurements are based on data from Gemini Observatory, McDonald Observatory, and the Sloan Digital Sky Survey.
For bulge-dominated galaxies the samples are 95\% complete 
in Perseus and Coma, and 74\% complete in A2029 and A2142, to a limit of $M_{\rm B,abs}\le -18.5$ mag.
The data serve as the local reference for our studies of galaxy populations in the higher redshift clusters
that are part of the Gemini/{\it HST} Galaxy Cluster Project (GCP).

We establish the scaling relations between line indices and velocity dispersions as reference for the GCP.
We derive stellar population parameters ages, metallicities [M/H], and abundance ratios from line indices, both averaged in 
bins of velocity dispersion, and from individual measurements for galaxies in Perseus and Coma. 
The zero points of relations between the stellar population parameters and the velocity dispersions
limit the allowed cluster-to-cluster variation of the four clusters to 
$\pm 0.08$ dex in age, $\pm 0.06$ dex in [M/H], $\pm 0.07$ dex in [CN/Fe], and $\pm 0.03$ dex in [Mg/Fe].

\end{abstract}


\section{Introduction \label{SEC-INTRO} }

Massive galaxy clusters with masses of $10^{14} M_{\sun}$ or larger are major building 
blocks of the large-scale structure of the Universe. 
The precursors of these and the galaxies residing in them can be traced back to $z \sim 2$ 
(e.g., Stanford et al.\ 2012; Gobat et al.\ 2013; Andreon et al.\ 2014; Newman et al.\ 2014; Daddi et al.\ 2017) at a time 
when the age of the Universe was about 25\% of its current age. 
As such, these galaxies are valuable time posts for studying galaxy evolution over a large fraction of the 
history of the Universe. 
The techniques used in investigations of galaxy evolution include photometric studies of luminosity functions
and color-magnitude diagrams (e.g., Ellis \& Jones 2004; Faber et al.\ 2007; Foltz et al.\ 2015), 
and investigations of morphological mixtures and galaxy sizes (e.g., Dressler et al.\ 1997; 
Delaye et al.\ 2014; Huertas-Company et al.\ 2013, 2016).
Detailed studies of the stellar populations using scaling relations for the spectroscopy,
the Fundamental Plane (Dressler et al.\ 1987; Djorgovski \& Davis 1987) and in some cases 
determinations of ages, metallicities, and abundance ratios cover many clusters at $z=0.2-1.0$, 
(e.g., Kelson et al.\ 2006; Moran et al.\ 2005, 2007; 
van Dokkum \& van der Marel 2007; S\'{a}nchez-Bl\'{a}zquez et al.\ 2009; Saglia et al.\ 2010,  
J\o rgensen et al 2005, 2006, 2017; Leethochawalit et al.\ 2018).
At $z>1$ the Lynx W cluster ($z=1.3$) has been studied by our group (J\o rgensen et al.\ 2014)
using data for 13 bulge-dominated galaxies,
while three $z=1.4-1.6$ clusters were studied by Beifiori et al.\ (2017) using data for a total
of 19 galaxies on the red sequence.
All these investigations rely on reference samples, defining the end-product of the galaxy evolution. 
To the extent that the cluster environment affects the evolution of the galaxies, the reference 
samples should match the cluster environment into which the higher redshift clusters will evolve.
There has been a lack of such data sets, exemplified by the fact that many researchers use the 
data for Coma cluster (J\o rgensen et al.\ 1995ab, 1996; J\o rgensen 1999) as their main, or only, 
low redshift reference (e.g., van Dokkum \& Franx 1996; Ziegler et al.\ 2001; 
Wuyts et al.\ 2004; van Dokkum \& van der Marel 2007; J\o rgensen et al 2005, 2006, 2007, 2017;  
Cappallari et al.\ 2009; Saglia et al. 2010; Saracco et al.\ 2014; Beifiori et al.\ 2017).
Limiting the low redshift reference to one cluster does not take into account the possible
cluster-to-cluster differences in the galaxy populations and their star formation histories.
In addition, the Coma cluster is in fact not massive enough to be the end-product of some of the 
massive $z=0.2-1.0$ clusters. In our project, the Gemini/{\it HST} Galaxy Cluster Project (GCP),
we study clusters whose masses are expected by $z \sim 0$ to reach close to $M_{\rm 500} = 10^{15} M_{\sun}$, which is more
than double that of the Coma cluster, cf.\ J\o rgensen et al.\ (2017, 2018). 
Other researchers have studied similarly massive clusters, 
e.g.\ Kelson et al.\ (2006), Moran et al.\ (2007), Beifiori et al.\ (2017).
The ongoing project {\it Gemini Observations of Galaxies in Rich Early Environments} (Balogh et al.\ 2017) 
includes three clusters at $z=1.0-1.5$ and with masses above $5\times 10^{14} M_{\sun}$, and will also benefit
from access to better reference data at low redshift.

In order to address this lack of in depth studies of very massive clusters at $z<0.1$, we selected for 
study the two most massive clusters within this redshift, Abell 2029 at $z=0.077$ and Abell 2142 at $z=0.089$.
Together with Coma and Perseus, these clusters will be used as our low redshift reference sample for 
future investigations of the GCP clusters.
In the present paper we present and analyze the spectroscopic data for the four clusters based on measurements
of velocity dispersions and absorption line indices. Our analysis includes determination of ages,
metallicities, and abundance ratios. Full spectrum fitting may be pursued in future analysis,
but is beyond the scope of the current paper.
A companion paper (J\o rgensen et al., in prep.) will present photometry for the galaxies, 
enabling studies of galaxy sizes and the Fundamental Plane. 

\begin{figure}
\epsfxsize 8.5cm
\epsfbox{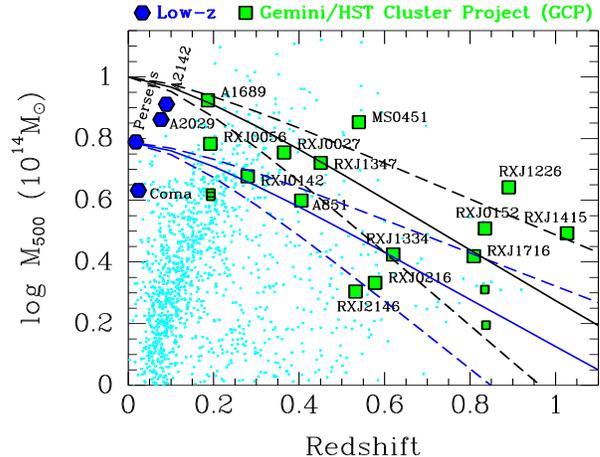}
\caption{The cluster masses, $M_{\rm 500}$, based on X-ray data versus the redshifts of the clusters,
adopted from J\o rgensen et al.\ (2018).
Blue -- local reference sample, and the topic of the present paper.
Green - $z=0.2-1$ GCP cluster sample.
The pairs of slightly smaller points at the same redshifts as RXJ0056.6+2622
and RXJ0152.7--1357 show the values for the sub-clusters of these binary clusters.
Small cyan points -- all clusters from Piffaretti et al.\ (2011) shown for reference.
Blue and black lines -- mass development of clusters based on numerical simulations
by van den Bosch (2002).
The black lines terminate at Mass=$10^{15} M_{\sun}$ at $z=0$ roughly matching the highest mass clusters at $z=0.1-0.2$.
The blue lines terminate at Mass=$10^{14.8} M_{\sun}$ at $z=0$ matching the mass of the Perseus cluster.
The dashed lines represent the typical uncertainty in the mass development represented by the numerical simulations.
\label{fig-M500redshift} }
\end{figure}

\begin{deluxetable*}{ll rrrrrrrrr}
\tablecaption{Cluster Properties and Parent Samples \label{tab-clusters} }
\tabletypesize{\scriptsize}
\tablewidth{0pc}
\tablecolumns{11}
%\tablenum{8}
\tablehead{
\colhead{Cluster} & \colhead{Redshift} & \colhead{$\sigma _{\rm cluster}$} & \colhead{$\Delta z$}& \colhead{$L_{500}$} & \colhead{$M_{500}$} & 
\colhead{$R_{500}$} & \colhead{Sample area} & \colhead{$g'_{\rm limit}$}  & \colhead{$A_g$} & \colhead{$k_g$} \\ 
 & & $\rm km~s^{-1}$ & & $10^{44} \rm{erg\,s^{-1}}$ & $10^{14}M_{\sun}$ & Mpc & arcmin $\times$ arcmin \\
\colhead{(1)}  & \colhead{(2)} & \colhead{(3)} & \colhead{(4)} & \colhead{(5)} & \colhead{(6)} & \colhead{(7)} & \colhead{(8)} & \colhead{(9)} & \colhead{(10)} & \colhead{(11)}}
\startdata
Perseus/Abell 426 & 0.0179 & $1368_{-69}^{+74}$ &  $0.006-0.030$ &  6.217 & 6.151 & 1.286 & 120 $\times$ 124& 15.5 &  0.555 & 0.03 \\
Coma/Abell 1656   & 0.0231 & $1154_{-56}^{+56}$ &  $0.011-0.035$ &  3.456 & 4.285 & 1.138 &  64 $\times$  70& 16.1 &  0.023 & 0.04 \\
Abell 2029        & 0.0780 & $1181_{-54}^{+42}$ &  $0.066-0.090$ &  8.727 & 7.271 & 1.334 &  30 $\times$  30& 18.7 &  0.134 & 0.17 \\
Abell 2142        & 0.0903 & $1190_{-48}^{+44}$ &  $0.078-0.103$ & 10.676 & 8.149 & 1.380 &  30 $\times$  30& 19.1 &  0.143 & 0.19 \\  
\enddata
\tablecomments{Column 1: Galaxy cluster. Column 2: Cluster redshift. Column 3: Cluster velocity dispersion, determined from
the available redshifts for the parent samples, see Section \ref{SEC-SAMPLES}.
Column 4: Redshift limits used for definition of cluster membership.
Column 5: X-ray luminosity in the 0.1--2.4 keV band within the radius $R_{500}$. X-ray data are from Piffaretti et al.\ (2011).
Column 6: Cluster mass within the radius $R_{500}$.
Column 7: Radius within which the mean over-density of the cluster is 500 times the critical density at the cluster redshift.
Column 8: Sample area, covering approximately $2 R_{\rm 500} \times 2 R_{\rm 500}$.
Column 9: Magnitude limit of sample in rest frame $g'$ band, corresponding to $M_{\rm B,abs}=-18.5$ mag (Vega magnitudes).
Column 10: Average Galactic extinction in the $g'$ band.
Column 11: Average {\it k}-correction for member galaxies on the red sequence, ie.\ for galaxies with $(g'-r')=0.8$.
}
\end{deluxetable*}

In Section \ref{SEC-CLUSTERS}, we provide the context for selecting the four clusters as our 
low redshift reference, and we summarize the properties of the clusters.
The parent galaxy samples are described in Section \ref{SEC-SAMPLES}. 
Section \ref{SEC-SPECTROSCOPY} gives an overview of the various data sources used to 
compile the most complete samples with measurements of velocity dispersions and absorption line indices. 
The section also details the calibration of parameters.
Appendix \ref{SEC-APPENDIX} contains additional detail.

Section \ref{SEC-SAMPLEMETHOD} presents the color-magnitude diagrams and defines the sub-samples of passive bulge-dominated
galaxies used in the remainder of the paper. 
Then we establish the scaling relations between the line indices and the velocity dispersions, Section \ref{SEC-INDICES}.
We determine ages, metallicities, and abundance ratios from line indices.
These measurements are used to study their correlations with the velocity dispersions, 
and to set limits on the cluster-to-cluster variations of the stellar populations, Section \ref{SEC-AGEMALPHA}.

In Section \ref{SEC-DISCUSSION} we compare our results with other
results for low redshift cluster galaxies and discuss these in the context of using 
the samples as reference samples for studies of higher redshift bulge-dominated galaxies,
and in the context of galaxy evolution in general.
The conclusions are summarized in Section \ref{SEC-CONCLUSION}.

Throughout this paper we adopt a $\Lambda$CDM cosmology with 
$H_{\rm 0} = 70\,{\rm km\,s^{-1}\,Mpc^{-1}}$, $\Omega_{\rm M}=0.3$, and $\Omega_{\rm \Lambda}=0.7$.
Magnitudes are given in the AB system, except where noted.

\section{Cluster Sample and Cluster Properties \label{SEC-CLUSTERS}}

Massive clusters at intermediate and high redshifts ($z=0.2-2.0$) continue to grow in mass as they evolve 
(e.g., van den Borsch 2002; Fakhouri et al.\ 2010).
Therefore, the descendants of such clusters at $z \sim 0$ will be significantly more massive.
If we want to ensure that we are comparing the galaxies in $z=0.2-2.0$ massive clusters to galaxies 
in $z \sim 0$ clusters that can be their descendants, then that low redshift reference sample needs to
contain the most massive known low redshift clusters.
We illustrate this in Figure \ref{fig-M500redshift}, which shows mass versus redshift for the GCP
clusters at $z=0.2-1.0$, and for reference all X-ray clusters from Piffaretti et al.\ (2011).
The solid and dashed lines on the figure represent example models for the mass evolution of clusters
as a function of redshift, cf.\ van den Borsch (2002).
The most massive GCP clusters at all redshifts are expected by $z\sim 0$ to evolve into clusters
significantly more massive than the Coma cluster, and even the Perseus cluster. 
Massive clusters at $z>1$, e.g., Khullar et al.\ (2018) and references therein, can also be expected to evolve into
clusters more massive than the Perseus cluster. 
A2029 and A2142 are the most massive clusters at $z<0.1$ and may
be more suitable as the low redshift reference clusters for studies of these high mass $z>0.2$ clusters.
We therefore selected these two clusters to use as local reference clusters for the GCP, together with
the more well-studied Coma and Perseus clusters.
All four clusters were included in the northern Abell catalog (Abell et al.\ 1989).
Table \ref{tab-clusters} summarizes the cluster masses, radii, and luminosities based on X-ray observations.
The clusters have masses of $M_{500} = 4.3-8.1 \times 10^{14} M_{\sun}$.
The Piffaretti et al.\ catalog contains a total of 20 clusters with $M_{500} \ge M_{500} {\rm (Coma)}$ and $z\le 0.1$.
Perseus and Coma are the two lowest redshift clusters of these, A2029 and A2142 are the two most massive.

\begin{figure}
\epsfxsize 8.5cm
\epsfbox{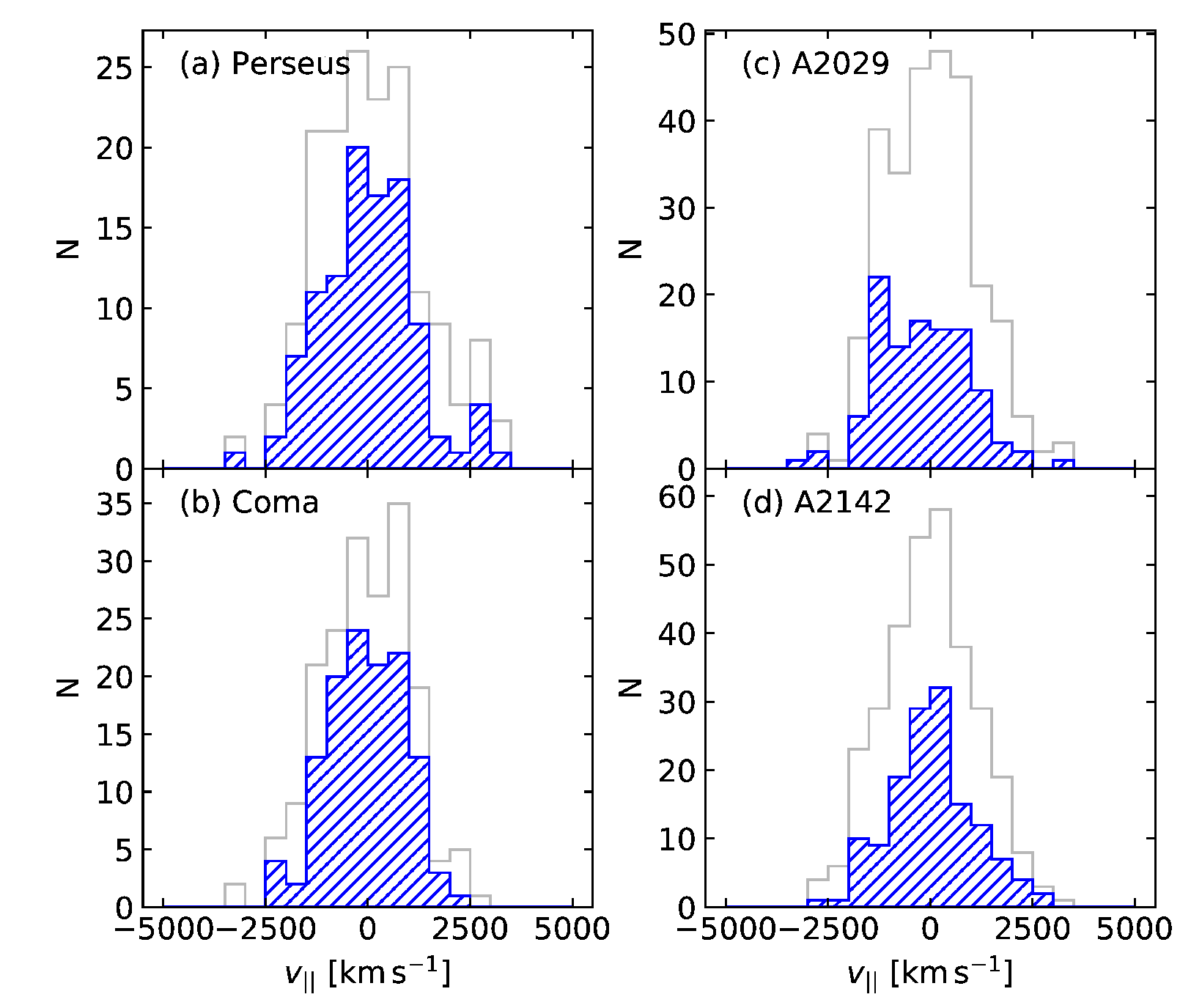}
\caption{Distribution of the radial velocities (in the rest frames of the clusters) 
relative to the cluster redshifts for cluster members, $v_{\|} = c (z - z_{\rm cluster}) / (1+z_{\rm cluster})$.
Blue hatched -- passive bulge-dominated members brighter than the adopted analysis limits.
For reference the grey histograms show all members in the parent samples. The parent samples have been
limited at magnitudes just below the adopted limits for the final samples, see Section \ref{SEC-SAMPLES}.
\label{fig-vradhist} }
\end{figure}

The large masses of the four selected clusters are also reflected in their high cluster velocity dispersions. 
In Figure \ref{fig-vradhist},
we show the distributions of the radial velocities of member galaxies included in the parent samples, 
see Section \ref{SEC-SAMPLES}, and the distributions of the passive bulge-dominated members, 
see Section \ref{SEC-SAMPLEMETHOD}. 
Kolmogorov-Smirnov tests give probabilities of 
45\%-93\% that for each cluster the distributions of the radial velocities of parent sample galaxies
and the passive bulge-dominated galaxies are drawn from the same parent distribution. Thus, there
are no significant differences in these distributions.
Using the biweight method from Beers et al.\ (1990), we determine the cluster velocity
dispersions, $\sigma _{\rm cl}$, from the member galaxies in the parent samples. 
The results are listed in Table \ref{tab-clusters}.
We find somewhat larger cluster velocity dispersions than found by other studies. In the classical
study by Zabludoff et al.\ (1990), velocity dispersions of $1277\, {\rm km\,s^{-1}}$ and $1010\, {\rm km\,s^{-1}}$
are found for Perseus and Coma, respectively, based on samples of 114 and 234 galaxies, respectively.
Sohn et al.\ (2017) finds $\sigma _{\rm cl}=947 \pm 31\, {\rm km\,s^{-1}}$ for Coma and 
$ 973 \pm 31\, {\rm km\,s^{-1}}$ for A2029, both based on samples of $\approx 1000$ galaxies.
Owers et al.\ (2011) find a similar velocity dispersion, $ 995 \pm 21\,{\rm km\,s^{-1}}$, for A2142, based on almost
1000 member galaxies.
It is possible, that our larger values are due to a selection effect of our smaller samples in Coma, A2029 and A2142. 
However, the differences are not significant for our analysis and results.

Below we briefly summarize the global properties of each of the clusters.
These summaries are by no means intended to be complete reviews of the substantial literature on each
of these clusters, but serve to put in context the role of the clusters as references
for studies of higher redshift massive clusters.

{\it Perseus:}
The cluster is included in the catalog by Zwicky (1942) and in the original Abell (1958) catalog.
The brightest cluster galaxy, NGC 1275, hosts a powerful active galactic nuclei (Conselice at al.\ 2001 and references therein). 
The cluster's X-ray emission shows that the cluster is a cooling flow cluster (Fabian 1994), though
more recent X-ray data indicate that the initial cooling rates were overestimated, see Rafferty et al.\ (2008) 
and references therein.
The cluster is generally considered relaxed and has no significant
optical substructure (cf., Girardi et al.\ 1997). 
However, earlier X-ray observations indicated that substructure may be present (Mohr et al.\ 1993).
Most recently {\it Chandra} data were
used to investigate the ``sloshing'' of the intracluster gas (Walker et al.\ 2017).
The properties and star formation history of NGC 1275 have been studied extensively, see e.g.\ Canning et al.\ (2014)
for a recent investigation of the young star clusters in the galaxy. 
Penny et al.\ (2009) investigated the dwarf galaxy population of the cluster based on {\it HST} imaging, and
argued that a large dark matter content of these was needed to avoid disruption by the cluster potential.
No recent detailed studies based on spectroscopy and line indices exist of the stellar populations of the member galaxies.
Fraix-Burnet et al.\ (2010) included the Perseus cluster in their study of the Fundamental Plane 
for the early-type galaxies, though their analysis contains no specific comments on differences 
among the clusters in the study.
                                                                         
{\it Coma:}
The cluster was first mentioned in the Curtis (1918) cluster catalog. It was included in 
the catalog by Zwicky (1942) and in the original Abell (1958) catalog, and has been studied
intensively since. 
X-ray observations from ROSAT (White et al.\ 1993) show that the Coma cluster has significant
sub-structure, most notably is the large sub-structure associated with NGC 4939 to the south-west of the cluster center.
Recent weak lensing studies and new X-ray observations confirm the presence of several sub-haloes within the 
cluster (Okabe et al.\ 2014; Sasaki, et al.\ 2016). 
Pimbblet et al.\ (2014) compared the Coma cluster to other massive low redshift 
clusters and concluded that kinematically the cluster is comparable to others, and that sub-clustering in
low redshift clusters is not unusual. However, the authors concluded
that Coma cluster galaxies have higher star formation rates for a given stellar mass, than found
for other clusters, but see also Tyler et al.\ (2013). 
In our discussion, we will return to the question of cluster-to-cluster differences of the 
stellar populations in the member galaxies.
Of the many studies of the stellar populations in the Coma cluster galaxies, most relevant 
for our analysis are the results by Harrison et al.\ (2011) who established scaling relations 
between line strengths and velocity dispersions, and investigated the ages, [M/H] and $\rm [\alpha /Fe]$
as a function of velocity dispersions and environment. These authors found no significant
dependency of the environment, reaching much larger cluster center distances than covered
in our analysis. We compare their other results with ours in the discussion (Section \ref{SEC-DISCUSSION}).

{\it Abell 2029:}
The cluster has been known since the original Abell (1958) catalog.
It is a cooling flow cluster, with cooling time comparable to that of the Perseus cluster 
(Allen 2000; Rafferty et al.\ 2008). 
Parekh et al.\ (2015) used {\it Chandra} data to study the X-ray morphology of several low redshift clusters. 
They classify A2029 as ``strong relaxed'', in agreement with the classification from Vikhlinin et al.\ (2009).
Sohn et al.\ (2017) investigated the luminosity function, the stellar mass function, and the velocity dispersion 
function for both A2029 and the Coma cluster. They find no significant differences between these two clusters.
Tyler et al.\ (2013) focused on the star forming galaxies and determined that A2029 contains star forming galaxies 
resembling those in the field, while the Coma cluster contains a populations of the star forming galaxies 
with significantly lower star formation rates for their stellar mass.

{\it Abell 2142:}
The cluster has been known since the original Abell (1958) catalog.
This is another cooling flow cluster (Allen 2000). 
Parekh et al.\ (2015) classify this cluster as ``non-relaxed'' based on their study of the X-ray morphology,
while Vikhlinin et al.\ (2009) regard the cluster as relaxed. Owers et al.\ (2011) identify sub-structures
in the cluster based on extensive redshift data and argue that these are the remnants of minor merging 
of groups into the cluster. Two of the identified sub-structures lie within the area we study in the present paper.
The earlier study of the {\it Chandra} data by Markevitch et al.\ (2000) also supports the presence of sub-structure
in the central part of the cluster. 
The spatial distribution of the galaxy types may be related to the infalling groups. 
In particular, Einasto et al.\ (2018) find that the central part within 0.7 Mpc of the 
cluster center is dominated by the passive older galaxies, while the star forming and recently quenched galaxies 
dominate at cluster center distances larger than 2.6 Mpc. 

\section{Parent Samples \label{SEC-SAMPLES}}

For each cluster we construct a parent sample of confirmed members within a square area centered on the cluster 
and with a side length of approximately $2\, R_{500}$. $R_{500}$ is the radius 
within which the mean over-density of the cluster is 500 times the critical density at the cluster redshift.
In our analysis, we use samples that are magnitude limited in the $g'$ band to a magnitude equivalent to an absolute B band 
magnitude of $M_{\rm B,abs}=-18.5$ mag (Vega magnitudes). 
This limit corresponds to a dynamical mass of $M_{\rm dyn} \approx 10^{10.3} M_{\sun}$ and a velocity
dispersion of approximately 100 $\rm km\,s^{-1}$ for galaxies on the Fundamental Plane. 
With current 8-meter class telescopes and instrumentation, $M_{\rm B,abs}=-18.5$ mag is a practical limit
for obtaining spectra of such galaxies to $z \approx 1$ and with sufficient S/N to study the stellar 
populations, e.g.\ J\o rgensen et al.\ (2017).
Thus, our local sample is a suitable reference for such studies. 

The parent samples are described in the following, while Section \ref{SEC-SPECTROSCOPY} details the spectroscopy. 
Table \ref{tab-clusters} summarizes the sample limits, areas and sizes. 
Our sample selection is based on the Sloan Digital Sky Survey (SDSS) photometry, which we correct for Galactic extinction using the 
values for the individual galaxies provided in SDSS. These originate from the calibration by Schlafly et al.\ (2011).
We calibrate the photometry for the cluster members to rest frame magnitudes using the {\it k}-corrections with color terms as 
described in Chilingarian et al.\ (2010).
Table \ref{tab-clusters} includes the mean Galactic extinction for the cluster members and the average {\it k}-correction 
for the $g'$ band for galaxies on the red sequence.
As part of the SDSS data processing, the galaxy profiles were fit with a linear combination of the best fit radial
$r^{1/4}$-profile and exponential profile. The magnitude from this fit is {\tt cmodelmag},
which we use as the measure of the total magnitudes of the galaxies.
As recommended on the SDSS web site, we use colors based on {\tt modelmag} (the total magnitude from the best 
fit $r^{1/4}$-profile or exponential profile) as these are consistent across the SDSS passbands. 

{\it Perseus / Abell 426}: The sample is based on the catalog by Paturel et al.\ (2003), 
in the following referred to using the catalog prefix for the galaxies as the Primary
Galaxy Catalog (PGC).
We crossmatch galaxies from the catalog with SDSS Data Release 14 (DR14) photometry. 
The area R.A.$_{\rm J2000}=49\fdg 8381 - 50\fdg 025$, Dec$_{\rm J2000} =41\fdg 597-41\fdg 48$ is not covered by DR14. 
Thus, we used Data Release 7 (DR7) photometry for sample selection in this area.
For the analysis, the parent sample was then limited to a rest frame magnitude of $g'_{\rm rest} = 15.5$ mag or brighter. 
The SDSS DR14 contains objects in the Perseus field typed as ``galaxies'' but not included in the PGC. 
Visual inspection of the SDSS images shows that these are saturated stars. Our inspection
confirms that the PGC provides a complete catalog to the magnitude limit relevant for our study.
The selected sample covers an area of 120 arcmin $\times$ 124 arcmin. 
Redshift data from SDSS, the NASA/IPAC Extragalactic Database (NED) and our own observations were 
used to identify cluster members. Only 8 galaxies with $g'_{\rm rest} \le 15.5$ mag do not 
have spectroscopic redshifts. The parent sample contains 166 spectroscopically confirmed members, 153 of which are brighter 
than the adopted magnitude limit for the analysis.

{\it Coma / Abell 1656}: The sample is based on the catalog from Godwin et al.\ (1983, GMP). 
We crossmatch galaxies from this catalog with SDSS DR14 photometry and use for the analysis the sample of confirmed 
members with $g'_{\rm rest} \le 16.1$ mag. 
The sample covers the same area as used in our previous publications on the cluster 64 arcmin $\times$ 70 arcmin 
(J\o rgensen \& Franx 1994; J\o rgensen 1999).
Redshift data from SDSS, NED and our own observations were used to identify cluster members. 
The parent sample contains 185 spectroscopically confirmed members, 152 of which are brighter than the adopted magnitude limit.
Only one galaxy brighter than the sample limit does not have a spectroscopic redshift.

\begin{deluxetable*}{lll}
\tablecaption{Instrumentation\label{tab-inst} }
\tabletypesize{\scriptsize}
\tablewidth{0pt}
\tablecolumns{3}
\tablehead{\colhead{Parameter} & \colhead{McD.\ 2.7-m, LCS} & \colhead{Gemini North, GMOS-N}  }
\startdata
CCDs            & TI1 $800\times 800$                         & 3 $\times$ E2V 2048$\times$4608             \\
r.o.n.          & 7.34 e$^-$, 11.23 e$^-$\tablenotemark{a}    & (3.5,3.3,3.0) e$^-$\tablenotemark{b}           \\
gain            & 3.30 e$^-$/ADU                              & (2.04,2.3,2.19) e$^-$/ADU\tablenotemark{b}   \\
Pixel scale     & $1\farcs 27 \,{\rm pixel}^{-1}$\tablenotemark{c} & $0\farcs 1454 \,{\rm pixel}^{-1}$\tablenotemark{c}   \\
Field of view   & $8\arcmin$\tablenotemark{d}                 & $5\farcm5\times5\farcm5$                    \\
Gratings        & \#43: 600 $\rm l\,mm^{-1}$, \#47: 1200 $\rm l\,mm^{-1}$  & B600\_G5303                                  \\
Wavelength range &  3530--4900\AA, 4890--5590\AA              & 3500--6500\AA\tablenotemark{e}              \\
\enddata
\tablenotetext{a}{First entry refers to observations in 1994, second entry to observations in 1995}
\tablenotetext{b}{Values for the three detectors in the array.}
\tablenotetext{c}{Pixel scale for detectors binned by two in the spatial direction as used for the observations.}
\tablenotetext{d}{Slit length.}
\tablenotetext{e}{For MOS observations, the exact wavelength range varies from slitlet to slitlet.}
\end{deluxetable*}

{\it Abell 2029}: The sample is based on the SDSS DR14 supplemented with DR7 for a small area at the very center of the 
cluster, where the photometry from DR14 appears to be systematically incorrect for many of the galaxies and other galaxies 
are missing. While we do not know the origin of this issue, it is possible it is due to the presence of the very large 
central cD galaxy. 
The area covers R.A.$_{\rm J2000}=227\fdg 715 - 227\fdg 751$ and Dec$_{\rm J2000} = 5\fdg 73417 - 5\fdg 78394$.
DR7 data were also used in the area R.A.$_{\rm J2000}=227\fdg 625-227\fdg 765$, Dec=$_{\rm J2000} = 5\fdg841-5\fdg 95$, 
which is not included in DR14. This was also noted by Sohn et al.\ (2017) in their use of DR12 for the cluster.
The SDSS type designation as ``galaxy'' was adopted for the initial selection.
With the sample limited to $g' \le 20$ mag (including all objects with possible useful spectroscopy), 
all objects were visually inspected on the SDSS images. 
A few artifacts and one saturated star were removed from the sample. 
The selected sample covers an area of 30 arcmin $\times$ 30 arcmin.
The sample for the analysis was then limited to $g'_{\rm rest} \le 18.7$ mag. 
Redshifts from SDSS, Sohn et al.\ (2017) and our own observations were used to assign membership. 
22 galaxies with $g'_{\rm rest} \le 18.7$ mag have no available redshift. 
Of these, 7 have photometric redshifts from SDSS of 0.06-0.1 and may be cluster members.
The parent sample contains 282 spectroscopically confirmed members, 189 of which are brighter than the adopted magnitude limit. 

{\it Abell 2142}: The sample is based on the SDSS DR14. 
The SDSS type designation as ``galaxy'' was adopted for the initial selection.
With the sample limited to $g' \le 20 $ mag, all objects were visually inspected 
on the SDSS images to ensure they were galaxies.
The selected sample covers an area of 30 arcmin $\times$ 30 arcmin. 
The sample for the analysis was then limited to $g'_{\rm rest} \le 19.1$ mag. 
Redshifts from SDSS, NED and our own observations were used to assign membership. 
Nine galaxies with $g'_{\rm rest} \le 19.1$ mag have no available redshift. None of them are 
expected to be cluster members based on their SDSS photometric redshifts.
The parent sample contains 313 spectroscopically confirmed members, 251 of which are brighter than the adopted magnitude limit.

\begin{deluxetable*}{llrrrrrrr}
\tablecaption{Summary of Spectroscopic Data \label{tab-spdata} }
\tabletypesize{\scriptsize}
\tablewidth{0pc}
\tablehead{
\colhead{Cluster} & \colhead{Telescope, spectrograph} & \colhead{Grating} & \colhead{Exposure time} & \colhead{FWHM} 
& \colhead{$\sigma _{\rm inst}$} & \colhead{Aperture} & \colhead{$N_{\rm member}$} & \colhead{S/N} \\ 
\colhead{(1)}  & \colhead{(2)} & \colhead{(3)} & \colhead{(4)} & \colhead{(5)} & \colhead{(6)} & \colhead{(7)} & \colhead{(8)} & \colhead{(9)} }
\startdata
Perseus & McD.\ 2.7-m, LCS & \#43 & 600--1800 & 1.2--2.5 & 2.35\AA, 134 $\rm km\,s^{-1}$ & $2.0\times 6.35$, 2.07 & 52 & 21.6 \\
Perseus & McD.\ 2.7-m, LCS & \#47 & 900--3600 & 2.5--3.0 & 1.03\AA,  59 $\rm km\,s^{-1}$ & $2.0\times 6.35$, 2.07 & 61 & 28.9 \\
Perseus & SDSS, SDSS spec   &     &      &               & 1.12\AA, 62 $\rm km\,s^{-1}$ & 1.5 & 119 & 80.2 \\ 
Coma\tablenotemark{a}  & McD.\ 2.7-m, LCS & \#47 & 900--3600 &  & 0.97\AA,  56 $\rm km\,s^{-1}$ & $2.0\times 6.35$, 2.07 & 44 & 28.3 \\
Coma\tablenotemark{a}  & McD.\ 2.7-m, FMOS & 300 $\rm l\,mm^{-1}$ & 1800--3600 &  & 4.25\AA, 246 $\rm km\,s^{-1}$ & 1.3 & 38 & 33.0 \\
Coma\tablenotemark{a}  & Literature & & & & & & 80 & \nodata \\
Coma  & SDSS, SDSS BOSS spec &    &      &               & 1.14\AA, 62 $\rm km\,s^{-1}$ & 1.5, 1.0 & 179 & 58.1 \\
A2029 & Gemini North, GMOS-N (MOS) & B600 & 2640 & 0.55--1.20 & 2.07\AA, 111 $\rm km\,s^{-1}$ & $1.0\times 2.4$, 0.90 & 49 & 36.8 \\ 
A2029 & Gemini North, GMOS-N (LS)  & B600 & 3600 & 0.52--2.17 & 1.66\AA, 89 $\rm km\,s^{-1}$ & $0.75\times 2.4$, 0.78 & 25 & 17.2 \\  
A2029 & SDSS, SDSS BOSS spec &     &      &                      & 1.18\AA, 62 $\rm km\,s^{-1}$ & 1.5, 1.0 & 117 & 29.0 \\
A2142 & Gemini North, GMOS-N (MOS) & B600 & 2640 & 0.66--1.20 & 2.07\AA, 110 $\rm km\,s^{-1}$ & $1.0\times 2.4$, 0.90 & 57 & 29.4 \\ 
A2142 & Gemini North, GMOS-N (LS)  & B600 & 3600 & 0.58--1.39 & 1.66\AA, 88 $\rm km\,s^{-1}$ & $0.75\times 2.4$, 0.78 & 11 & 17.6 \\ 
A2142 & SDSS, SDSS BOSS spec &     &      &                      & 1.19\AA, 62 $\rm km\,s^{-1}$ & 1.5, 1.0 & 140 & 18.4 \\
\enddata
\tablecomments{Column 1: Cluster name. 
Column 2: Telescope and spectrograph used for the observations. The Gemini North data were obtained under Gemini program IDs GN-2014A-Q-27 (MOS data) and
GN-2014A-Q-104 (Longslit data). The SDSS data were obtained from the SDSS data archive.
Column 3: Grating.
Column 4: Exposures times in seconds. 
Column 5: Image quality for observations at McDonald Observatory were measured as the full-width-half-maximum (FWHM) in arcsec of stars observed at the start of each night.
Image quality for observations at Gemini Observatory were measured as the FWHM in arcsec of 2-3 stars in the acquisition images.
Column 6: Median instrumental resolution derived as sigma in Gaussian fits to the sky lines of the stacked
LCS and GMOS-N spectra. For the SDSS spectra we list the resolution of the convolved spectra, see text.
The second entry is the equivalent resolution in $\rm km\,s^{-1}$ at 5175{\AA} in the rest frame of the cluster.
Column 7: Aperture size in arcsec. For the LCS and GMOS-N spectra, the first entry is the rectangular extraction aperture 
(slit width $\times$ extraction length). The second entry is the radius in an equivalent 
circular aperture, $r_{\rm ap}= 1.025 (\rm {length} \times \rm{width} / \pi)^{1/2}$, cf.\ J\o rgensen et al.\ (1995b).
For the SDSS spectra we list the radius of the SDSS spectrograph fibers ($1\farcs 5$) and the BOSS spectrograph fibers ($1\arcsec$).
For the Fiber Multi-Object Spectrograph (FMOS) spectra we list the radius of the spectrograph fibers.
Column 8: Number of member galaxies with data from this mode, including galaxies fainter than the adopted magnitude limits for the analysis.
Column 9: Median S/N per {\AA}ngstrom for the cluster members, in the rest frame of the clusters. For the GMOS-N and SDSS data derived in the wavelength
interval 4100--5250 {\AA}. For the LCS data 4100--4750 {\AA} and 4900--5400 {\AA} were used for grating \#43 and \#47, respectively.
}
\tablenotetext{a}{Information from J\o rgensen (1999).}
\end{deluxetable*}

\section{Spectroscopy \label{SEC-SPECTROSCOPY}}

\subsection{Data Sources \label{SEC-DATASOURCES} }
The spectroscopic data were assembled from four main sources.

\begin{enumerate}
\item
We previously published spectroscopic data for 116 Coma cluster members (J\o rgensen 1999). 
These data were assembled from our own observations with the Large Cassagrain Spectrograph (LCS) and 
the Fiber Multiple-Object Spectrograph (FMOS) on the McDonald Observatory 2.7-m telescope. 
The derived spectroscopic parameters were supplemented with literature data available at the time 
and calibrated to consistency. The reader is referred to J\o rgensen (1999) for full details on 
the observations, the processing and the calibration to consistency. 

\item
Observations of 61 Perseus cluster members were obtained with the LCS on the McDonald Observatory 2.7-m telescope.
These data were previously used as part of the reference sample in our GCP analysis papers 
(J\o rgensen et al.\ 2005, Barr et al.\ 2005, J\o rgensen \& Chiboucas 2013; J\o rgensen et al.\ 2014, 2017). 
However, the data have not previously been published.

\item
Observations of A2029 and A2142 cluster members were obtained with the 
Gemini Multi-object Spectrograph (GMOS-N, Hook et al.\ 2004) on Gemini North.
The observations cover 69 and 63 galaxies in A2029 and A2142, respectively.

\item
We use SDSS spectra for all four clusters and derive the relevant spectroscopic
parameters from these spectra. 
The data cover 336 member galaxies with no data from the other data sources, 222 of these 
are passive bulge-dominated galaxies brighter than our analysis limit of 
$M _{\rm B,abs} = -18.5$ mag (Vega magnitudes).
The SDSS spectra also provide all line indices blue-wards of H$\beta$ for the Coma cluster galaxies.

\end{enumerate}

We refer to the previous data for the Coma and Perseus cluster galaxies, items (1) and (2),
as the {\it Legacy Data}. In the following sections we describe the processing of the data
and the derivation of the spectral parameters. Appendix \ref{SEC-APPENDIX} contains 
additional information.

\subsection{Perseus Observations \label{SEC-PERSEUS}}

Observations of galaxies in Perseus were carried out with the LCS on the McDonald Observatory 2.7-m telescope 
in periods 1994 October 27 -- November 2, and 1995 October 25-30. Spectra were obtained in two configurations, 
see Table \ref{tab-inst}, such that the full wavelength coverage is 3500--5600 {\AA}.
The spectra obtained with grating \#47 had sufficient spectral resolution for determination
of velocity dispersions, as well as line indices. The spectra obtained with grating \#43 had
lower spectral resolution and were used for determination of line indices in the blue. 
The slit was aligned with the major axis of the galaxies.

The LCS spectroscopic observations were processed using the methods described in 
J\o rgensen (1999). The processing involved the standard steps of bias and dark subtraction, 
correction for scattered light, flat fielding, correction for the slit function,
wavelength calibration using argon lamp spectra, and sky subtraction. 
The spectra were cleaned for signal from cosmic-ray-events using the technique originally 
described in J\o rgensen et al.\ (1995b). 
Observations of the spectrophotometric standard stars BD+284211, Feige 110, and Hiltner 600 
were used to calibrate the spectra to a relative flux scale.
One-dimensional spectra were extracted with a resulting aperture size of $2\arcsec \times 6\farcs 35$, 
see Table \ref{tab-spdata}.

\subsection{Abell 2029 and Abell 2142 Observations \label{SEC-A2029A2142} }

Galaxies in Abell 2029 and Abell 2142 were observed with GMOS-N during semester 2014A, using either the
longslit (LS) or the multi-object spectroscopy (MOS) mode. The main purpose of the observations were 
to obtain deeper spectroscopy for faint cluster members without SDSS spectra, and to
provide sufficient overlap with brighter galaxies with SDSS spectra to facilitate consistent 
calibration of parameters derived from the two data sources.

Table \ref{tab-inst} summarizes the instrumention,
while Table \ref{tab-spdata} summarizes the observations.
All observations were obtained with the detector binned by two in both spatial and spectral direction.
Four MOS fields were observed in each cluster, covering 11--18 galaxies each.
Each longslit pointing covered two or three sample galaxies.

The data were processed in a standard fashion using tasks from the Gemini GMOS package. 
The processing includes bias subtraction, flat fielding, sky subtraction and wavelength calibration using
CuAr lamp spectra. 
Observations of the spectrophotometric standard star Wolf 1346 and HZ44 were used to calibrate
the spectra to a relative flux scale.
One-dimensional spectra were extracted with a resulting aperture size of 
$1\farcs 0\times 2\farcs 4$ and $0\farcs 75\times 2\farcs 4$ 
for the MOS and longslit observations, respectively, see Table \ref{tab-spdata}.

\subsection{SDSS spectra \label{SEC-SDSS} }

We primarily use SDSS spectra from DR14. As described in Section \ref{SEC-SAMPLES}, some areas of the Perseus cluster and Abell 2029 
are missing in DR14. For these we used spectra from DR10.
The majority of the spectra were obtained with the original SDSS spectrograph, while a few were obtained with
the Baryon Oscillation Spectroscopic Survey (BOSS) spectrograph. 
Table \ref{tab-spdata} summarizes the available data relevant for our samples.
The wavelength scale of the spectra were first transformed from vacuum wavelengths, $\lambda _{\rm vac}$, 
to air wavelengths, $\lambda _{\rm air}$ using the transformation provided on the SDSS DR13 web pages (cf.\ Morton 1991):
\begin{eqnarray}
\lambda _{\rm air} = \lambda _{\rm vac}  ( 1.0 + 2.735182\cdot 10^{-4} \nonumber \\ + 131.4182\, \lambda _{\rm vac}^{-2} 
+ 2.76249\cdot 10^8 \, \lambda _{\rm vac}^{-4}  ) ^{-1}
\end{eqnarray}
The spectra were then resampled onto a linear wavelength scale and convolved with a variable
kernel Gaussian to achieve spectra with a constant resolution. For use in determination of the 
velocity dispersions we use minimal convolution to achieve a resolution of 1.0786{\AA} in the rest frame
of the clusters. This matches the resolution of the single stellar population models from
Maraston \& Str\"{o}mb\"{a}ck (2011), which we use in the determination of the velocity dispersions
(see Section \ref{SEC-SPECPARAM}).

\subsection{Spectroscopic parameters \label{SEC-SPECPARAM} }

The spectroscopic parameters were determined using the same methods as described in J\o rgensen et al.\ (2005, 2017) 
and J\o rgensen \& Chiboucas (2013).
In particular, the redshifts and velocity dispersions were determined from the LCS, GMOS-N, and SDSS data
by fitting the galaxy spectra with template spectra, using software made available by Karl Gebhardt 
(Gebhardt et al.\ 2000, 2003). 

The kinematics fitting of LCS grating \#47 spectra obtained for the Perseus galaxies used only one 
template star, an observation of the K0III star HD52071, which was obtained
with the same instrumental configuration as the galaxies. 
The velocity dispersions determined from these fits were used in the GCP papers 
(J\o rgensen et al.\ 2005; Barr et al.\ 2005; J\o rgensen \& Chiboucas 2013; J\o rgensen et al.\ 2014; 
J\o rgensen et al.\ 2017).
The spectra were fit in the wavelength range 4900-5400 \AA\ in the rest frame.
All galaxies observed with the LCS are passive galaxies dominated by metal lines in this wavelength interval.
The fits have $\chi ^2 \approx 1$, see Table \ref{tab-perseuslcs} in Appendix \ref{SEC-SPECPARAMAPP}, showing that any template 
mismatch from using only one template spectrum for these fits is minimal.

For the GMOS-N data and SDSS data, the fits were limited to a wavelength range of 3750-5500 \AA\ as for 
our $z=0.2-0.5$ GCP spectra (cf.\ J\o rgensen et al.\ 2017). 
We use three single stellar population (SSP) models from Maraston \& Str\"{o}mb\"{a}ck (2011) as template spectra. 
The models have (age, Z) = (1\,Gyr, 0.01), (5\,Gyr, 0.02), and (15\,Gyr, 0.04) and a Salpeter (1955) initial mass function (IMF).
The choice of IMF is not critical for the fits as the lines in the fitted wavelength range are not sensitive to the IMF.
These three models adequately span the spectral types of the galaxies, and lead to velocity dispersions unaffected by template mismatch. 
This is similar to the situation for our $z=0.2-0.9$ GCP data, for which we used three template stars spanning similar
spectral properties, see J\o rgensen \& Chiboucas (2013) for discussion.
The Maraston \& Str\"{o}mb\"{a}ck SSP models are based on the MILES library 
(Medium-resolution Isaac Newton Telescope Library of Empirical Spectra, Vazdekis et al.\ 2010) and we adopt a spectral resolution of 
$\sigma = 1.0786$ {\AA} as found by Maraston \& Str\"{o}mb\"{a}ck. 

Absorption line indices were determined using the Lick/IDS passband definitions (Worthey et al.\ 1994).
In addition, we derive the indices CN3883 and CaHK (Davidge \& Clark 1994), D4000 (Bruzual 1983), 
the higher order Balmer line indices H$\delta _{\rm A}$ and H$\gamma _{\rm A}$ (Worthey \& Ottaviani 1997), the H$\beta _{\rm G}$ index 
defined by J\o rgensen (1997), and the high order Balmer line index H$\zeta _{\rm A}$ (Nantais et al.\ 2013).
In all cases, we first convolve the spectra to the Lick/IDS resolution in the rest frame of the galaxies, cf.\ Worthey \& Ottaviani (1997).

At the redshift of A2029, the 5577 {\AA} sky line falls within the passband for the Mg$b$ index. 
The SDSS spectra have very strong residuals from inadequate subtraction of this skyline. In order to obtain reliable
Mg$b$ indices for as many of the A2029 galaxies as possible we first interpolate across the residuals from the 5577 {\AA} sky line.
We evaluate the effect of this on the resulting measurements by interpolating the SDSS spectra of Coma cluster galaxies 
in the same way, affecting the same wavelengths in the rest frame as affected in the A2029 spectra. 
We then compare Mg$b$ measured from the interpolated galaxy spectra with the values from the original Coma spectra. 
As expected, the Mg$b$ measurements are strongly affected if the interpolation is done across the center of the strongest 
of the magnesium triplet lines, weakening the measurement by more than 0.2 dex. However, only eight
out of the 118 A2029 members with SDSS spectra are affected to this extent. For these we choose not to measure
Mg$b$. For the remainder of the galaxies, our test shows that the interpolation contributes about 0.03 to the 
uncertainty on $\log {\rm Mg}b$, which is only half of the typical internal uncertainties.

\begin{figure*}
\epsfxsize 14cm
\begin{center}
\epsfbox{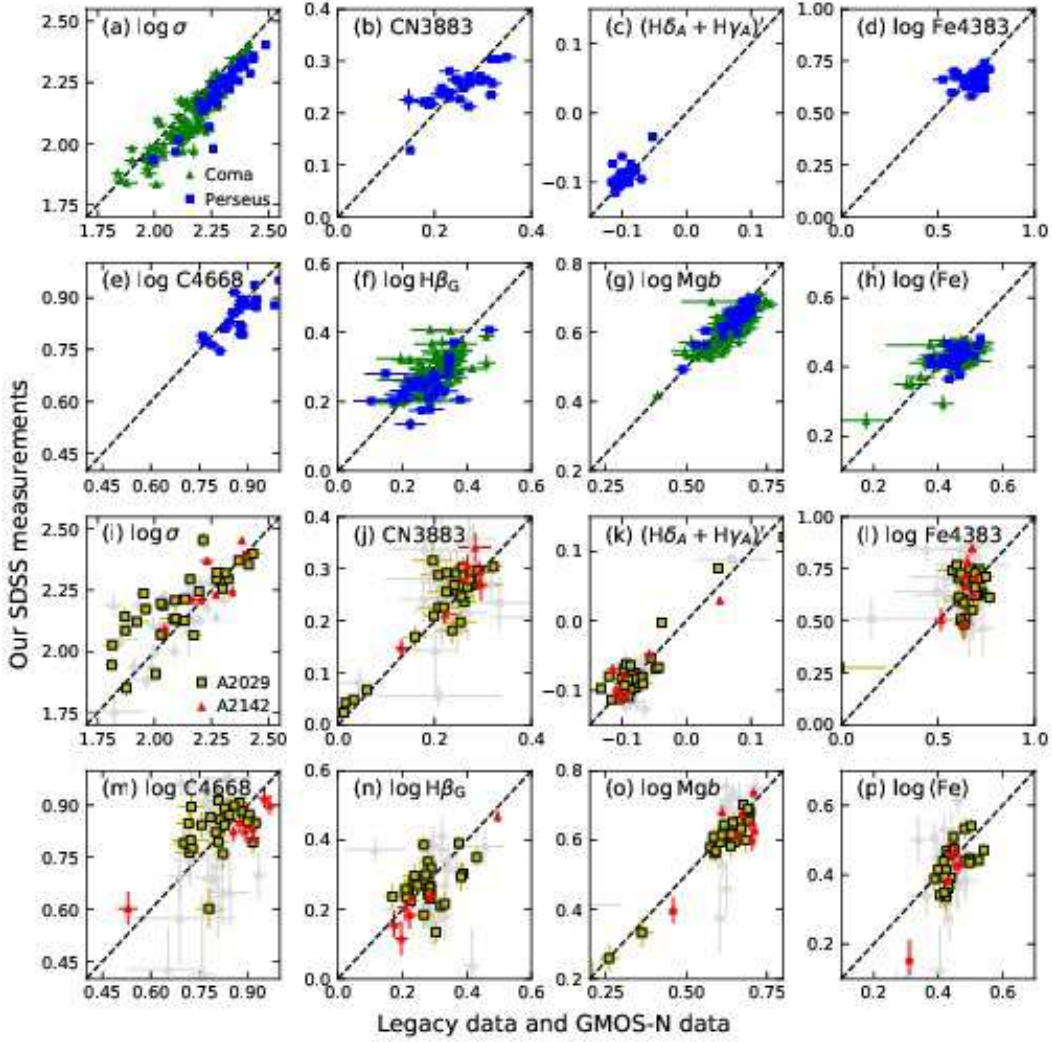}
\end{center}
\caption{Comparison of previous data with measurements from the SDSS spectra.
Panels (a)--(h) show the {\it Legacy Data} for Perseus and Coma. Blue squares -- Perseus; green triangles -- Coma.
The {\it Legacy Data} are on the X-axis, our SDSS based determinations on the Y-axis.
Panels (i)--(p) show the GMOS-N data for Abell 2029 and Abell 2142. Yellow squares -- A2029; red triangles -- A2142. 
Grey points -- data from spectra with S/N $< 20 {\rm \AA}^{-1}$.
Dashed lines mark the one-to-one relations.
The GMOS-N data are on the X-axis, our SDSS based determinations on the Y-axis.
Only comparisons for the velocity dispersions and the line indices used in the analysis are included on the figure.
Table \ref{tab-speccomp} summarizes all the comparisons.
\label{fig-comp_allprevdata} }
\end{figure*}

The line indices were corrected to zero velocity dispersion, see J\o rgensen et al.\ (2005) for details
and typical sizes of these corrections. 
The velocity dispersions and the absorption line indices indices were then aperture corrected
to a standard aperture diameter of 3.4 arcsec at the distance of the Coma cluster.
Correction coefficients are listed in J\o rgensen et al.\ (2005), except for H$\zeta _{\rm A}$ for
which we adopt a zero correction (cf.\ J\o rgensen et al.\ 2014).
The aperture sizes for the various data used are listed in Table \ref{tab-spdata}.
Velocity dispersions were also corrected for systematic effects based on simulations, see Appendix \ref{SEC-SIM}.

Duplicate GMOS-N observations of A2029 and A2142 galaxies and duplicate SDSS observations of galaxies in 
the Perseus and Coma clusters were used to assess the uncertainties on the derived parameters.
The details are provided in Appendix \ref{SEC-COMPREPEAT}.
In general, the Monte Carlo simulations used in the kinematics fitting to obtain uncertainty estimates give
reliable estimates. The uncertainties on the line indices are in general larger than estimated from
the S/N of the spectra. 
Tables \ref{tab-repeat} and \ref{tab-uncadopt} in Appendix \ref{SEC-COMPREPEAT} list the scaling 
factors for the uncertainties and adopted typical uncertainties on the final measurements.
Measurements from repeat observations with GMOS-N or repeat SDSS spectra were averaged. Only the average
values are used in the following.

The spectra were inspected for emission in [\ion{O}{2}], [\ion{O}{3}], and/or H$\beta$, and significant 
emission was noted. Measurements of emission line equivalent widths are available through the Portsmouth group's work
(Thomas et al.\ 2013) and were not repeated here. We use our emission line flags to omit 
galaxies with strong emission lines from the analysis. Our flags correspond to
equivalent widths of approximately 5 {\AA} and 2 {\AA} for [\ion{O}{2}] and H$\beta$, respectively.

Tables \ref{tab-perseuslcs}-\ref{tab-sdss} in Appendix \ref{SEC-SPECPARAMAPP} list the results from the 
template fitting and the measured absorption line indices, for each of the data sets. 

\begin{deluxetable}{l rrr rrr}
\caption{Comparisons of {\it Legacy}, GMOS-N, and SDSS Data \label{tab-speccomp} }
\tabletypesize{\scriptsize}
\tablecolumns{7}
\tablehead{ 
& \multicolumn{3}{c}{\it Legacy Data} & \multicolumn{3}{c}{GMOS-N Data}\\
\colhead{Parameter} & \colhead{$N$} & \colhead{$\Delta$} & \colhead{rms} & \colhead{$N$} & \colhead{$\Delta$} & \colhead{rms} \\
\colhead{(1)} & \colhead{(2)} & \colhead{(3)} & \colhead{(4)} & \colhead{(5)} & \colhead{(6)} & \colhead{(7)}
}
\startdata
Redshift      & 147 & 0.00002 & 0.0007     & 62 & 0.00008 & 0.00009 \\
$\log \sigma$ & 147 & {\it 0.056} & 0.052  & 62 & {\it --0.044} & 0.117 \\
CN3883        & 27  & 0.005& 0.038         & 34 & --0.012 & 0.035  \\
$\log {\rm H}\zeta _{\rm A}$ & 18 & --0.048 & 0.197 & 33 & --0.032 & 0.285 \\
log CaHK      & 27  & {\it --0.017} & 0.027  & 38 & {\it 0.016} & 0.042 \\
D4000         & 26  & -0.026 & 0.108        & 34 & --0.013 & 0.113 \\
H$\delta _{\rm A}$ & 27 & --0.110 &  0.980   & 38 & --0.349 & 1.326 \\
H$\gamma _{\rm A}$ & 27 & --0.243 &  0.543   & 38 & 0.253 & 0.833 \\
$({\rm H}\delta _{\rm A} + {\rm H}\gamma _{\rm A})'$        & 27 & --0.002 &  0.014        & 38 & --0.001 & 0.020 \\
CN$_2$        & 27 & {\it 0.013} & 0.017    & 38 & {\it --0.008} & 0.028 \\
log G4300     & 27 &{ \it 0.017} & 0.043          & 36 & {\it --0.017} & 0.078 \\
log Fe4383    & 27 & 0.016 & 0.062          & 37 & 0.017 & 0.152 \\
log C4668     & 27 & {\it 0.016} & 0.038          & 38 & {\it --0.006} & 0.169 \\
$\log {\rm H}\beta _{\rm G}$ & 123  & {\it 0.028} & 0.053 & 35 & {\it 0.018} & 0.061 \\
$\log {\rm H}\beta$ & 34  & {\it 0.036} & 0.091 & 35 & {\it 0.020} & 0.088 \\
log Mg$b$     & 146 & {\it 0.021}  & 0.031  & 38 & {\it 0.017} & 0.037 \\
log Fe5270    & 103 & {\it 0.006}  & 0.044  & 35 & {\it 0.017} & 0.180 \\
log Fe5335    & 103 & {\it 0.038}  & 0.060  & 34 & {\it 0.033} & 0.090 \\
$\log \langle {\rm Fe} \rangle$        & 103 & {\it 0.024}  & 0.041  & 34 & {\it 0.023} & 0.076 \\
\enddata
\tablecomments{Column 1: Spectroscopic parameter. 
Column 2: Number of galaxies included in the comparisons between the {\it Legacy Data} and our SDSS measurements.  
Column 3: Median of differences derived as ``{\it Legacy Data}'' -- ``SDSS measurements''. 
Offsets in italics are applied to reach consistent calibration, see text.
Column 4: Scatter of the comparison between the {\it Legacy Data} and the SDSS measurements .
Columns 5--7: Same information for the comparisons between the GMOS-N data and the SDSS measurements. 
Differences are derived as ``GMOS-N data'' -- ``SDSS measurements''.
Comparisons for redshift and $\log \sigma$ include data for which the SDSS S/N $\ge 10 {\rm \AA}^{-1}$. 
Comparisons for line indices include data for which the SDSS S/N $\ge 20 {\rm \AA}^{-1}$.}
\end{deluxetable}

\subsection{Calibration of the Spectroscopic Parameters \label{SEC-SPECCALIB} }

The main purpose of this section is to establish consistently calibrated velocity dispersions
and absorption line indices, such that the samples can reliably be used as the low redshift
reference samples for studies of higher redshift galaxies. As described in Section \ref{SEC-INTRO}, 
our previous Coma cluster data (J\o rgensen et al.\ 1995ab, 1996; J\o rgensen 1999) 
have been used widely in the literature as the low redshift reference sample.
We have used the {\it Legacy Data} for the Coma and Perseus clusters
in our previous analysis of $z=0.2-1.3$ cluster galaxies as part of the GCP
(J\o rgensen et al.\ 2005, 2014, 2017; J\o rgensen \& Chiboucas 2013).
We know from our previous work with these data, that the two samples are consistent, as will also
be confirmed in the following. 
For these reasons, we aim to calibrate all other data to consistency with the {\it Legacy Data}
and proceed as follows:
\begin{enumerate}
\item
We first establish offsets between our measurements from the SDSS spectra and the {\it Legacy Data}.
\item
We then establish offsets between our measurements from the SDSS spectra and the GMOS-N data for A2029 and A2142.
\item
We apply the adopted offsets to our measurements from the SDSS spectra and to measurements from the GMOS-N spectra, such
that all individual measurements are consistent with the {\it Legacy Data}.
\item
We then average available measurements for each galaxy, to obtain the best average parameters consistent with the {\it Legacy Data}.
\end{enumerate}

\begin{deluxetable*}{l rrr rrr rrr rrr rrr}
\caption{Comparisons of Final Data with SDSS Data \label{tab-comp_portmpajhu} }
\tabletypesize{\scriptsize}
\tablehead{ 
& \multicolumn{3}{c}{All clusters\tablenotemark{a}} & \multicolumn{3}{c}{Perseus}
& \multicolumn{3}{c}{Coma} & \multicolumn{3}{c}{A2029} & \multicolumn{3}{c}{A2142} \\
\colhead{Parameter} & \colhead{$N$} & \colhead{$\Delta$} & \colhead{rms} & \colhead{$N$} & \colhead{$\Delta$} & \colhead{rms}
& \colhead{$N$} & \colhead{$\Delta$} & \colhead{rms} & \colhead{$N$} & \colhead{$\Delta$} & \colhead{rms} & \colhead{$N$} & \colhead{$\Delta$} & \colhead{rms} \\
\colhead{(1)} & \colhead{(2)} & \colhead{(3)} & \colhead{(4)} & \colhead{(5)} & \colhead{(6)} & \colhead{(7)} & \colhead{(8)} 
& \colhead{(9)} & \colhead{(10)} & \colhead{(11)} & \colhead{(12)} & \colhead{(13)} & \colhead{(14)} & \colhead{(15)} & \colhead{(16)}
}
\startdata
Redshift & 457 & -0.00002 & 0.0002 & 119 & -0.00003 & 0.0004 & 179 & -0.00002 & 0.0001 & 107 & -0.00003 & 0.0001 & 52 & -0.00002 & 0.0001 \\
$\log \sigma$ & 456 & 0.014 & 0.139 & 119 & 0.013 & 0.104 & 178 & 0.022 & 0.165 & 107 & 0.005 & 0.146 & 52 & 0.010 & 0.064 \\
D4000 & 353 & {\it 0.018} & 0.033 & 84 & 0.074 & 0.030 & 119 & 0.002 & 0.002 & 99 & 0.018 & 0.026 & 51 & 0.022 & 0.024 \\
$({\rm H}\delta _{\rm A} + {\rm H}\gamma _{\rm A})'$ & 376 & 0.003 & 0.017 & 98 & 0.003 & 0.012 & 128 & 0.003 & 0.009 & 99 & 0.002 & 0.015 & 51 & 0.001 & 0.034 \\
CN$_2$ & 376 & {\it -0.014} & 0.023 & 98 & -0.015 & 0.020 & 128 & -0.016 & 0.025 & 99 & -0.013 & 0.026 & 51 & -0.007 & 0.017 \\
log G4300 & 372 & -0.013 & 0.111 & 98 & -0.016 & 0.065 & 126 & -0.014 & 0.044 & 97 & -0.012 & 0.197 & 51 & 0.001 & 0.057 \\
log Fe4383 & 375 & {\it -0.065} & 0.044 & 97 & -0.066 & 0.025 & 128 & -0.067 & 0.052 & 99 & -0.065 & 0.047 & 51 & -0.055 & 0.041 \\
log C4668 & 374 & 0.004 & 0.036 & 98 & 0.005 & 0.054 & 126 & 0.003 & 0.015 & 99 & 0.004 & 0.037 & 51 & 0.009 & 0.028 \\
$\log {\rm H}\beta _{\rm G}$ & 350 & {\it 0.018} & 0.049 & 91 & 0.020 & 0.035 & 118 & 0.015 & 0.041 & 92 & 0.017 & 0.068 & 49 & 0.023 & 0.048 \\
log Mg$b$ & 371 & 0.010 & 0.058 & 98 & 0.009 & 0.010 & 128 & 0.010 & 0.018 & 95 & -0.004 & 0.109 & 50 & 0.023 & 0.029 \\
$\log \left < {\rm Fe} \right >$ & 375 & -0.008 & 0.035 & 98 & -0.016 & 0.028 & 127 & -0.016 & 0.029 & 99 & -0.007 & 0.042 & 51 & 0.028 & 0.035 \\
\enddata
\tablecomments{Column 1: Spectral parameter. Column 2: Total number of measurements in comparison. 
Column 3: Median difference, all differences are derived as ``This paper''--``SDSS data'', where SDSS data refer to redshifts from DR14, velocity dispersions
from the Portsmouth group (Thomas et al.\ 2013), and line indices from the MPA-JHU group. Differences significant at the $5\sigma$ level or higher
are shown in italics. Column 4: Scatter of the comparison.
Columns 5--7: Number of measurements, median difference, and scatter of comparison for Perseus galaxies.
Columns 8--10: Number of measurements, median difference, and scatter of comparison for Coma galaxies.
Columns 11--13: Number of measurements, median difference, and scatter of comparison for A2029 galaxies.
Columns 14--16: Number of measurements, median difference, and scatter of comparison for A2142 galaxies.}
\end{deluxetable*}

Figure \ref{fig-comp_allprevdata} and Table \ref{tab-speccomp} summarize the comparisons for items (1) and (2).
For the redshift and velocity dispersion comparisons, only observations with S/N $\ge 10 {\rm \AA}^{-1}$ were included,
while for the line index comparisons we required S/N $\ge 20 {\rm \AA}^{-1}$.
We estimate the uncertainties on the median offsets as $\sigma _{\Delta} \approx {\rm rms}\, N^{-0.5}$,
where rms is the scatter of the comparisons, and $N$ is the number of measurements in the comparisons.
We apply offset to those parameters where at least one of the comparisons have an offset with 
3$\sigma$ or larger significance. 
Those offsets are shown in italics in Table \ref{tab-speccomp}.
The apparently large and disparate offsets for H$\delta _{\rm A}$ and H$\gamma _{\rm A}$ are
not statistically significant, and only amounts to $\approx 5$\% of the range of these indices for low redshift passive galaxies.
The comparisons also showed that offsets between the velocity dispersions determined from the SDSS spectra and 
the {\it Legacy Data} for Coma and Perseus are the same within the uncertainties for the two clusters, 
confirming the consistency of these two data sets of {\it Legacy Data}.

Final average measurements are listed in Table \ref{tab-all} in Appendix \ref{SEC-SPECPARAMAPP}.
This table also lists the effective S/N per {\AA}ngstrom in the rest frame of the galaxies.
Where two measurements were averaged, we list S/N values added in quadrature. 
Complete S/N information for the Coma {\it Legacy Data} does not exist as part of the data
were compiled from the literature without such information (J\o rgensen 1999). 
We adopted the median S/N $= 28\rm \AA ^{-1}$ for our own data
as the typical for all Coma {\it Legacy Data}.

In the analysis we use the Balmer line indices H$\beta _{\rm G}$ and
$({\rm H\delta _A + H\gamma _A})' \equiv -2.5~\log \left ( 1.-({\rm H\delta _A + H\gamma _A})/(43.75+38.75) \right )$
(Kuntschner 2000), the
iron indices Fe4383 and $\rm \langle Fe \rangle \equiv (Fe5270+Fe5335)/2$, and the indices Mg$b$, CN3883 and C4668.
These indices can generally be reliably measured from spectra with S/N$\ge 15 {\rm \AA}^{-1}$, and are therefore
realistic to use for studies of $z \ge 0.2$ galaxies. 
According to stellar population models (e.g., Thomas et al.\ 2011), the indices are also sufficient to 
derive ages, [M/H] and abundance ratios, making studies of the variation and evolution of these parameters possible. 
Calibrated values for other Lick/IDS indices are included Table \ref{tab-all}, but not used in the present analysis.

\begin{figure*}
\epsfxsize 14cm
\begin{center}
\epsfbox{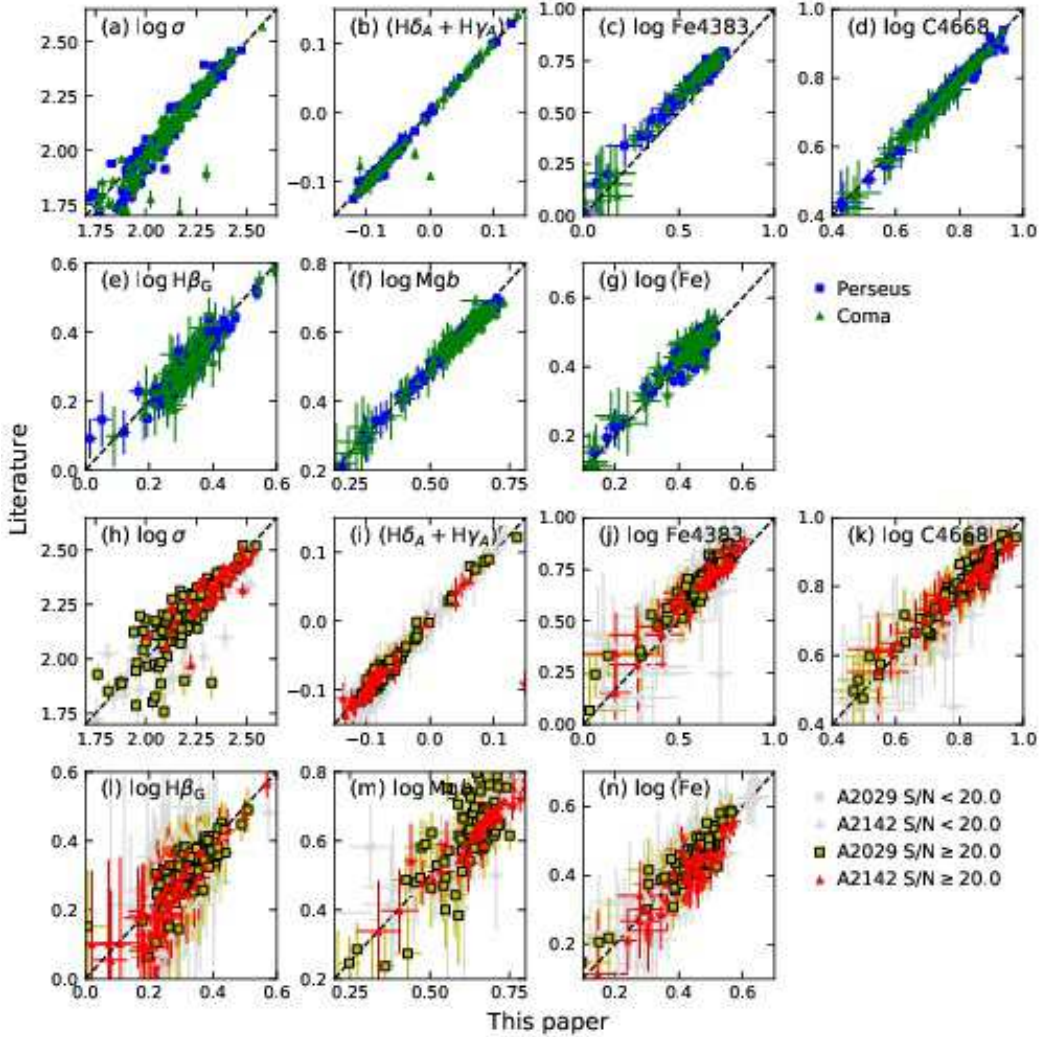}
\end{center}
\caption{Comparison of our final average measurements for Perseus and Coma with 
(a) the Portsmouth group's velocity dispersion measurements based on the SDSS spectra, and (b)--(g) line indices from
the MPA-JHU group based on SDSS spectra. 
Blue squares -- Perseus; green triangles -- Coma.
Comparison of our final average measurements for A2029 and A2142 with 
(h) the Portsmouth group's velocity dispersion measurements based on the SDSS spectra, and (i)--(n) Line indices from
the MPA-JHU group based on SDSS spectra. 
Yellow squares -- A2029; red triangles -- A2142. Grey symbols -- measurements from spectra with S/N $< 20 {\rm \AA}^{-1}$.
Dashed lines mark the one-to-one relations.
All panels have our measurements on the X-axis, and the Portsmouth or MPA-JHU data on the Y-axis.
Only comparisons for the velocity dispersions and the line indices used in the analysis are included on the figure.
Table \ref{tab-comp_portmpajhu} summarizes all the comparisons.
\label{fig-comp_portmpajhu} }
\end{figure*}

\subsection{Comparison with SDSS, Portsmouth and MPA-JHU measurements}

We compare the final calibrated measurements with (1) redshifts from SDSS DR14, (2) the Portsmouth group's measurements 
of velocity dispersions (Thomas et al.\ 2013) and (3) the Max Planck Institute for Astrophysics and 
the Johns Hopkins University (MPA-JHU) group's measurements of line indices. 
The methods for the various measurements from the MPA-JHU group are described in 
Brinchmann et al.\ (2004), Kauffmann et al.\ (2003), and Tremonti et al.\ (2004). 
All measurements are available through the SDSS DR14, though the Portsmouth group used DR12 and the MPA-JHU group's
methods were last run on DR8 spectra. 
The Portsmouth velocity dispersions were not aperture corrected (D.\ Thomas, personal communication). Thus, we first
aperture correct these measurements to our standard size aperture.
It is not clear if the MPA-JHU measurements were aperture corrected, or corrected to zero velocity dispersion. 
Since it is stated on the SDSS DR14 web pages that the measurements are on the Lick system, 
we assume they have been corrected to zero velocity dispersion.
We apply our aperture correction to the MPA-JHU measurements before comparing them to our fully calibrated measurements.
The comparisons are summarized in Figure \ref{fig-comp_portmpajhu} and Table \ref{tab-comp_portmpajhu}. 

As expected, the redshift measurements are in agreement with SDSS DR14 values. 
The comparisons of the velocity dispersions show offsets of 0.005--0.022 in $\log \sigma$ for each of the clusters. 
These are all within our previous estimates of internal consistency of 0.026 (J\o rgensen et al.\ 2005). 
Assuming that the data from Thomas et al.\ are internally consistent, these comparisons confirm that
the final measurements of $\log \sigma$ are consistent between the four clusters.
We note that small offsets between measurements performed using different techniques are not unusual. 
Thomas et al.\ performed more extensive comparisons with previous velocity dispersion measurements based on
SDSS spectra, and found offsets of 4--7 percent, with scatter of 16--19 percent.

The comparisons of the line indices show significant offsets at the 5$\sigma$ or larger level 
only for D4000, CN$_2$, $\log {\rm Fe4383}$, and $\log {\rm H}\beta _{\rm G}$ (shown in italics in Table \ref{tab-comp_portmpajhu}). 
In absolute terms, these offsets are quite small. They may originate from small differences in the flux calibration
applied by SDSS in the earlier data releases compared to DR14. 
As can be seen from the error bars on the figures, the scatter in all the comparisons are within expectations
based on the uncertainties, with exception of the comparison of Mg$b$ for the A2029 galaxies.
The larger scatter in this comparison most likely originates from our corrections for 
the 5577 {\AA} sky line residuals in those spectra, as we assume such a correction was not done by the MPA-JHU group.

\section{The Galaxy Samples and Methods \label{SEC-SAMPLEMETHOD} }

Our aim is to characterize the stellar populations in the passive bulge-dominated cluster galaxies by 
(1) establishing the scaling relations between the velocity dispersions and
the absorption line indices, and 
(2) establishing the relations between the velocity dispersions and the ages, metallicities,
and abundance ratios of the galaxies.
To achieve the latter, we use the individual measurements for Perseus and Coma galaxies from
spectra with S/N $\ge 50 {\rm \AA}^{-1}$, as well as luminosity weighted averages of the absorption line indices and
the velocity dispersions for all four clusters.

In this section, we establish the samples of passive bulge-dominated cluster galaxies 
and the completeness of these samples in each cluster, and
describe the determination of the luminosity weighted parameters. 
We briefly recap the adopted methods and stellar population models, all consistent with our approach in our 
recent GCP publications for $z=0.2-0.9$ clusters (J\o rgensen \& Chiboucas 2013; J\o rgensen et al.\  2017).

\subsection{The Samples}

We base our selection of the passive bulge-dominated galaxies on a combination of
colors in $(g'-r')$ and the modeling parameters {\tt fracdev} from SDSS.
This parameter is the fraction of the luminosity modeled by the $r^{1/4}$-profile. 
In practice, the measurements of {\tt fracdev} are quite noisy, especially for galaxies at the 
distances of A2029 and A2142, where comparison of {\tt fracdev} in the $g'$, $r'$ and $i'$ bands
indicate an uncertainty of about 0.07. 
The measurements for galaxies in Perseus and Coma have uncertainties of about 0.03.
Using similar comparisons, we estimate that the uncertainties on {\tt fracdev} in $u'$ and $z'$ bands
are approximately a factor four and two larger, respectively.
To identify bulge-dominated galaxies, we use the product of {\tt fracdev} in the $g'$, $r'$ and $i'$ bands,
$f_{g} \times f_{r} \times f_{i}$.

\begin{figure}
\epsfxsize 8.5cm
\epsfbox{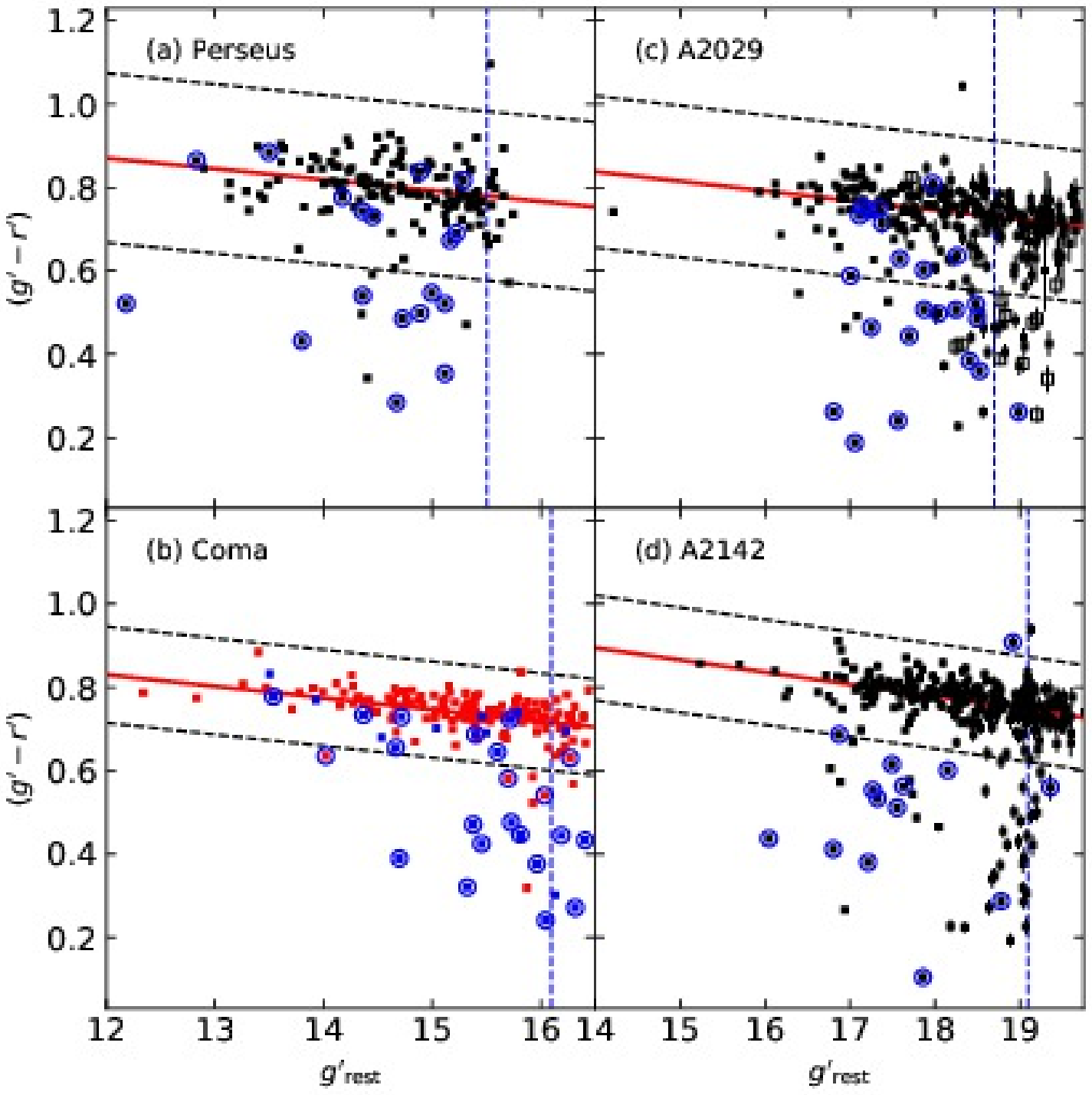}
\caption{The color-magnitude diagrams for the four clusters shown as rest frame $(g'-r')$ versus $g'_{\rm rest}$.
Solid black squares -- known members of Perseus, A2029, and A2142. 
Open black squares (panels c and d) -- galaxies in A2029 and A2142 with photometric redshifts that indicate possible membership of the clusters.
Red solid squares on panel (b) -- bulge-dominated members of Coma, classifications from Dressler (1980).
Blue solid squares on panel (b) -- disk-dominated members of Coma, classifications from Dressler (1980).
Blue circles -- emission line galaxies. 
Solid red lines -- best fits to the red sequences, Table \ref{tab-CM}.
Dashed black lines -- mark the location of $\pm 3.5$ the scatter relative to the best fit red sequence.
Vertical dashed blue lines -- adopted magnitude limit for the main samples, corresponding to 
$M _{\rm B,abs} = -18.5$ mag (Vega magnitudes).
\label{fig-CM} }
\end{figure}

\begin{figure}
\epsfxsize 8.5cm
\epsfbox{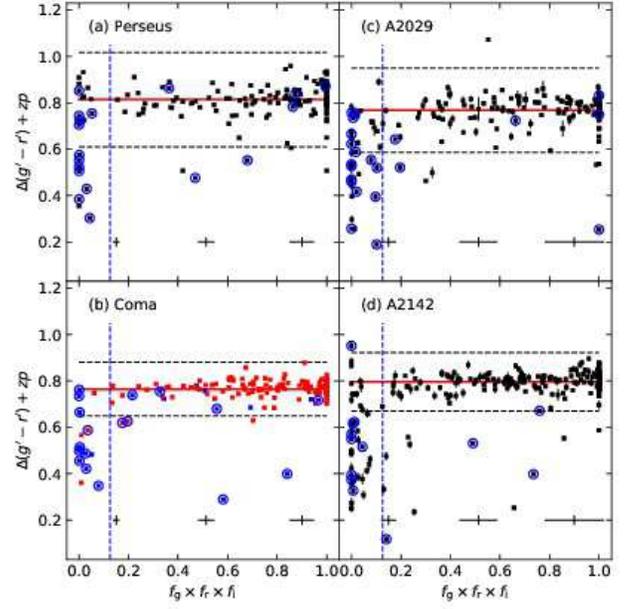}
\caption{The color residuals $\Delta (g'-r')$, offset to the zero points listed in Table \ref{tab-CM}, versus $f_{g} \times f_{r} \times f_{i}$. 
Symbols and lines as in Figure \ref{fig-CM}, except the vertical dashed blue lines mark the adopted limit of $f_{g} \times f_{r} \times f_{i} = 0.125$
between the disk- and bulge-dominated galaxies, see text.
\label{fig-classes} }
\end{figure}

We establish the best fit to the red sequence in the $g'_{\rm rest}$ versus $(g'-r')$ color-magnitude diagrams.
The relations for the four clusters are shown in Figure \ref{fig-CM} and summarized in Table \ref{tab-CM}.
The total $g'_{\rm rest}$ magnitude is {\tt cmodelmag} in the $g'$ band from SDSS, corrected for Galactic
Extinction and $k$-corrected to the rest frame of the galaxies. The $k$-correction was done using
the calibration from Chilingarian et al.\ (2010) available through their web interface.
The colors are based on {\tt modelmag} from SDSS, and have also been corrected to the rest frames. 
All four clusters have tight red sequences, though the scatter of the Perseus and A2029 galaxies is
almost a factor two larger than found for Coma and A2142. 
In Figure \ref{fig-classes}, we show the residuals $\Delta (g'-r')$ relative to the red sequence fits,
offset to the zero points listed in Table \ref{tab-CM}, versus $f_{g} \times f_{r} \times f_{i}$.
To identify the galaxies that can be considered passive and bulge-dominated we proceed as follows.
We consider galaxies with $f_{g} \times f_{r} \times f_{i} < 0.125$ disk dominated, corresponding to
the exponential disk contributing at least 50\% of the flux in all three filters. 
Galaxies with $(g'-r')$ more than 3.5$\sigma$ below the red sequence are also considered disk dominated, independent
of the value of $f_{g} \times f_{r} \times f_{i}$. 
The remainder are considered bulge-dominated.
We then visually inspected images of all the galaxies using the SDSS interface to false-color images of
the galaxies. This resulted in 38 galaxies (4\% of the parent sample) being re-classed from bulge-dominated to disk-dominated,
or visa-versa, see Table \ref{tab-classes}.

As is common for studies of intermediate and high redshift galaxies, our sample selection in
J\o rgensen et al.\ (2017) relied on the S\'{e}rsic (1968) index, adopting $n_{\rm ser} \ge 1.5$ as bulge-dominated galaxies.
Profile fitting of the galaxies in the present low redshift reference sample will be covered in our companion
photometry paper (J\o rgensen et al., in prep.).
However, for the $\approx 170$ Coma and Perseus galaxies for which we have completed the fitting, the 
preliminary results show disagreements between the two methods for only four galaxies for which 
$f_{g} \times f_{r} \times f_{i} < 0.125$ and $n_{\rm ser} \ge 1.5$. 
All four of these were re-classed as bulge-dominated as a result of our visual inspection of the SDSS images. 
Additional discussion of the consistency of the sample selection methods will be included in the photometry paper.

\begin{deluxetable*}{l rrr}
\caption{Color-Magnitude Relations \label{tab-CM} }
\tabletypesize{\scriptsize}
\tablehead{ 
\colhead{Cluster} & \colhead{Relation} & \colhead{$N$} & \colhead{rms} \\
\colhead{(1)} & \colhead{(2)} & \colhead{(3)} & \colhead{(4)}
}
\startdata
Perseus & $(g'-r') = (-0.026 \pm 0.007) (g'-14.0) + (0.815 \pm 0.007)$ & 148 & 0.058  \\
Coma    & $(g'-r') = (-0.028 \pm 0.003) (g'-14.6) + (0.765 \pm 0.003)$ & 162 & 0.033 \\
A2029   & $(g'-r') = (-0.023 \pm 0.004) (g'-17.2) + (0.769 \pm 0.005)$ & 240 & 0.059  \\
A2142   & $(g'-r') = (-0.029 \pm 0.003) (g'-17.6) + (0.796 \pm 0.003)$ & 259 & 0.036  \\
\enddata
\tablecomments{Column 1: Cluster. 
Column 2: Best fit relation. 
Column 3: Number of galaxies included in the fit.
Column 4: Scatter of the fit.
}
\end{deluxetable*}

\begin{deluxetable*}{l rr}
\caption{Modifications to Classifications\label{tab-classes} }
\tabletypesize{\scriptsize}
\tablehead{ 
\colhead{Cluster} & \colhead{From disk- to bulge-dominated} & \colhead{From bulge- to disk-dominated}  \\
\colhead{(1)} & \colhead{(2)} & \colhead{(3)} 
}
\startdata
Perseus & 12253, 12290, 12474, 12193, 12287, 12434 & 
          12081, 12497, 12537, 12627, 12392, 12780, 2197137, 2185837, 2174899, 12119   \\
Coma    & 2940, 4679, 3664, 4156 & 
          2355, 2374, 2431, 2441, 3238, 4522, 4597, 4933, 5038 \\
A2029   & 2555, 968, 189 & 4252  \\
A2142   & 447, 1514, 2661, 326, 2788 & \nodata  \\
\enddata
\tablecomments{Column 1: Cluster. 
Column 2: IDs for galaxies reclassified from disk- to bulge-dominated. 
Column 3: IDs for galaxies reclassified from bulge- to disk-dominated.
}
\end{deluxetable*}

\begin{figure}
\epsfxsize 8.5cm
\epsfbox{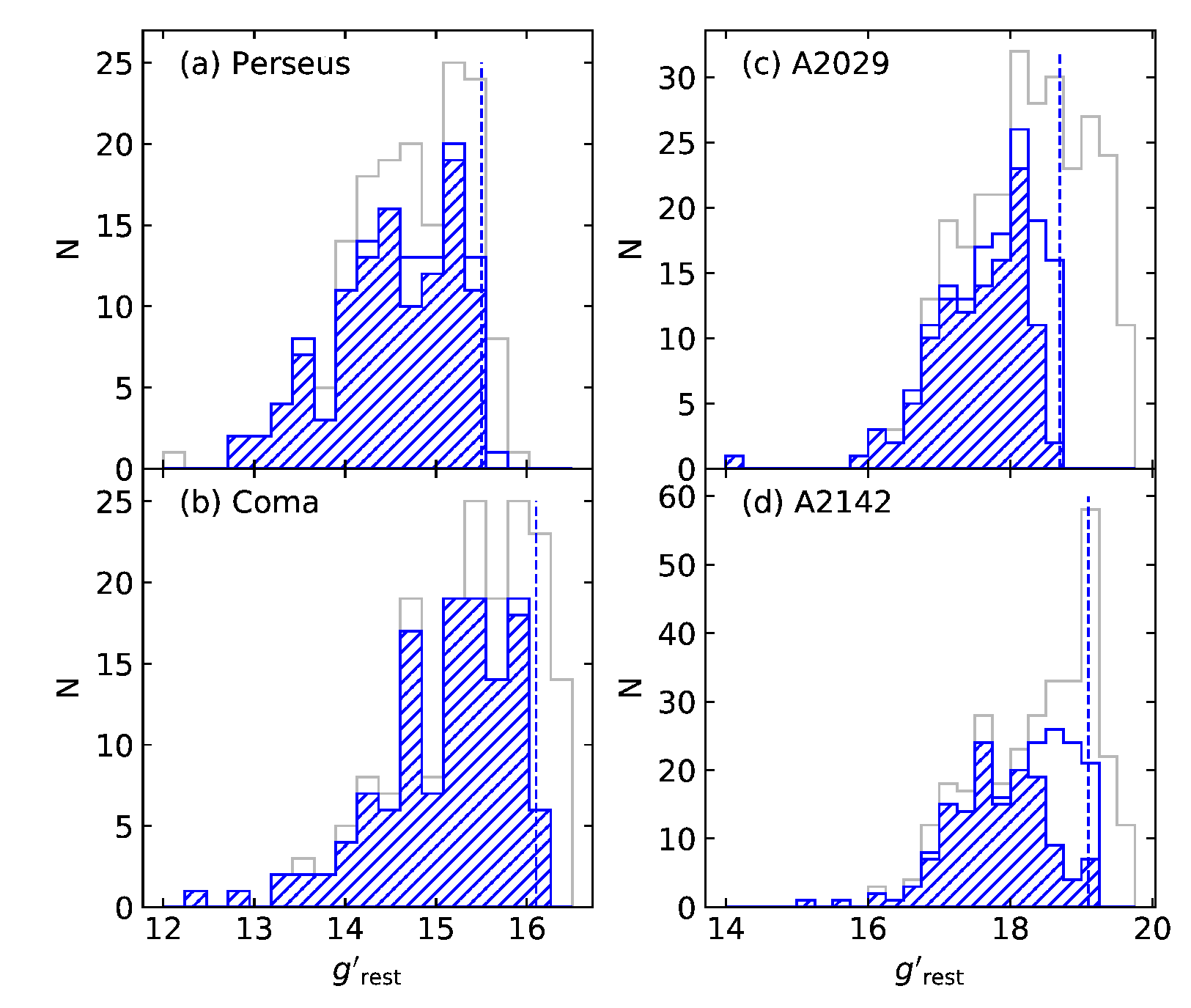}
\caption{Distributions of $g'_{\rm rest}$ illustrating the completeness of the samples. 
Blue open -- bulge-dominated galaxies on the red sequence brighter than the adopted sample limits.
Blue hatched -- bulge-dominated galaxies on the red sequence brighter than the adopted sample limits and with spectroscopy.
For reference the grey histograms show all members in the parent samples. The parent samples have been
limited at magnitudes just below the adopted limits for the final samples, see Section \ref{SEC-SAMPLES}.
\label{fig-maghist} }
\end{figure}

Finally we divide the samples according to the target limiting magnitude, and divide the 
bulge-dominated sample into passive and emission line galaxies.
Table \ref{tab-sub-samples} summarizes the number of galaxies in each sub-sample. Our main sample for the analysis
consists of the passive bulge-dominated galaxies, 
the number of which is listed in column $N_{\rm passive}$ in Table \ref{tab-sub-samples}. 
In Figure \ref{fig-maghist}, we show the distributions of $g'_{\rm rest}$ for the four clusters. 
The blue hatched histograms on the figure compares the distribution of the bulge-dominated galaxies 
with spectroscopy to all bulge-dominated members of the clusters (blue open histogram). 
The completeness of the spectroscopy for the bulge-dominated galaxies is 92\% and 99\% for 
Perseus and Coma, respectively. A2029 and A2142 have lower completeness, reaching 77\% and 71\%, respectively.

\begin{deluxetable}{l rrr rrr r}
\caption{Sub-Samples \label{tab-sub-samples} }
\tabletypesize{\scriptsize}
\tablehead{ 
\colhead{Cluster} & \colhead{$N_{\rm member}$} & \colhead{$N_{\rm faint}$} & \colhead{$N_{\rm disk}$} 
                  & \colhead{$N_{\rm bulge}$}   & \colhead{$N_{\rm emis}$} & \colhead{$N_{\rm passive}$} & \colhead{$C$} \\
\colhead{(1)} & \colhead{(2)} & \colhead{(3)} & \colhead{(4)} & \colhead{(5)} & \colhead{(6)} & \colhead{(7)} & \colhead{(8)}
}
\startdata
Perseus & 166 & 13 & 33 & 120 & 5 & 105 & 92\% \\
Coma    & 185 & 33 & 26 & 126 & 2 & 123 & 99\% \\
A2029   & 282 & 93 & 42 & 147 & 4 & 109 & 77\% \\
A2142   & 313 & 62 & 51 & 200 & 1 & 141 & 71\% \\
\enddata
\tablecomments{Column 1: Cluster. 
Column 2: Number of member galaxies in the parent sample. 
Column 3: Number of galaxies fainter than the sample limit of $M _{\rm B,abs} = -18.5$ mag (Vega magnitudes).
Column 4: Number of disk or blue galaxies brighter than the sample limit.
Column 5: Number of bulge-dominated galaxies brighter than the sample limit, not all of these have available spectroscopy.
Column 6: Number of emission line bulge-dominated  galaxies brighter than the sample limit.
Column 6: Number of passive bulge-dominated galaxies brighter than the sample limit.
Column 7: Completeness of spectroscopic sample of bulge-dominated galaxies derived as 
$C = (N_{\rm emis} + N_{\rm passive})/N_{\rm bulge}$.
}
\end{deluxetable}

\subsection{Average Parameters \label{SEC-AVERAGEPARAM} }

In our determination of ages, metallicities and abundance ratios, 
we use line indices averaged in bins of 0.05 in $\log \sigma$. The averages are 
luminosity weighted and only measurements from spectra with S/N $\ge 20 {\rm \AA}^{-1}$ are included.
For a few bins in A2029 and A2142 neighboring bins were merged to ensure all bins contain
at least 4 galaxies. The median number of galaxies in each bin is 11. 
The velocity dispersions for the galaxies in each bin were averaged the same way.
Average parameters are listed in Appendix \ref{SEC-SPECPARAMAPP} Table \ref{tab-average}.

\subsection{The Methods and Stellar Population Models}

Our technique for establishing the scaling relations and associated uncertainties on
slopes and zero points is the same as we used in 
J\o rgensen et al.\ (2005) and J\o rgensen \& Chiboucas (2013).
Briefly, we establish the scaling relations using a fitting technique that
minimizes the sum of the absolute residuals,
determines the zero points as the median, and uncertainties on the slopes using
a boot-strap method.
The technique is very robust to the effect of outliers.
In the discussion of the zero point differences we use both the 
random uncertainties as established from the scatter relative to the relations,
and the systematic uncertainties on the zero point differences. 
The latter is expected to be dominated by the possible inconsistency in the calibration of the 
velocity dispersions. 
Based on the comparison of the velocity dispersions with data from Thomas et al.\ (2013),
we adopt an upper limit on the systematic differences of 0.022 in $\log \sigma$.

For determination of luminosity weighted mean ages, metallicities, and abundance
ratios, we use the SSP models from 
Thomas et al.\ (2011) for a Salpeter (1955) IMF, and adopt the methods used in J\o rgensen et al.\ (2017).
The models assume that the abundance ratios for carbon and nitrogen track those of the $\alpha$-elements, 
specifically the magnesium abundance ratios.

\begin{deluxetable*}{lrrr rrr rrr rrr}
\tablecaption{Scaling Relations \label{tab-relations} }
\tablewidth{0pc}
\tabletypesize{\scriptsize}
%\tablenum{8}
\tablehead{
\colhead{Relation} & \multicolumn{3}{c}{Perseus} & 
  \multicolumn{3}{c}{Coma} &\multicolumn{3}{c}{A2029} & \multicolumn{3}{c}{A2142}   \\
 & \colhead{$\gamma$} & \colhead{$N_{\rm gal}$} & \colhead{rms} 
 & \colhead{$\gamma$} & \colhead{$N_{\rm gal}$} & \colhead{rms} 
 & \colhead{$\gamma$} & \colhead{$N_{\rm gal}$} & \colhead{rms} 
 & \colhead{$\gamma$} & \colhead{$N_{\rm gal}$} & \colhead{rms}  \\
\colhead{(1)} & \colhead{(2)} & \colhead{(3)} & \colhead{(4)} 
& \colhead{(5)} & \colhead{(6)} & \colhead{(7)} & \colhead{(8)} & \colhead{(9)} & \colhead{(10)} 
& \colhead{(11)} & \colhead{(12)} & \colhead{(13)} 
}
\startdata
$({\rm H}\delta _{\rm A} + {\rm H}\gamma _{\rm A})' = (-0.064 \pm 0.006) \log \sigma + \gamma$ &  0.053 & 102 & 0.015 &  0.049 & 119 & 0.014 &  0.059 & 102 & 0.024 &  0.055 & 83 & 0.019   \\   
$\log \rm{C4668}            = (0.29 \pm 0.04) \log \sigma + \gamma$    &  0.204 & 103 & 0.052 &  0.174 & 119 & 0.055 &  0.207 & 103 & 0.073 &  0.158 & 84 & 0.093   \\   
$\log \rm{Fe4383}           = (0.183 \pm 0.017) \log \sigma + \gamma$  &  0.265 & 103 & 0.052 &  0.256 & 119 & 0.042 &  0.235 & 103 & 0.084 &  0.254 & 84 & 0.122   \\   
$\rm{CN3883}                = (0.198 \pm 0.012) \log \sigma + \gamma$  & -0.207 &  99 & 0.034 & -0.200 & 118 & 0.022 & -0.201 & 101 & 0.037 & -0.193 & 84 & 0.051   \\   
$\log \rm{H}\beta _{\rm G}  = (-0.194 \pm 0.026) \log \sigma + \gamma$ &  0.729 & 101 & 0.052 &  0.743 & 122 & 0.041 &  0.750 & 103 & 0.063 &  0.715 & 83 & 0.063   \\   
$\log \rm{Mg}b              = (0.243 \pm 0.016) \log \sigma + \gamma$  &  0.109 & 105 & 0.038 &  0.104 & 123 & 0.030 &  0.107 &  98 & 0.055 &  0.117 & 84 & 0.043   \\   
$\log \left < \rm{Fe} \right > = (0.097 \pm 0.018) \log \sigma + \gamma$ &0.238 & 105 & 0.033 &  0.235 & 122 & 0.036 &  0.237 & 103 & 0.049 &  0.220 & 83 & 0.064   \\ 
\enddata
\tablecomments{Column 1: Scaling relation. Column 2: Zero point for the Perseus sample. Column 3: Number of galaxies
included from the Perseus sampkle. Column 4: Scatter, rms, in the Y-direction of the scaling relation for the Perseus sample.
Columns 5--7: Zero point, number of galaxies, rms in the Y-direction for the Coma sample.
Columns 8-10: Zero point, number of galaxies, rms in the Y-direction for the A2029 sample.
Columns 11-13: Zero point, number of galaxies, rms in the Y-direction for the A2142 sample.
}
\end{deluxetable*}

\begin{figure}
\epsfxsize 8.5cm
\epsfbox{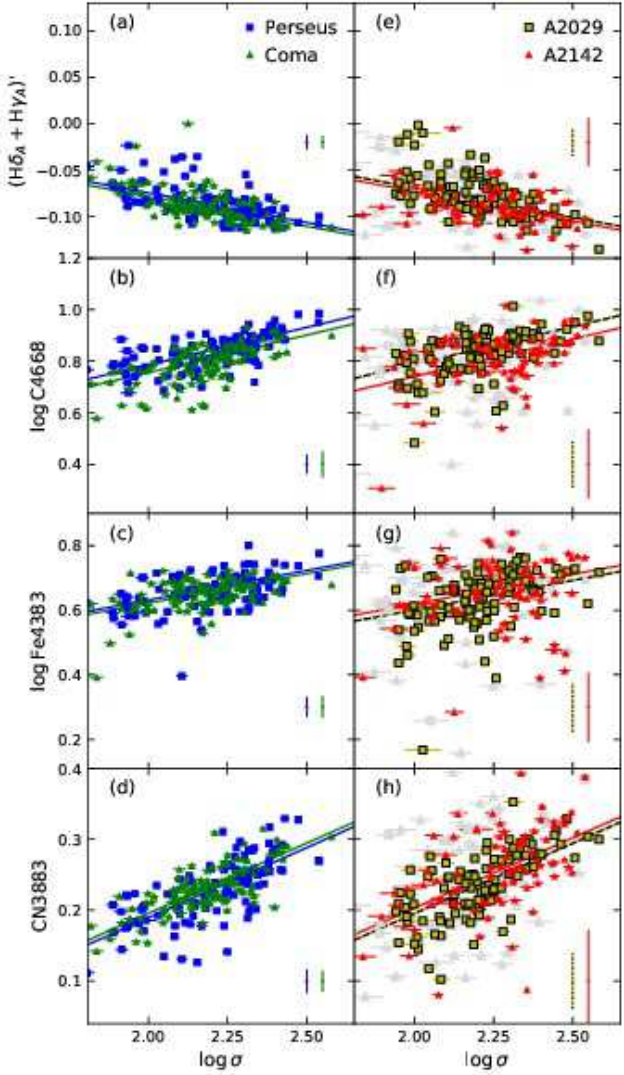}
\caption{ 
Absorption lines in the blue,
$({\rm H}\delta _{\rm A} + {\rm H}\gamma _{\rm A})'$, CN3883, Fe4383, and C4668, versus velocity dispersions. 
Blue squares -- Perseus; green triangles -- Coma; yellow squares -- A2029; red triangles -- A2142.
Grey points on panels (e)--(h) show measurements from spectra with S/N $< 20 {\rm \AA}^{-1}$.
The best fit relations offset to the median zero points for each cluster are overlaid, color coded
as the points.
Typical uncertainties are shown on the lower right of each panel.
\label{fig-lsigma_blueline_all} }
\end{figure}
 
\begin{figure}
\epsfxsize 8.5cm
\epsfbox{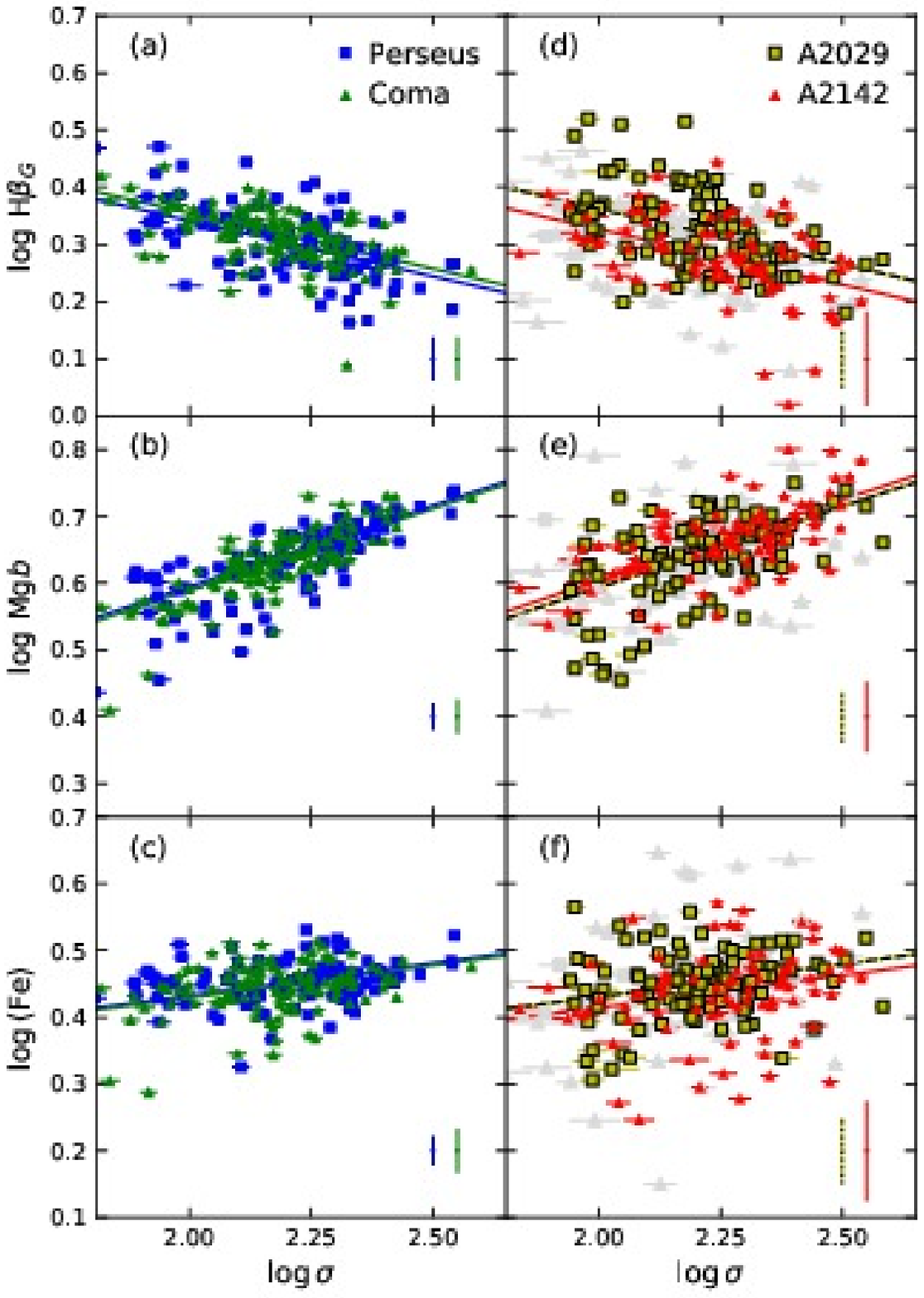}
\caption{ 
Absorption lines in the visible, H$\beta _{G}$, Mg$b$ and $\langle {\rm Fe} \rangle$,
versus velocity dispersions. Symbols as on Fig.\ \ref{fig-lsigma_blueline_all}.
The best fit relations offset to the median zero points for each cluster are overlaid.
Typical uncertainties are shown on the lower right of each panel.
\label{fig-lsigma_visline_all} }
\end{figure}

\section{Scaling Relations for Absorption Lines \label{SEC-INDICES} }

In Figures \ref{fig-lsigma_blueline_all} and \ref{fig-lsigma_visline_all}, 
we show the absorption line indices versus the velocity dispersions for 
the passive bulge-dominated galaxies. 
We concentrate on those absorption line indices that we use in our analysis of 
the $z=0.2-0.9$ GCP clusters (J\o rgensen et al.\ 2017), specifically the indices at blue wavelengths
($({\rm H}\delta _{\rm A} + {\rm H}\gamma _{\rm A})'$, C4668, Fe4383, CN3883) and 
the indices at visible wavelengths (H$\beta _{\rm G}$, Mg$b$, $\langle {\rm Fe} \rangle$).
In the following we refer to these indices as the ``blue'' and ``visible'' indices, respectively. 

We first fit the relations to each cluster separately. Measurements originating from
spectra with S/N$<20 {\rm \AA}^{-1}$ are omitted. This affects only A2029 and A2142.
Within the uncertainties, the slopes of these fits are the same for the four clusters. 
Thus, we determined the best fit relations using all four cluster samples together, requiring common slopes
but allowing different zero points for the clusters. 
Table \ref{tab-relations} lists the relations shown on the figures, including the 
zero points and the scatter for each of the cluster samples. The residuals were
minimized perpendicular to the relations, except for $({\rm H}\delta _{\rm A} + {\rm H}\gamma _{\rm A})'$
for which the residuals were minimized in the direction of the Y-axis. 

The larger scatter seen for the galaxies in A2029 and A2142 compared to those in Perseus and Coma 
is completely explained by the larger measurement uncertainties.
Subtracting off the measurement uncertainties in quadrature, we find that the internal scatter 
is $\approx 0.015$ for the ($({\rm H}\delta _{\rm A} + {\rm H}\gamma _{\rm A})' - \log \sigma$ relation
and 0.02--0.03 for all other relations.  

\begin{figure}
\epsfxsize 8.5cm
\epsfbox{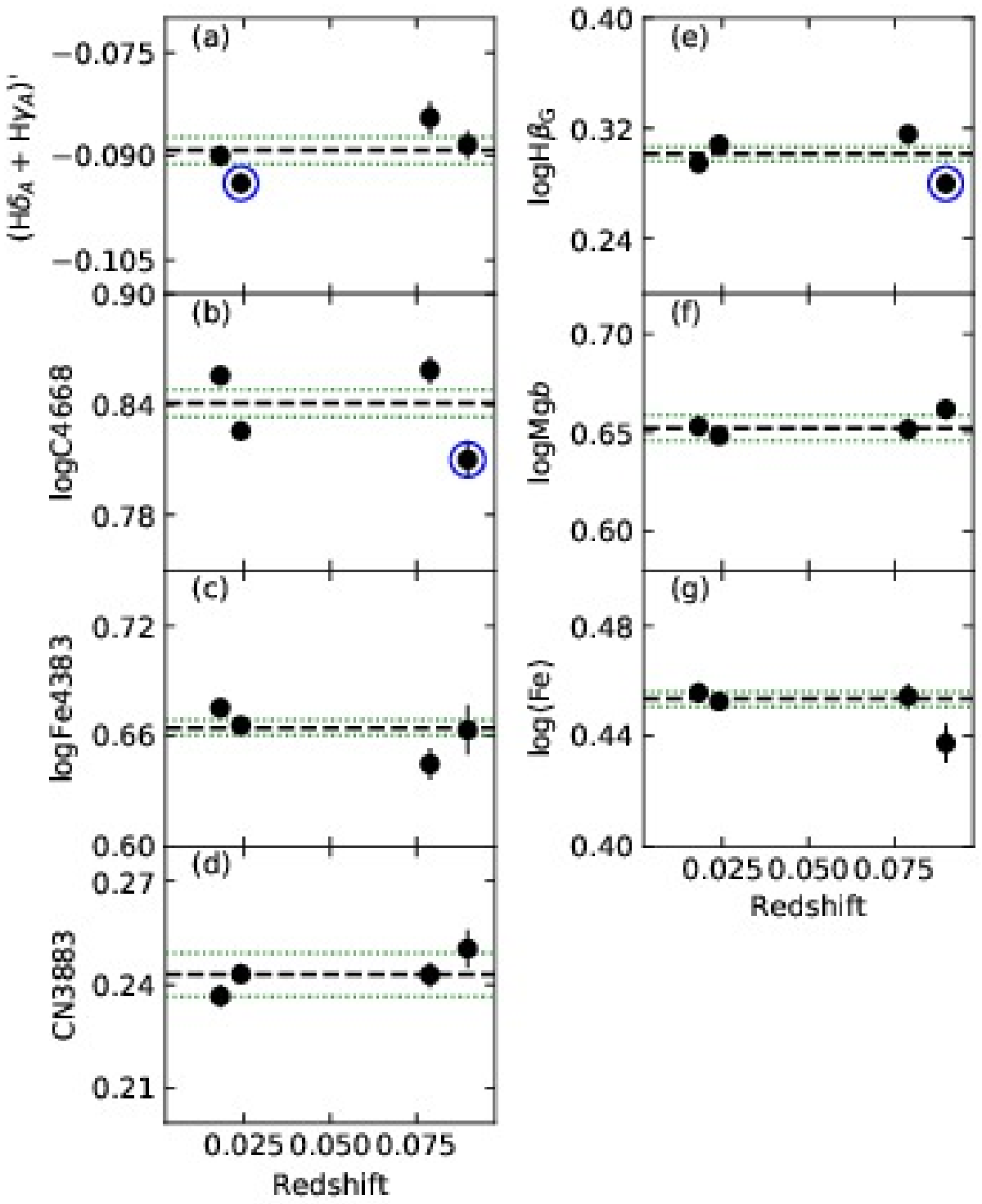}
\caption{ 
Scaling relation zero points shown as the index value at a velocity dispersion of $\log \sigma =2.24$, versus redshift.
Dashed black lines -- median zero point of the four clusters.
Dotted green lines -- limits on zero point variations due to systematic errors on the velocity dispersions, see text.
Open blue circles -- highlight the measurements deviating from the median zero points with 2-3$\sigma$. There are no
deviations larger than 3$\sigma$.
\label{fig-linezp} }
\end{figure}

The differences between the zero points are very small. In Figure \ref{fig-linezp}, we illustrate
this by showing the line index values for the fits at $\log \sigma =2.24$ for each of the clusters, as a function of cluster redshift. 
The values at $\log \sigma =2.24$ are taken as representative for the clusters, following the convention from J\o rgensen et al.\ (2017).
No significant redshift dependence is expected, the choice of X-axis on the plot is purely to separate 
the clusters to visualize the zero point differences. The random uncertainties are shown as error bars derived 
as ${\rm rms}\, N_{\rm gal}^{-0.5}$,
where $N_{\rm gal}$ is the number galaxies included in each fit.
The upper limit on systematic uncertainties due to the possible systematic differences of $\log \sigma$ of 0.022
are shown as green dotted lines offset from the median values marked by black lines. We adopt these 
as marking the possible scatter due to systematic errors.
In almost all cases, the clusters are within 2$\sigma$ of the lines marking scatter possible
due to systematic errors, and no clusters deviate more than 3$\sigma$. The three cases of deviations of 2-3$\sigma$
are marked with blue circles. 
These very small zero point differences set limits on the cluster-to-cluster variation of the
ages, metallicities, abundance ratios.  
However, since all the indices depend on all three physical quantities, we opt to proceed
with determination of these parameters directly before discussing the possible cluster-to-cluster variation.

\begin{figure*}
\epsfxsize 13cm
\begin{center}
\epsfbox{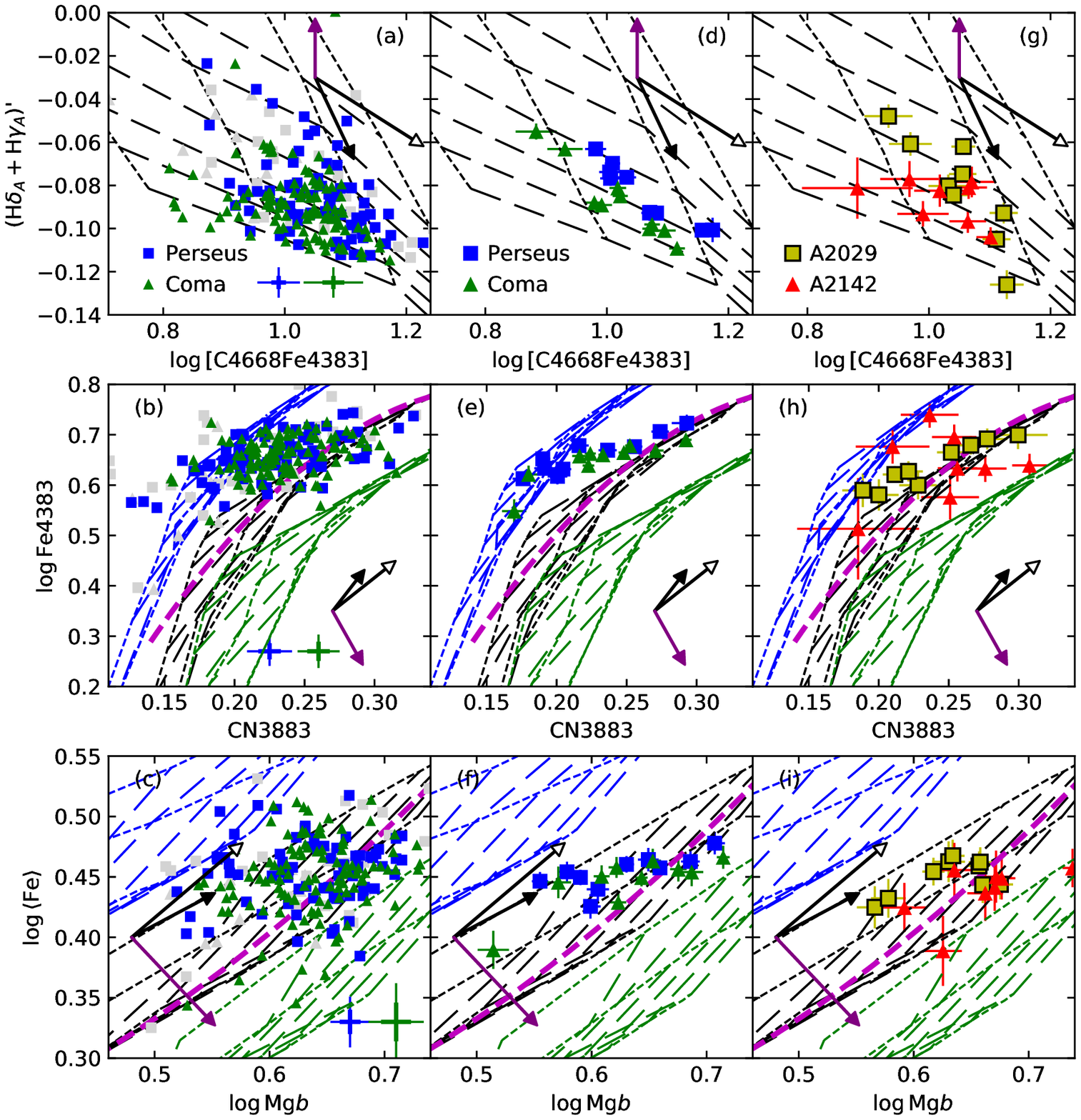}
\end{center}
\caption{ 
Absorption line strengths versus each other. 
Panels (a)--(c) show the measurements for individual galaxies in the Perseus and Coma clusters. 
Blue squares -- galaxies in Perseus with S/N $\ge 50 {\rm \AA}^{-1}$.
Green triangles -- galaxies in Coma with S/N $\ge 50 {\rm \AA}^{-1}$.
Grey points -- galaxies with S/N $< 50 {\rm \AA}^{-1}$.
Typical error bars on the panels show the measurement uncertainties as thick lines, and the scaled
measurement uncertainties as thinner larger error bars, cf.\ Table \ref{tab-uncadopt}. 
Panels (d)--(f) show the average measurements for galaxies in Perseus (blue squares) and Coma (green triangles), 
binned by velocity dispersion. 
Panels (g)--(i) show the average measurements for galaxies in A2029 (yellow squares) and A2142 (red triangles), 
binned by velocity dispersion. 
The error bars on panels (d)--(i) show the measurement uncertainties on the averages, scaled with
the scaling factors from Table \ref{tab-uncadopt}.
Thomas et al.\ (2011) SSP model grids are overlaid.
Short-dashed lines -- constant metallicities $\rm [M/H] = -0.33$, 0, 0.35, and 0.67. 
Long-dashed lines -- constant ages from 2 to 15 Gyr. 
Blue grid -- $\rm [\alpha/Fe]=0.0$; black grid -- $\rm [\alpha/Fe]=0.3$; green grid -- $\rm [\alpha/Fe]=0.5$. 
The arrows on each panel show the effect on the line indices from changes in age (filled black), metallicity [M/H] (open black), and 
$\rm [\alpha/Fe]$ (purple). In all cases, we show the effect of a 0.3 dex change. 
Purple lines in panels (e), (f), (h), and (i) are second-order fits to the models for $\rm [\alpha/Fe]=0.3$, 
used in the process of determining the abundance ratios, see text.
\label{fig-line_line} }
\end{figure*}

\section{Ages, Metallicities and Abundance Ratios \label{SEC-AGEMALPHA} }

To determine ages, metallicities [M/H], and abundance ratios, we use both the individual
measurements for the Perseus and Coma cluster galaxies with S/N $\ge 50 {\rm {\AA}^{-1}}$, 
and for all four clusters the luminosity weighted average indices for sub-samples binned in $\log \sigma$.

Figure \ref{fig-line_line} shows the Balmer line index $({\rm H}\delta _{\rm A} + {\rm H}\gamma _{\rm A})'$  
versus the combination index [C4668Fe4383], 
and the iron indices versus CN3883 and $\log {\rm Mg}b$. Model grids from Thomas et al.\ (2011) are overlaid. 
The model values for CN3883 are derived from CN$_2$ as described in J\o rgensen \& Chiboucas (2013).
The metal combination index is defined to minimize its dependence on the abundance ratios $[\alpha/ {\rm Fe}]$,
\begin{equation}
\rm {[C4668\,Fe4383] \equiv C4668 \cdot (Fe4383)^{1/3} }
\end{equation}
see J\o rgensen \& Chiboucas (2013).
 
We proceed as in J\o rgensen \& Chiboucas (2013) and J\o rgensen et al.\ (2017). 
We determine (age, [M/H]) by linearly interpolating between the models from Thomas et al.\ in
the $({\rm H}\delta _{\rm A} + {\rm H}\gamma _{\rm A})'$ -- log [C4668\,Fe4383] space to identify the (age,[M/H]) value
matching each galaxy's line indices.
The abundance ratios are derived from the iron indices versus CN3883 and $\log {\rm Mg}b$.
This is done by fitting second order polynomials to the $[\alpha/ {\rm Fe}]=0.3$ models in the parameter spaces
of (log Fe4383, CN3883) and (log $\langle {\rm Fe} \rangle$, log Mg$b$). 
The abundance ratio is then determined from the distance between the polynomial fit
and the measured parameters, measured along the lines of the $[\alpha/ {\rm Fe}]$ change as indicated by the purple arrows 
on the figures. 
In the following we refer to these two determinations as [CN/Fe] and [Mg/Fe], respectively, or $[\alpha/ {\rm Fe}]$ collectively.
Note that the Thomas et al.\ models assume that these abundance ratios are identical. 
Uncertainties are in all cases estimated from the extreme points of the uncertainties on the line indices.

We chose to use $({\rm H}\delta _{\rm A} + {\rm H}\gamma _{\rm A})'$ versus [C4668\,Fe4383]
as age and metallicity indicators, because the uncertainties on the ages are a factor 2-2.5 smaller
than if determined from H$\beta _{\rm G}$ versus [MgFe] (see Gonz\'{a}lez 1993 for original definition of the [MgFe] index).
This is because the uncertainties on our H$\beta _{\rm G}$ measurements are higher
relative to the index's age dependency, compared to those of our $({\rm H}\delta _{\rm A} + {\rm H}\gamma _{\rm A})'$ measurements. 
Further, to enable use of the results as reference for higher redshift studies, where measurements of the Mg$b$ and $\langle {\rm Fe} \rangle$ 
indices often are impossible, we use [C4668\,Fe4383] rather than [MgFe].
The uncertainties on [M/H] are also $\approx 30$\% larger if we use the visible versus the blue indices.

\begin{figure}
\epsfxsize 8.5cm
\begin{center}
\epsfbox{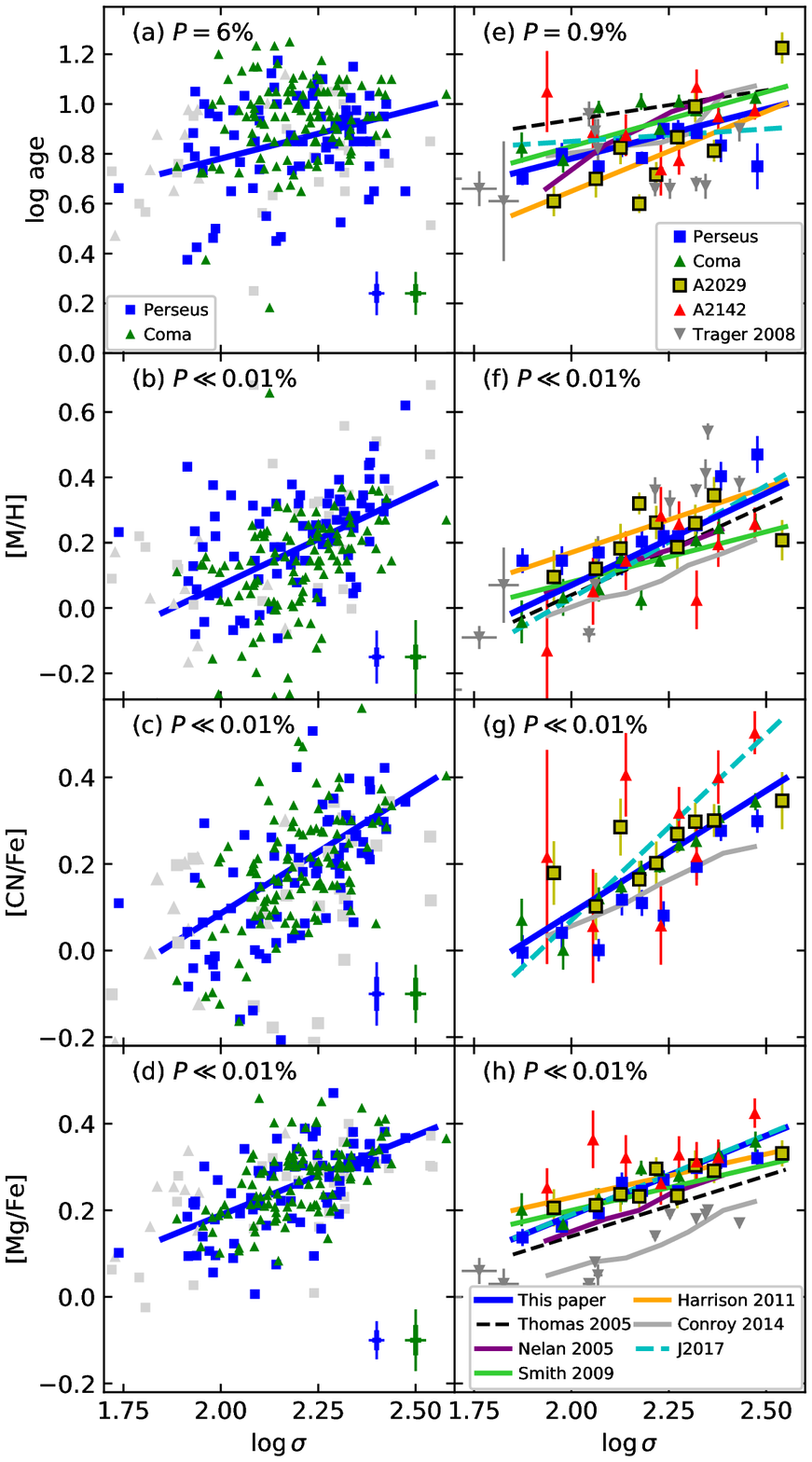}
\end{center}
\caption{ 
Ages, [M/H], and abundances ratios derived from the line indices,
shown as a function of the velocity dispersions. 
Our measurements and error bars are shown with the same symbols as on Fig.\ \ref{fig-line_line}.
Each panel is labeled with the probability that no correlation is present, as determined from a Spearman rank order correlation test.
Panels (a)--(d): Parameters derived for the individual galaxies in the Perseus and Coma cluster.
Panels (e)--(h): Parameters derived from average indices.
Grey triangles on panels (e), (f) and (h) -- Measurements for individual galaxies in the Coma cluster from Trager et al.\ (2008).
Solid blue lines on all panels -- best fit relations to the average measurements and shown at the median zero point for the four clusters.
Black dashed lines on panels (e), (f) and (h) -- Thomas et al.\ (2005).
Purple lines on panels (e), (f) and (h) -- Nelan et al.\ (2005).
Light green lines on panels (e), (f) and (h) -- Smith et al.\ (2009b).
Orange lines in panels (e), (f) and (h) -- Harrison et al.\ (2011).
Grey lines on panels (e)--(h) -- Conroy et al.\ (2014).
Cyan dashed lines on panels (e)--(h) -- J\o rgensen et al.\ (2017).
\label{fig-lsigmaagemetal_all_a030} }
\end{figure}

In Figure \ref{fig-lsigmaagemetal_all_a030}, we show ages, [M/H], [CN/Fe] and [Mg/Fe] versus the velocity dispersions.
Best fit relations are determined as least squares fits to the parameters derived from the average line indices.
There is no significant differences in the slopes for the four clusters. 
Thus, the slopes are determined by fitting all four clusters together, requiring a common slope, but allowing the zero points to vary.
The residuals were minimized in the direction of the Y-axis.
The best fits are shown on Figure \ref{fig-lsigmaagemetal_all_a030} at the median zero point for the clusters (blue lines).
We then determine median zero points for all four clusters relative to these fits. The results are summarized in Table \ref{tab-agemetal}.
The individual measurements (Figure \ref{fig-lsigmaagemetal_all_a030}a--d) follow the same relations as the measurements based on the average line indices,
but with a higher scatter due to the measurement uncertainties. 
The correlation between the ages and the velocity dispersions is quite weak, a Spearman rank order correlation test gives a probability
$P=0.9\%$ that no correlation is present when using measurements from the average line indices. If using the individual measurements the 
probability of no correlation is $P=6\%$.

For the individual measurements of ages, [M/H], [CN/Fe] and [Mg/Fe], we tested for possible dependencies on the 
cluster environment.
We used the cluster center distances, $R_{\rm cl}/R_{500}$, and the radial velocity of the galaxies
relative to the clusters, $v_{\rm ||}/\sigma _{\rm cl}$ in this test,
but found no significant correlations between the residuals for the relations on \ref{fig-lsigmaagemetal_all_a030}a--d
and $R_{\rm cl}/R_{500}$, or with the phase-space parameter
$| v_{\rm ||}|/\sigma _{\rm cl} \cdot R_{\rm cl}/R_{500}$, which is expected to be related to the accretion epoch of a
 galaxy onto a cluster, cf.\ Haines et al.\ (2012, 2015).
Since Smith et al.\ (2012) found environmental effects in the Coma cluster for the low mass galaxies (equivalent to $\log \sigma \lesssim 1.9$), 
we repeated the tests including only galaxies with velocity dispersion below the median for our sample. None of the tests
showed any significant dependency on the cluster center distance or the phase-space parameter. However, we note
that our samples only reach cluster center distances of $\approx 1.2 R_{500}$ (1.3-1.6 Mpc in Perseus and Coma) and 
do not include a significant number of galaxies with $\log \sigma < 1.9$. Thus, we do not expect to see the 
environmental dependency detected by Smith et al.\ for low mass galaxies.

\begin{figure}
\epsfxsize 8.5cm
\epsfbox{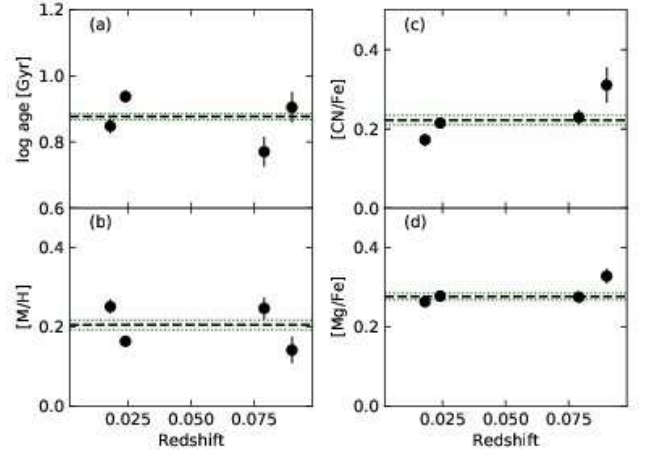}
\caption{ 
Ages, [M/H], and abundance ratios for each cluster at a velocity dispersion
of $\log \sigma = 2.24$ shown versus redshift, cf.\ Table \ref{tab-agemetal}.
No significant dependency on the redshift is expected.
Dashed black lines -- median value of the four clusters.
Dotted green lines -- limits on variations due to systematic errors on the velocity dispersions, see text.
\label{fig-agemetalalpha} }
\end{figure}

\begin{figure}
\epsfxsize 8.5cm
\epsfbox{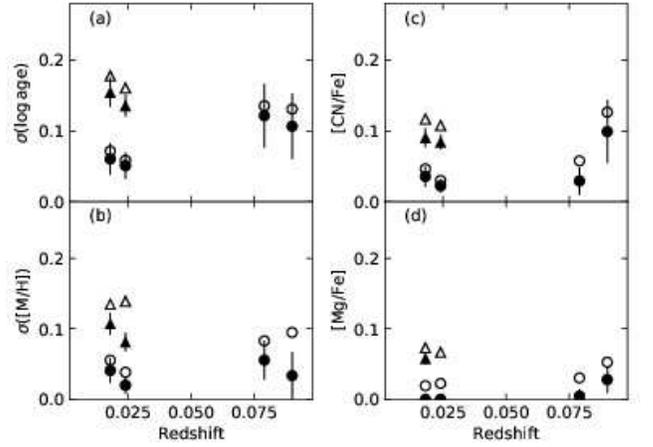}
\caption{ 
The scatter in the relations between velocity dispersions and ages, [M/H], and abundance ratios versus redshift.
No significant dependency on the redshift is expected.
Open circles -- measured scatter for the relations fit to parameters based on the average line indices.
Solid circles -- internal scatter for the relations fit to parameters based on the average line indices.
Open triangles -- measured scatter of parameters based on the individual line index measurements
for galaxies in Perseus and Coma.
Solid triangles -- internal scatter of parameters based on the individual line index measurements
for galaxies in Perseus and Coma.
The internal scatter is derived from the measured scatter by subtracting off in quadrature the measurement uncertainties
of the parameters.
In the cases where the measurement uncertainties are larger than the measured scatter, the internal scatter is shown at zero.
\label{fig-scatteragemetalalpha} }
\end{figure}

\begin{figure}
\epsfxsize 6.4cm
\begin{center}
\epsfbox{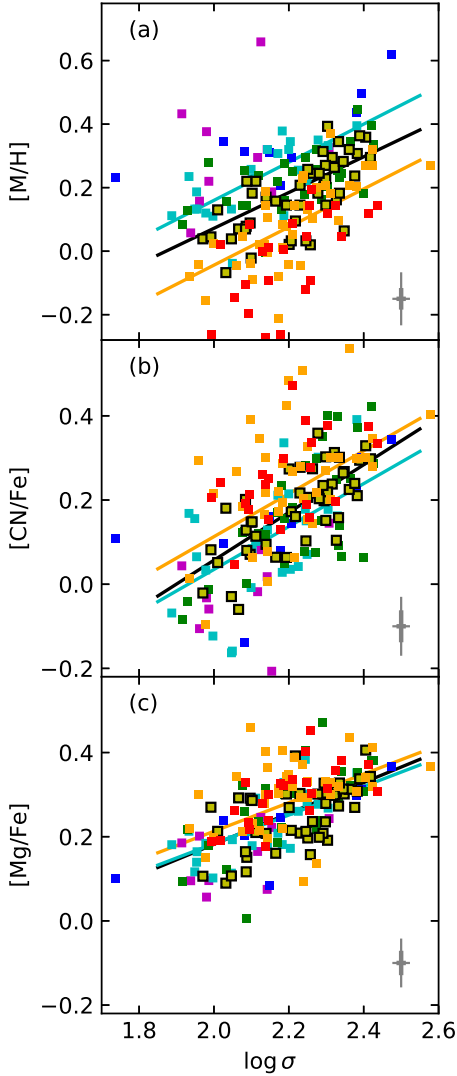}
\end{center}
\caption{ 
Metallicity [M/H], and abundance ratios [CN/Fe] and [Mg/Fe] versus velocity dispersions. 
The measurements are for individual galaxies in Perseus and Coma. 
Only measurements based on spectra with S/N $\ge 50 {\rm \AA}^{-1}$ are shown.
The symbols are color coded matching bins in log age:
Magenta -- log age $<$ 0.6;
blue -- log age = 0.6--0.7;
cyan -- log age = 0.7--0.8;
green -- log age = 0.8--0.9;
yellow -- log age = 0.9--1.0;
orange -- log age = 1.0--1.1;
red -- log age $>$ 1.1.
Black line -- best fits without age terms, as listed in Table \ref{tab-agemetal}. 
Cyan and orange lines -- best fits with age terms, shown for log age = 0.75 (cyan)
and log age = 1.05 (orange) matching the similarly colored points.
Typical error bars on the panels show the measurement uncertainties as thick lines, and the scaled
measurement uncertainties as thinner larger error bars.
\label{fig-metalagelsigma} }
\end{figure}

\begin{deluxetable*}{lrrrr rrrr rrr rrr}
\tablecaption{Age, Metallicity and Abundance Ratio Relations \label{tab-agemetal} }
\tablewidth{0pc}
\tabletypesize{\scriptsize}
\tablehead{
\colhead{Relation} & \multicolumn{4}{c}{Perseus} & 
  \multicolumn{4}{c}{Coma} &\multicolumn{3}{c}{A2029} & \multicolumn{3}{c}{A2142}   \\
 & \colhead{$\gamma$} & \colhead{rms$_{\rm avg}$} & \colhead{rms$_{\rm ind}$}& \colhead{$\gamma _{2.24}$}
 & \colhead{$\gamma$} & \colhead{rms$_{\rm avg}$} & \colhead{rms$_{\rm ind}$}& \colhead{$\gamma _{2.24}$}
 & \colhead{$\gamma$} & \colhead{rms$_{\rm avg}$} & \colhead{$\gamma _{2.24}$}
 & \colhead{$\gamma$} & \colhead{rms$_{\rm avg}$} & \colhead{$\gamma _{2.24}$} \\
\colhead{(1)} & \colhead{(2)} & \colhead{(3)} & \colhead{(4)} & \colhead{(5)} & \colhead{(6)} & \colhead{(7)} & \colhead{(8)} & \colhead{(9)}  
& \colhead{(10)} & \colhead{(11)} & \colhead{(12)} & \colhead{(13)} & \colhead{(14)} & \colhead{(15)}
}
\startdata
$\log {\rm age}     = (0.40 \pm 0.16) \log \sigma + \gamma$ &  -0.045 &  0.072 & 0.178 & 0.848    &  0.044 & 0.058 & 0.161 & 0.938 & -0.123 & 0.136 & 0.771 &  0.011 & 0.131 & 0.905 \\   
${\rm [M/H]}        = (0.56 \pm 0.16) \log \sigma + \gamma$ &  -1.010 &  0.055 & 0.135 & 0.251    & -1.098 & 0.038 & 0.139 & 0.163 & -1.014 & 0.083 & 0.247 & -1.120 & 0.095 & 0.141 \\   
${\rm [CN/Fe]} = (0.57 \pm 0.09) \log \sigma + \gamma$      &  -1.098 &  0.046 & 0.117 & 0.172    & -1.056 & 0.030 & 0.108 & 0.215 & -1.041 & 0.058 & 0.229 & -0.960 & 0.127 & 0.310 \\  
${\rm [Mg/Fe]}  = (0.37 \pm 0.06) \log \sigma + \gamma$     &  -0.559 &  0.019 & 0.073 & 0.263    & -0.544 & 0.023 & 0.067 & 0.278 & -0.547 & 0.030 & 0.275 & -0.494 & 0.053 & 0.328 \\   
\enddata
\tablecomments{Column 1: Scaling relation. Column 2: Zero point for the Perseus sample. 
Column 3: Scatter, rms, in the Y-direction of the scaling relation for the Perseus sample, based on average line indices.
Column 4: Scatter, rms, in the Y-direction of the scaling relation for the Perseus sample, based on individual line indices.
Column 5: Value of the parameter (log age, M/H], [CN/Fe], or [Mg/Fe]) at a velocity dispersion of $\log \sigma =2.24$ for the Perseus sample.
Columns 6--9: Values for the Coma sample.
Columns 10--12: Values for the A2029 sample, only rms for measurements based on average line indices are listed.
Columns 13--15: Values for the A2142 sample, only rms for measurements based on average line indices are listed.
}
\end{deluxetable*}

To illustrate possible differences between the clusters, Figure \ref{fig-agemetalalpha} shows ages, [M/H], [CN/Fe], and [Mg/Fe] at $\log \sigma =2.24$
as determined by the relations. We use the redshift of the clusters as the X-axis of the plot only to separate the measurements.
No significant dependence on redshift is expected to be detectable. 
The measurements for the Perseus and Coma clusters are in general consistent within 2$\sigma$ and also consistent with the median
for the four clusters. A2029 and A2142 exhibit differences at the $2-2.5\sigma$ level.
Based on maximum differences between the four clusters, we can quantify to what extent the data allow cluster-to-cluster variations 
in ages, [M/H], [CN/Fe], and [Mg/Fe]. 
For the four clusters, variations of $\pm 0.08$ dex in median ages are possible, 
while variations in [M/H] and [CN/Fe] are within $\pm 0.06$ dex and $\pm 0.07$ dex, respectively.
The abundance ratio [Mg/Fe] is restricted to $\pm 0.03$ dex.

Figure \ref{fig-scatteragemetalalpha} shows the scatter in ages, [M/H], [CN/Fe], and [Mg/Fe] at fixed velocity dispersion.
The scatter determined from the relations based on average line indices should be understood as lower limits. 
More realistic estimates are achieved by using the individual measurements for the Perseus and Coma cluster galaxies
(triangles on Figure \ref{fig-scatteragemetalalpha}). Subtracting off the measurement uncertainties in quadrature, 
we find an internal scatter of 0.15, 0.1, 0.09, and 0.06 dex in ages, [M/H], [CN/Fe], and [Mg/Fe], respectively.

Using the individual measurements for the Perseus and Coma galaxies, we find that the residuals
for the [M/H]--velocity dispersion relation are correlated with the ages. Fitting all three parameters together gives
\begin{equation}
{\rm [M/H]} = (0.60\pm 0.05) \log \sigma - (0.68\pm 0.07) \log {\rm age} - 0.53
\end{equation}
The residuals were minimized in [M/H]. 
This relation is illustrated in Figure \ref{fig-metalagelsigma}a, where the points are color coded in bins of age. 
The scatter of the relation is 0.092 in [M/H], while a 
fit to [M/H] as a function of only the velocity dispersion, requiring a common zero point for the Perseus and Coma samples, 
has a scatter of 0.14. The reduction in the scatter by including the age term is significant at the 5$\sigma$ level, and
the relation has no significant internal scatter.
Figure \ref{fig-metalagelsigma}b and c show similarly age color-coded versions of [CN/Fe] and [Mg/Fe] versus 
the velocity dispersions for the Perseus and Coma galaxies.
Only an insignificant reduction in scatter is achieved by inclusion of an age term in these relations, 
though formally the age coefficients are significant at the 2.5-5 sigma level.
We find
\begin{equation}
{\rm [CN/Fe]} = (0.51\pm 0.06) \log \sigma - (0.26\pm 0.05) \log {\rm age} - 1.18
\end{equation}
and
\begin{equation}
{\rm [Mg/Fe]} = (0.34\pm 0.04) \log \sigma - (0.10\pm 0.04) \log {\rm age} - 0.57
\end{equation}
with scatter of 0.10 and 0.067, respectively. The scatter of fits without the age term
is 0.11 and 0.070 for the two relations.

\section{Discussion \label{SEC-DISCUSSION} }

\subsection{Comparison of Scaling Relations with Previous Results}

We want to ensure that our results are consistent with previous results
for massive low redshift clusters. We used smaller samples of galaxies in Perseus and Coma
as our reference samples in previous GCP papers, e.g., J\o rgensen \& Chiboucas (2013)
and J\o rgensen et al.\ (2017). The scaling relations established in those papers agree 
with our results based on the larger samples in the present paper. Thus, using the 
larger low redshift reference sample established in the present paper will not significantly affect our results
for the $z=0.2-0.9$ GCP clusters.

Nelan et al.\ (2005) and Smith et al.\ (2006) established scaling relations for clusters in the NOAO 
Fundamental Plane (FP) survey.
The main emphasis of Smith et al.\ was an investigation of the possible effects of the 
cluster environment at very large cluster center distances.
Since Nelan et al.\ and Smith et al.\ determine the relations in the form ${\rm index} = a_1 \log \sigma + a_2$, 
rather than using the logarithm of the indices, we overlay their relations on our results, and
compare the slope and zero points at $\log \sigma =2.2$. We omit the cluster center distance terms
from the Smith et al.\ relations.
Before comparison, we convert H$\beta$ to H$\beta _{\rm G}$ using the transformation from J\o rgensen (1997).
We combine the literature relations for H$\delta _{\rm A}$ and H$\gamma _{\rm A}$ to a relation for 
$({\rm H}\delta _{\rm A} + {\rm H}\gamma _{\rm A})'$. Similarly, we combine the literature relations
for Fe5270 and Fe5335 to a relation for $\langle {\rm Fe} \rangle$.
Our slopes agree within the uncertainties with those from Smith et al.
The slopes from Nelan et al.\ are marginally steeper for $({\rm H}\delta _{\rm A} + {\rm H}\gamma _{\rm A})'$, 
C4668, H$\beta _{\rm G}$, Mg$b$ and $\langle {\rm Fe} \rangle$, and marginally shallower for Fe4383.
In all cases, the zero points agree with our zero point for the Coma cluster within $\pm 0.03$ dex at $\log \sigma =2.2$.
This agreement is similar to the agreement in our results for the four clusters studied in the present paper, 
cf.\ Table \ref{tab-relations}.

Harrison et al.\ (2011) used data for four nearby clusters, including the Coma cluster, to establish 
scaling relations between line indices and velocity dispersions. These authors used indices converted to magnitudes. 
Thus, we compare our results to their cluster sample results by comparing slopes and zero points at $\log \sigma =2.2$.
In all cases, the zero points agree with our zero point for the Coma cluster within $\pm 0.03$ dex.
The slope from Harrison et al.\ (2011) for the Mg$b$-velocity dispersion relation is marginally steeper at $\log \sigma =2.2$
than our result, but for higher velocity dispersion galaxies flattens to agreement with our determination.

\subsection{Ages, Metallicities, and Abundance ratios}

There are numerous studies in the literature aimed at establishing the relations between
the velocity dispersions (or masses) and the stellar population ages, metallicities and abundance ratios,
see Harrison et al.\ (2011) for an overview. 
Here we compare to a few selected studies, spanning different techniques and sample sizes.

Thomas et al.\ (2005) used an earlier version (Thomas et al.\ 2003) of the same models used in the present paper.
They base their study on available literature data for 124 galaxies and derive ages, [M/H], and $\rm [\alpha /Fe]$
using the indices in the visible. 
The slopes of the relations established by Thomas et al.\ (2005; dashed lines on Figure \ref{fig-lsigmaagemetal_all_a030} panels e, f, and h) agree with our results.
A later study by Thomas et al.\ (2010) uses the same techniques, but a much larger sample of SDSS spectra, 
and results in a slightly steeper age-velocity dispersion relation, than in the 2005 study.
The Thomas et al.\ relations for age and $\rm [\alpha /Fe]$ are offset from our results with 
approximately +0.1 dex and $-0.1$ dex, respectively, possibly due to small differences in the adopted models.

Nelan et al.\ (2005) stacked the NOAO FP survey spectra in five bins by velocity dispersion in order to
achieve high S/N spectra for their study of ages, [M/H], and $\rm [\alpha /Fe]$. They derive the 
parameters from the visible indices (H$\beta$, Mg$b$, $\langle {\rm Fe} \rangle$), using the Thomas
et al.\ (2003) models. We show their result on Figure \ref{fig-lsigmaagemetal_all_a030} (panels e, f, and h) as purple lines.
The slopes of their relations for [M/H] and $\rm [\alpha /Fe]$ agree with ours, while they find an age-velocity 
dispersion relation slightly steeper than our result. The $\rm [\alpha /Fe]$ is offset to slight lower values relative to ours,
again presumably due to small differences in the assumed models.

Smith et al.\ (2009b) studied the stellar populations of massive galaxies in the Shapley concentration.
They used the indices ($H\gamma _F$, Mg$b$, Fe5015) combined with models from Thomas et al.\ (2003) to derive 
ages, metallicities and abundance ratios of the galaxies. We include their results on 
Figure \ref{fig-lsigmaagemetal_all_a030} (panels e, f, and h) as light green lines.
Their relations for [M/H] and [Mg/Fe] are slightly shallower than our results, while their result for 
the age-velocity dispersion relation agree with our results, within the uncertainties. 
Smith et al.\ (2009a) list shallower slopes for all the relations for the Shapley sample, 
and find slightly steeper [M/H] and [Mg/Fe] relations for low mass galaxies in the Coma cluster, 
though presumably these differences are due to the range of masses (velocity dispersions) sampled
and not due to a cluster-to-cluster difference.

Trager et al.\ (2008) studied a small sample of Coma cluster galaxies using high S/N spectra.
They determined ages, [M/H], and $\rm [\alpha /Fe]$ using the visible indices (H$\beta$, Mg$b$, $\langle {\rm Fe} \rangle$).
We show their results overlaid on Figure \ref{fig-lsigmaagemetal_all_a030} (grey triangles on panels e, f, and h). 
In general their results are in agreement with
ours, except there is a systematic offset in $\rm [\alpha /Fe]$ of about 0.15 dex, with the values from 
Trager et al.\ being smaller than ours. This offset can be traced back to the difference in the adopted SSP models.
Trager et al.\ used models from Worthey (1994) with revised response functions to model the non-solar abundance ratios.
Comparing their Figure 4 model grids for different $\rm [\alpha /Fe]$ with our Figure \ref{fig-line_line} model grids
illustrates that the models give different abundance ratios. 
For a typical galaxy with (Mg$b$, $\langle {\rm Fe} \rangle$) $\approx$ (4., 2.8) 
(GMP 3483 in the Trager et al.\ sample) Trager et al.\ find $\rm [\alpha /Fe]$=0.08, while our method gives 0.23, 
confirming an offset of 0.15 dex. 

Harrison et al.\ (2011) derived ages, [M/H] and $\rm [\alpha /Fe]$  from several absorption line indices.
They use models from Thomas et al.\ (2003). We show their results overlaid on Figure \ref{fig-lsigmaagemetal_all_a030} (orange lines on panels e, f, and h).
Their age-velocity dispersion relation is slightly steeper than our result, while their [M/H]-velocity dispersion
relation is slightly shallower than our result.
At $\log \sigma=2.24$, this leads to $\approx 0.1$ dex lower ages, and $\approx 0.06$ higher [M/H] than our results.
Their results for $\rm [\alpha /Fe]$ agree with ours, within the uncertainties.

Conroy et al.\ (2014) used very high S/N stacks of SDSS spectra for their investigation. They performed full-spectrum
fitting with models that allowed variations in the individual abundance ratios. In order to compare their
results with ours we make the following assumptions. (1) Conroy et al.\ [Mg/Fe] can be used as stand-in
for $\rm [\alpha /Fe]$ determined from the (Mg$b$, $\langle {\rm Fe} \rangle$) diagram. (2) The average of 
Conroy et al.\ [C/Fe] and [N/Fe] can be used as stand-in
for $\rm [\alpha /Fe]$ determined from the  (CN3883, Fe4383) diagram, keeping in mind that the underlying
assumption for the SSP models we use is that carbon and nitrogen track the $\alpha$-element abundances.
(3) We can convert Conroy et al.\ [Fe/H] to total metallicity [M/H] using the conversion from 
Thomas et al.\ (2003), $\rm [M/H] = [Fe/H] + 0.94 [\alpha/Fe]$. 
With these assumptions and conversions, we then overlay the results from Conroy et al.\ on 
Figure \ref{fig-lsigmaagemetal_all_a030} (grey lines on panels e--h).
The ages from Conroy et al.\ are in agreement with our results.
The metallicities from Conroy et al.\ are about 0.1 dex below our results.
Our [CN/Fe] determinations are $\approx 0.05$ dex higher than the average of [C/Fe] and [N/Fe] from Conroy et al.,
while our [Mg/Fe] values about 0.15 dex higher than [Mg/Fe] from Conroy.
It is possible that the differences are due to differences in the models, though a direct comparison
of index-index model grids is not possible.
 
Finally, we compare to our previous results for the $z=0.2-0.9$ GCP clusters in J\o rgensen et al.\ (2017), shown in
cyan on Figure \ref{fig-lsigmaagemetal_all_a030}e--g. The age-velocity dispersion relation presented in J\o rgensen et al.\ (2017)
was flat when all clusters were fit together. 
The data for the $z \approx 0.2$ clusters (A1689 and RXJ0056.2+2622) give a slope of $\approx 0.2$. In all cases,
the correlation between age and velocity dispersion is very weak, as also found here for the four low redshift clusters.
The [M/H]-velocity dispersion relation for the $z=0.2-0.9$ GCP clusters is almost identical to our result
for the low redshift clusters, while the [CN/Fe]-velocity dispersion relation is significantly steeper.
This steeper relation also leads to the median [CN/Fe] value at $\log \sigma =2.24$ being 0.06 dex higher
than we find for the low redshift clusters, with the difference increasing with velocity dispersion.
We augment the J\o rgensen et al.\ (2017) results
with the [Mg/Fe]-velocity dispersion relation for the $z=0.2-0.9$ GCP clusters using the data from that paper.
The result is shown in cyan on Figure \ref{fig-lsigmaagemetal_all_a030}h. 
The relation is in fact identical to that of the four low redshift clusters.

In summary, the slopes for the age-velocity dispersion relations established here are within the range found
by other studies (Thomas et al.\ 2005, 2010; Nelan et al.\ 2005; Harrison et al.\ 2011; Conroy et al. 2014). 
However, as also noted by Harrison et al.\ (2011) there is substantial variation in the determinations of the slope
of the age-velocity dispersion relation, and also, but to a lesser extent, in the determinations of the slope
of the [M/H]-velocity dispersion relation. 
It is possible that the steep relations found for some low redshift samples is due to the inclusion of very young low-mass 
galaxies in some of the studies (e.g.\ McDermid et al.\ 2015). However, it does not 
fully explain the steep relation found by Nelan et al.\ (2005).

There is agreement in the literature and with our new results that the 
slope of the $\rm [\alpha /Fe]$-velocity dispersion relation is $0.3-0.35$, when the determinations
are based on the visible indices, ie.\ the [Mg/Fe]-velocity dispersion relation. 
We find a somewhat steeper relation from the blue indices, ie.\ the [CN/Fe]-velocity dispersion relation.
The behavior is in general agreement with the results for carbon and nitrogen from Conroy et al.\ (2014),
though the difference in slopes found by these authors is smaller than seen from our data. 
The fact that the [Mg/Fe]-velocity dispersion relation does not appear to have any redshift dependence,
while the [CN/Fe]-velocity dispersion relation is steeper at higher redshift, may 
be related to the timescale for the formation of the $\alpha$-elements (magnesium in this case) versus carbon and nitrogen,
and may reflect a real difference in the enrichment of magnesium versus carbon and nitrogen.

Our result for the combined [M/H]-age-velocity dispersion relation is similar to the results by Johansson et al.\ (2012).
A similar relationship was also noticed in the study of the Coma cluster galaxies by J\o rgensen (1999).
In agreement with our result of only a very weak age dependence of [Mg/Fe] (if any), 
Johansson et al.\ note that [Mg/Fe] does not depend on age. 
These results support the idea that [Mg/Fe] appears to be set very early
in the evolution of the galaxies, while later star formation primarily leads to lower mean ages and
higher total metallicities of the galaxies' stellar populations.

\subsection{Cluster-to-Cluster Variations}

The agreement between the result for the four clusters covered in the present paper may be used to 
set limits on the allowed cluster-to-cluster differences in passive cluster galaxies.
To recap, for the four clusters in our study median ages, [M/H], [CN/Fe], and [Mg/Fe] may vary by $\pm 0.08$ dex, 
$\pm 0.06$ dex, $\pm 0.07$ dex, and $\pm 0.03$ dex, respectively.
In our study of the $z=0.2-0.9$ GCP clusters (J\o rgensen et al.\ 2017 and references therein), we 
find that possible cluster-to-cluster differences in [M/H] are within $\pm 0.10$ dex. Thus, such
galaxies can plausibly evolve passively to galaxies similar to those studied in our four low redshift
reference clusters, though may deviate at the 2$\sigma$ level from our low redshift reference clusters. 

On the other hand, we found two cases of large $\rm [\alpha /Fe]$ deviations from the median,
RXJ0152.7--1357 with $\rm [\alpha /Fe]$ $0.25$ dex higher than the median, and 
RXJ1347.5--1154 with $\rm [\alpha /Fe]$ $0.16$ dex lower than the median. 
RXJ0152.7--1357 is a binary cluster at redshift $z=0.83$, possibly in the process of merging
(Maughan et al.\ 2003; Jones et al.\ 2004).
RXJ1347.5--1154 at $z=0.45$ has been the topic of debate regarding its mass, as its 
high X-ray luminosity seemed at odds with mass estimates based on the cluster velocity dispersion (Cohen \& Kneib (2002). 
However, improved measurements of the cluster velocity dispersion
(Lu et al.\ 2010) and correction of the X-ray mass for diffuse substructure (Ettori et al.\ 2004)
bring agreement between the X-ray properties and the cluster velocity dispersion (J\o rgensen et al.\ 2017).
If the galaxies in these two clusters evolve passively, they will maintain their $\rm [\alpha /Fe]$, and clusters' 
galaxy populations will at the present epoch not resemble those of our four reference clusters.
These abundance ratio measurements were based on CN3883 versus Fe4383, and should therefore be compared 
to [CN/Fe] in the present paper. 
The deviations are significant at the $3.5\sigma$ and $2.3\sigma$ level. 
If in fact no $z\sim 0$ clusters exist which deviate to that extent from the median [CN/Fe], 
then this continues to be a challenge to a simple 
passive evolution model for the bulge-dominated passive galaxies in these clusters, see J\o rgensen et al.\ (2017).
We stress that our reference sample contains four out of twenty known clusters at $z <0.1$ and 
at least as massive as the Coma cluster. It cannot be ruled out that more variation
in [CN/Fe] exists among these twenty clusters, than found from the four we have studied.
However, in our study of nine $z=0.2-0.9$ clusters we found two clusters with significantly deviating
[CN/Fe]. 
If a similar frequency (20\%-25\%) of clusters with unusual [CN/Fe] is present (randomly) among the 20 massive
$z<0.1$ clusters, then we would have a 60\%-70\% chance of detecting at least one of them in our sample of four clusters.

\section{Summary and Conclusions \label{SEC-CONCLUSION}}

We have presented consistently calibrated velocity dispersions and absorption line indices for large homogeneous samples of galaxies
in the four low redshift massive clusters Perseus, Coma, Abell 2029 and Abell 2142. 
The samples are magnitude limited to an absolute B band magnitude of $M_{\rm B,abs}=-18.5$ mag (Vega magnitudes), 
and between 71\% and 99\% complete. 
The systematic errors in velocity dispersions are estimated from external comparisons to be $\le 0.022$ in $\log \sigma$.
The data presented here form the low redshift reference sample for the Gemini/{\it HST} Galaxy Cluster Project (GCP)
aimed at studying the evolution of $z=0.2-2.0$ cluster galaxies, and is suitable as a low reference sample
for other studies of medium to high-redshift galaxies.

We used luminosity weighted average absorption line indices to derive ages, metallicities [M/H],
and abundance ratios for the galaxies binned in velocity dispersion. We also
derived these parameters from the individual line index measurements for Perseus and Coma galaxies, 
limiting this study to spectra with S/N$\ge 50 {\rm \AA}^{-1}$.

We select the sub-sample of passive bulge-dominated galaxies based on a combination of colors and the SDSS
provided information on the fraction of the luminosity modeled by an $r^{1/4}$ profile.
Our main conclusion from the analysis of the properties of this sub-sample are as follows:

\begin{enumerate}
\item
The galaxies in the four clusters follow the same scaling relations between absorption line indices and 
velocity dispersions. All zero point differences are within $3\sigma$. The internal scatter is very low at 
0.015 for the high order Balmer lines, and 0.02-0.03 for all other indices.
The relations are in agreement with recent literature results for low redshift clusters.

\item
The clusters follow relations between ages, [M/H], abundance ratios and velocity dispersions 
with no cluster-to-cluster variations in the slope. Any zero point differences in ages, [M/H], [CN/Fe] and [Mg/Fe] are limited to 
$\pm 0.08$ dex, $\pm 0.06$ dex, $\pm 0.07$ dex, and $\pm 0.03$ dex, respectively.
The limit on [CN/Fe] variations poses a challenge for
our previous results from $z=0.2-0.9$ clusters, which showed two cases of [CN/Fe] deviating 0.16-0.25 dex
from the median value. 
We find an internal scatter in ages at a fixed velocity dispersion of 0.15 dex, while the scatter in [M/H], [CN/Fe], and [Mg/Fe]
is $0.06-0.10$ dex. 

\item 
The [M/H]-age-velocity dispersion relation has significantly lower measured scatter in [M/H] 
than found for the [M/H]-velocity dispersion relation, and no significant internal scatter.
A similar reduction in the scatter is not seen if including an age term in the relations
for [CN/Fe] or [Mg/Fe]. We speculate that the reason for this may be related to the epoch at which
the abundance ratios are established for the stellar populations.

\end{enumerate}

\acknowledgments
Karl Gebhardt is thanked for making his kinematics software available.
The Gemini TACs and the McDonald Observatory TACs are thanked for generous time allocations
to carry out these observations.
The staff at McDonald Observatory are thanked for assistance during the observations
with the 2.7-m and the 0.8-m telescopes.
Bo Milvang-Jensen and Gary Hill are thanked for their contributions to the planning and execution
of the observations at McDonald Observatory.
Marcel Bergmann is thanked for his contributions to this project during planning and execution
of the Gemini observations.

Based on observations from the McDonald Observatory operated by the University of Texas at Austin.
Based on observations obtained at the Gemini Observatory, which is operated by the
Association of Universities for Research in Astronomy, Inc., under a cooperative agreement
with the NSF on behalf of the Gemini partnership: the National Science Foundation (United
States), the National Research Council (Canada), CONICYT (Chile), the Australian Research Council
(Australia), Minist\'{e}rio da Ci\^{e}ncia e Tecnologia (Brazil) 
and Ministerio de Ciencia, Tecnolog\'{i}a e Innovaci\'{o}n Productiva  (Argentina)
The data presented in this paper originate from the Gemini programs GN-2014A-Q-27 and GN-2014A-Q-104.

This research made use of the ``{\it k}-corrections calculator'' service available at http://kcor.sai.msu.ru/

The paper makes use of photometry and spectroscopy from Sloan Digital Sky Survey (SDSS). 
Funding for SDSS-IV has been provided by the Alfred P. Sloan Foundation, 
the U.S. Department of Energy Office of Science, and the Participating Institutions. SDSS-IV acknowledges
support and resources from the Center for High-Performance Computing at
the University of Utah. The SDSS web site is www.sdss.org.

%\clearpage

\appendix

\section{Spectroscopic Data \label{SEC-APPENDIX}}

\subsection{Spectroscopic parameters: Systematic Effects in Derived Velocity Dispersions \label{SEC-SIM}}

\begin{figure}
\epsfxsize 8.5cm
\begin{center}
\epsfbox{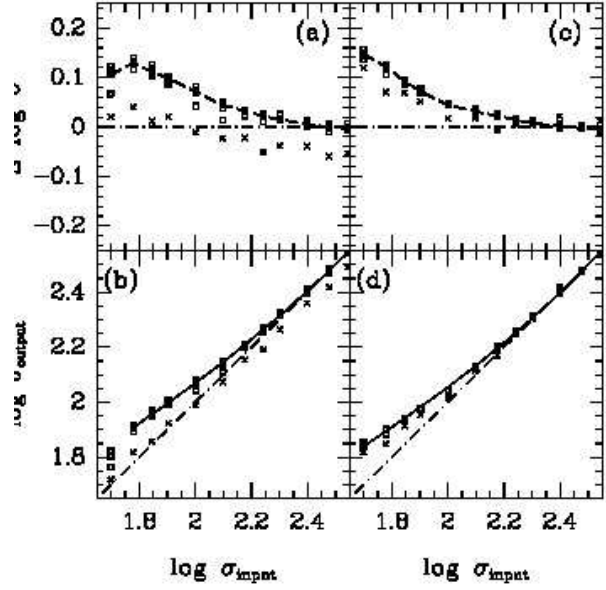}
\end{center}
\caption{Results from simulations of the velocity dispersions. 
Squares -- simulations with S/N $\ge 10 {\rm \AA}^{-1}$.
Crosses -- simulations with S/N $= 6 {\rm \AA}^{-1}$.
The top row of panels show the systematic error on $\log \sigma$,
$\Delta \log \sigma = \log \sigma _{\rm output} - \log \sigma _{\rm input}$ as a function of the
input values $\log \sigma _{\rm input}$. The bottom row of panels show input versus
output values. Simulations apply to the data as follows: (a) and (b) GMOS observations of A2029 and A2142,
(c) and (d) SDSS data for all four clusters.
Dashed lines on panels (a) and (c) -- the median values of $\Delta \log \sigma$.
Dot-dashed lines on panels (b) and (d) -- one-to-one relations.
Solid lines on panels (b) and (d) -- the adopted correction
for the systematic errors, see Eq.\ \ref{eq-sigmasys1} and \ref{eq-sigmasys2}.
\label{fig-velsim} }
\end{figure}

Following the techniques used in J\o rgensen \& Chiboucas (2013) and J\o rgensen et al.\ (2017), we used simuations
to quantify any systematic effects on the derived velocity dispersions. Model spectra were made using the average 
of the two oldest SSP models from Maraston \& Str\"{o}mb\"{a}ck (2011) used for our template fitting. 
These are the models with (age, Z) = (5 Gyr, 0.02) and (age, Z) = (15 Gyr, 0.04). We then used information from the 
real noise spectra to add random noise to the model spectra. The simulations cover velocity dispersions between 
50 and 350 $\rm km\,s^{-1}$ and S/N $=6-50 {\rm \AA}^{-1}$. 
Simulations were performed for both the GMOS-N data and the SDSS data. The results are shown on Figure \ref{fig-velsim}.
The behavior of the two sets of simulations are very similar, though not identical. 
At S/N $= 6 {\rm \AA}^{-1}$ (crosses on Figure \ref{fig-velsim}), the simulations deviate significantly from the behavior at higher S/N.
Thus, these simulations are omitted from the analysis and for the real data we require S/N $\ge 10 {\rm \AA}^{-1}$ in order for
the velocity dispersion and other parameters to be included in our final measurements.
We then fit second order polynomials to the difference between input and output, as a function of the output velocity dispersions. 
The correction for the GMOS-N data was derived omitting the lowest velocity dispersion simulations as these deviate significantly from the remainder
of the simulations.
The resulting correction for the GMOS-N data is
\begin{equation}
\log \sigma _{\rm corrected} =\log \sigma _{\rm out} - 0.060 + 0.296(\log \sigma _{\rm out}-2.1) - 0.367(\log \sigma _{\rm out}-2.1)^2
\label{eq-sigmasys1}
\end{equation}
while the correction for the SDSS data is
\begin{equation}
\log \sigma _{\rm corrected} =\log \sigma _{\rm out} - 0.041 + 0.274(\log \sigma _{\rm out}-2.1) - 0.422(\log \sigma _{\rm out}-2.1)^2
\label{eq-sigmasys2}
\end{equation}

\subsection{Comparison of Measurements from Repeat Observations \label{SEC-COMPREPEAT} }

\begin{figure*}
\epsfxsize 14cm
\begin{center}
\epsfbox{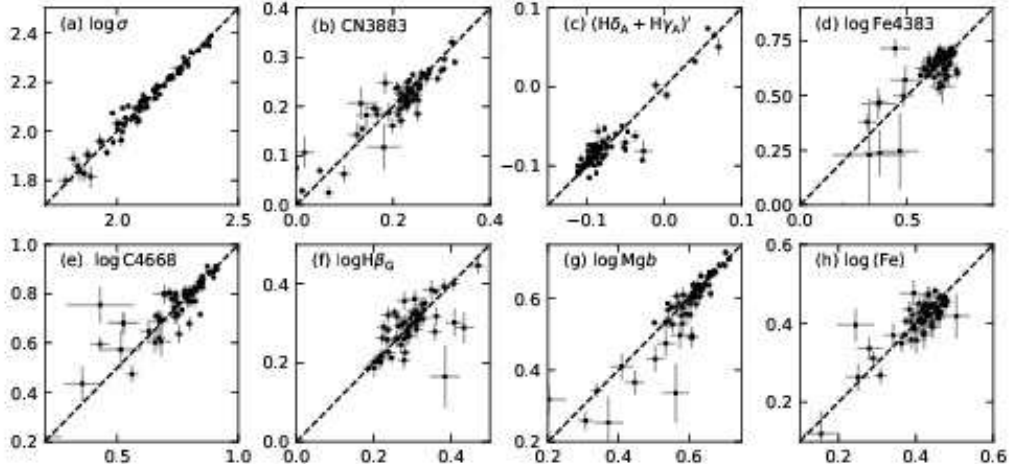}
\end{center}
\caption{Comparison of measurements from SDSS repeat observations of galaxies in the Perseus and Coma clusters.
Only the velocity dispersions and the line indices used in the analysis are included on the figure.
All comparisons are summarized in  Table \ref{tab-repeat}.
\label{fig-msdss_comp} }
\end{figure*}

\begin{figure*}
\epsfxsize 14cm
\begin{center}
\epsfbox{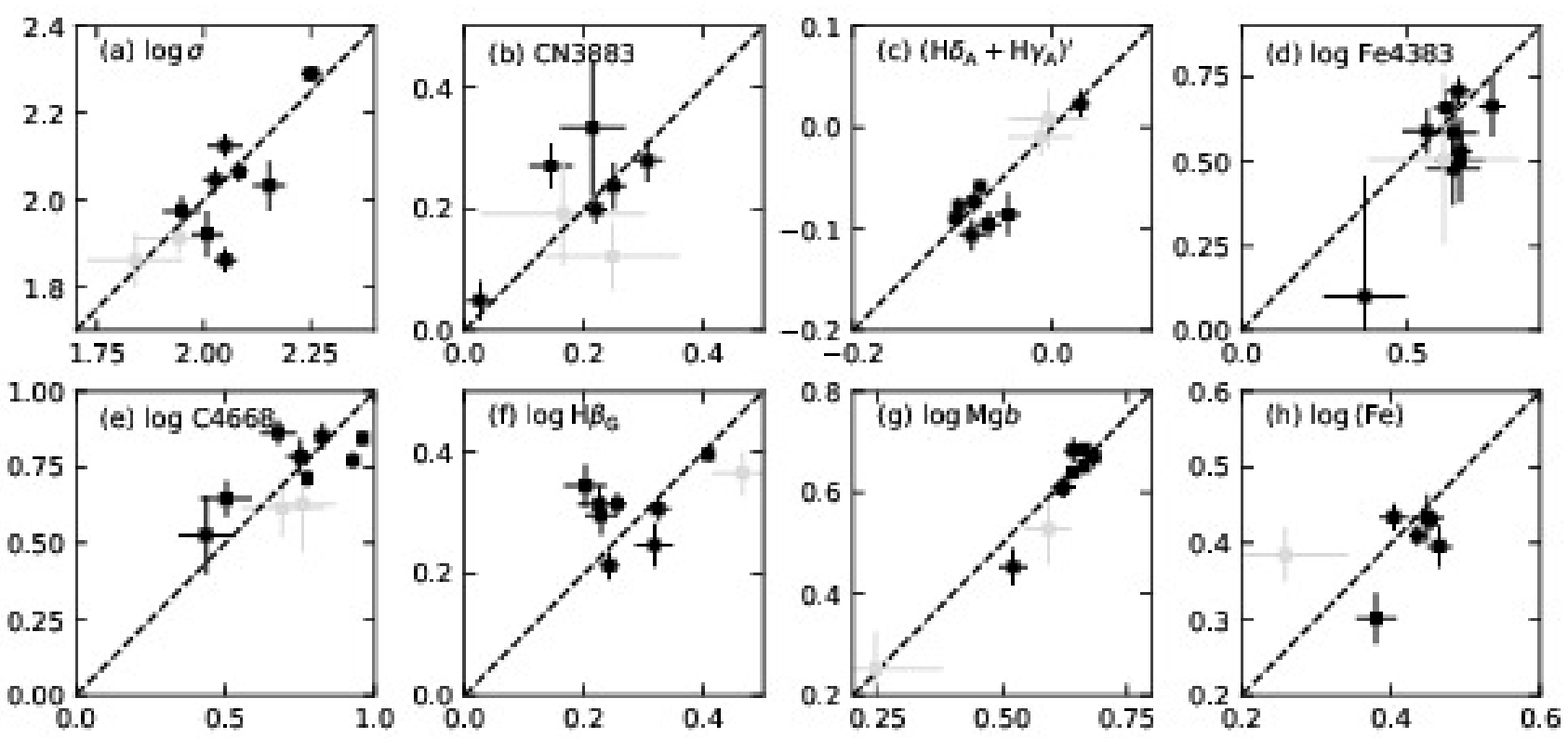}
\end{center}
\caption{Comparison of measurements from GMOS-N repeat observations of galaxies in A2029 and A2142. 
Only the velocity dispersions and the line indices used in the analysis are included on the figure.
All comparisons are summarized in  Table \ref{tab-repeat}.
Grey points -- measurements from spectra with S/N $< 15 {\rm \AA}^{-1}$. These measurements are 
omitted from the statistics in Table \ref{tab-repeat}.
\label{fig-gmos_comp} }
\end{figure*}

SDSS and GMOS-N repeat observations are compared in Figures \ref{fig-msdss_comp} and \ref{fig-gmos_comp}.
Table \ref{tab-repeat} summarizes the comparison. The table lists the ratios between the median 
internal measurement uncertainties based on the S/N of the spectra and the scatter in the comparisons.
The ratios are correlated, with the SDSS ratios in general being larger than the GMOS-N ratios, possibly due to
inadequate accounting for all uncertainties in the error spectra of the SDSS spectra. 
We estimate uncertainties for samples of galaxies by deriving the median values of the 
internal measurement uncertainties and then for each line index scale with average of the two ratios from Table \ref{tab-repeat}.
The results for the bulge-dominated passive galaxies in each cluster (our main samples for the analysis) are summarized in Table \ref{tab-uncadopt}.
These uncertainties are shown as typical error bars for the line indices on the figures in the main text. 

\begin{deluxetable*}{lrrrr rrrr}
\tablecaption{Comparison of Repeat Observations\label{tab-repeat} }
\tabletypesize{\scriptsize}
\tablewidth{0pc}
\tablehead{ 
  & \multicolumn{4}{c}{SDSS data} & \multicolumn{4}{c}{GMOS-N data} \\
\colhead{Parameter} & \colhead{N} & \colhead{rms} & \colhead{ratio} & \colhead{$\sigma _{\rm int}$}  
                    & \colhead{N} & \colhead{rms} & \colhead{ratio} & \colhead{$\sigma _{\rm int}$} \\ 
\colhead{(1)}  & \colhead{(2)} & \colhead{(3)} & \colhead{(4)} & \colhead{(5)} & \colhead{(6)} 
               & \colhead{(7)} & \colhead{(8)} & \colhead{(9)} }
\startdata
$\log \sigma$     &    70 & 0.029  &  1.6 & 0.013  &    9 & 0.089  &  2.2 & 0.028  \\
CN3883     &    67 & 0.030  &  2.2 & 0.010  &    6 & 0.070  &  1.6 & 0.036  \\
$\log {\rm H}\zeta _{\rm A}$    &    70 & 0.170  &  1.4 & 0.076  &    6 & 0.436  &  1.2 & 0.192  \\
log CaHK      &    70 & 0.034  &  2.6 & 0.009  &    6 & 0.056  &  1.4 & 0.034  \\
D4000      &    63 & 0.058  &  6.2 & 0.007  &    6 & 0.215  &  6.9 & 0.023  \\
$({\rm H}\delta _{\rm A} + {\rm H}\gamma _{\rm A})'$     &    70 & 0.016  &  3.3 & 0.003  &    8 & 0.022  &  1.5 & 0.010  \\
CN$_2$        &    70 & 0.015  &  1.8 & 0.006  &    9 & 0.033  &  1.3 & 0.018  \\
log G4300     &    70 & 0.077  &  3.5 & 0.015  &    9 & 0.118  &  1.7 & 0.042  \\
log Fe4383    &    70 & 0.090  &  2.4 & 0.024  &    9 & 0.111  &  1.1 & 0.072  \\
log C4668     &    70 & 0.110  &  3.9 & 0.020  &    9 & 0.114  &  2.2 & 0.038  \\
$\log {\rm H}\beta _{\rm G}$   &    65 & 0.051  &  2.2 & 0.017  &    8 & 0.072  &  2.0 & 0.025  \\
$\log {\rm Mg}b$       &    70 & 0.046  &  2.6 & 0.013  &    8 & 0.031  &  1.3 & 0.017  \\
$\log \left < {\rm Fe} \right >$        &    70 & 0.035  &  1.3 & 0.016  &    7 & 0.070  &  2.7 & 0.018  \\
\enddata
\tablecomments{Column 1: Spectroscopic parameter. Column 2: Number of galaxies in comparisons of the SDSS data.
Column 3: Scatter of the comparisons for the SDSS data. Column 4: The ration between the scatter of the comparison
and the expected scatter based on the measurement uncertainties on the SDSS data. Column 5: Median measurement uncertainties
of the measurements included in the comparisons of the SDSS data.
Columns 6--9: Number of galaxies, scatter, ration and median measurement uncertainties for the GMOS-N data.
}
\end{deluxetable*}

\begin{deluxetable*}{lrrr rrr rrr rrr}
\tablecaption{Adopted Uncertainties\label{tab-uncadopt} }
\tabletypesize{\scriptsize}
\tablewidth{0pc}
\tablehead{
  & \multicolumn{3}{c}{Perseus} & \multicolumn{3}{c}{Coma} & \multicolumn{3}{c}{A2029} & \multicolumn{3}{c}{A2142}\\
\colhead{Parameter} & \colhead{N} & \colhead{$\sigma _{\rm int}$} & \colhead{$\sigma _{\rm adopt}$}
                    & \colhead{N} & \colhead{$\sigma _{\rm int}$} & \colhead{$\sigma _{\rm adopt}$}
                    & \colhead{N} & \colhead{$\sigma _{\rm int}$} & \colhead{$\sigma _{\rm adopt}$}
                    & \colhead{N} & \colhead{$\sigma _{\rm int}$} & \colhead{$\sigma _{\rm adopt}$} \\ 
\colhead{(1)}  & \colhead{(2)} & \colhead{(3)} & \colhead{(4)} 
               & \colhead{(5)} & \colhead{(6)} & \colhead{(7)}
               & \colhead{(8)} & \colhead{(9)} & \colhead{(10)} 
               & \colhead{(11)} & \colhead{(12)} & \colhead{(13)}}
\startdata
$\log \sigma$  & 105 & 0.011 & 0.024 &  123 & 0.015 & 0.033 &  109 & 0.018 & 0.040 &  141 & 0.023 & 0.051  \\
CN3883 &  99 & 0.009 & 0.020 &  118 & 0.008 & 0.018 &  107 & 0.021 & 0.044 &  140 & 0.038 & 0.084  \\
$\log {\rm H}\zeta _{\rm A}$ & 94 & 0.081 & 0.113 &  119 & 0.064 & 0.090 &  101 & 0.146 & 0.202 &  110 & 0.235 & 0.329  \\
log CaHK & 103 & 0.009 & 0.023 &  119 & 0.007 & 0.018 &  108 & 0.019 & 0.049 &  141 & 0.030 & 0.078  \\
D4000 & 94 & 0.006 & 0.041 &  115 & 0.005 & 0.034 &  107 & 0.013 & 0.090 &  140 & 0.025 & 0.172  \\
$({\rm H}\delta _{\rm A} + {\rm H}\gamma _{\rm A})'$ & 103 & 0.003 & 0.010 &  119 & 0.003 & 0.010 &  108 & 0.006 & 0.020 &  140 & 0.011 & 0.036  \\
CN$_2$ & 103 & 0.006 & 0.011 &  119 & 0.005 & 0.009 &  109 & 0.011 & 0.020 &  141 & 0.019 & 0.034  \\
log G4300 & 103 & 0.013 & 0.045 &  119 & 0.011 & 0.038 &  109 & 0.026 & 0.091 &  141 & 0.039 & 0.140  \\
log Fe4383 & 103 & 0.018 & 0.043 &  119 & 0.019 & 0.046 &  109 & 0.043 & 0.106 &  141 & 0.062 & 0.149  \\
log C4668 & 103 & 0.012 & 0.047 &  119 & 0.017 & 0.066 &  109 & 0.029 & 0.117 &  141 & 0.044 & 0.172  \\
$\log {\rm H}\beta _{\rm G}$  & 101 & 0.018 & 0.040 &  122 & 0.018 & 0.040 &  109 & 0.024 & 0.055 &  140 & 0.040 & 0.088  \\
$\log {\rm Mg}b$ & 105 & 0.010 & 0.026 &  123 & 0.013 & 0.034 &  104 & 0.020 & 0.052 &  141 & 0.027 & 0.073  \\
$\log \left < {\rm Fe} \right >$  & 105 & 0.011 & 0.030 &  122 & 0.017 & 0.046 &  109 & 0.025 & 0.068 &  140 & 0.038 & 0.103  \\
\enddata
\tablecomments{Column 1: Spectroscopic parameter. 
Column 2: Number of passive bulge-dominated galaxies with the parameter measured, in Perseus.
Column 3: Median measurement uncertainty based on the S/N of the spectra, for galaxies in Perseus.
Column 4: Adopted scaled measurement uncertainty, for galaxies in Perseus, see text.
Columns 5--7: Number of galaxies, median measurement uncertainty, adopted scaled uncertainty, for Coma galaxies.
Columns 8--10: Number of galaxies, median measurement uncertainty, adopted scaled uncertainty, for A2029 galaxies.
Columns 11--13: Number of galaxies, median measurement uncertainty, adopted scaled uncertainty, for A2142 galaxies.
}
\end{deluxetable*}

\noindent
\begin{minipage}{18cm}
\vspace{0.5cm}
\subsection{Spectroscopic Parameters \label{SEC-SPECPARAMAPP} } 

Table \ref{tab-perseuslcs} lists the spectral parameters from the LCS data of Perseus galaxies.
Table \ref{tab-gmos} lists the spectral parameters from the GMOS-N data of A2029 and A2142 galaxies.
Table \ref{tab-sdss} lists the spectral parameters derived from the SDSS spectra of all four clusters.
Table \ref{tab-all} gives the final calibrated spectral parameters for all four clusters.
Only a portion of each of these tables are shown here, the tables are available in their entirety as machine-readable tables.
The velocity dispersions and line indices have been corrected to a standard size aperture equivalent to a
circular aperture with diameter of 3.4 arcsec at the distance of the Coma cluster.
The velocity dispersions from the GMOS-N data and SDSS data have been corrected for systematic effects as explained
in Section \ref{SEC-SIM}. 
The line indices have been corrected to zero velocity dispersion.
The calibration of all data to consistency is detailed in Section \ref{SEC-SPECCALIB}.
Table \ref{tab-columns} explains the columns in these tables other than the line indices and matching uncertainties.

Table \ref{tab-average} gives the average velocity dispersions and line indices for bins in velocity dispersion.
The table also lists ages, [M/H], [CN/Fe], and [Mg/Fe] derived from these.
\end{minipage}

% this table ought to go after the data tables and before the last table with average parameters, but it makes a mess of the formatting
% Instead I have forced the numbering on the rest of the tables to have them numbered in the order that they occur in the text
\begin{deluxetable*}{lrrrr l}
\tablecaption{Columns in Data Tables\label{tab-columns} }
\tabletypesize{\scriptsize}
\tablewidth{0pc}
\tablenum{17}
\tablehead{
\colhead{Column header} & \colhead{Table \ref{tab-perseuslcs}} & \colhead{Table \ref{tab-gmos}} & \colhead{Table \ref{tab-sdss}} &\colhead{Table \ref{tab-all}} & \colhead{Description}
}
\startdata
Cluster/ID    &  1 & 1 & 1 & 1 & Cluster name if more than one cluster in table, galaxy ID number \\
R.A.\ (J2000) &  \nodata & \nodata & \nodata & 2 & Right Ascension (J2000) in degrees \\
Dec (J2000)   &  \nodata & \nodata & \nodata & 3 & Declination (J2000) in degrees \\
Sample        &  \nodata & \nodata & \nodata & 4 & Sample number, 1=passive bulge-dominated brighter than analysis limit, \\
              &          &         &         &   & 2=emission line bulge-dominated brighter than analysis limit \\
              &          &         &         &   & 3=bulge-dominated fainter than analysis limit \\
              &          &         &         &   & 4=disk-dominated brighter than analysis limit \\
              &          &         &         &   & 5=disk-dominated fainter than analysis limit \\              
$F$           &  \nodata & \nodata & \nodata & 5 & Product of {\tt fracdev} from SDSS, $f_g \times f_r \times f_i$ \\
$g'_{\rm rest}$ & \nodata & \nodata & \nodata & 6 & SDSS {\tt cmodelmag} in $g'$, calibrated to rest frame \\
$(g'-r')_{\rm rest}$ & \nodata & \nodata & \nodata & 7 & SDDS color based on {\tt modelmag}, calibrated to rest frame \\
Redshift      & 2 & 2 & 2 & 8 & Redshift \\
$\log \sigma$ & 3 & 3 & 3 & 9 & Logarithm of the velocity dispersion in $\rm km\,s^{-1}$, calibrated to standard aperture diameter \\
              &   &   &   &   & of 3.4 arcsec at the distance of the Coma cluster and corrected for systematic effects \\
$\sigma _{\log \sigma}$ & 4 & 4 & 4 & 10 & Uncertainty on $\log \sigma$ \\
$f_{\rm 1Gyr}$ & \nodata & 5 & 5 & \nodata & Fraction of SSP model (age,Z) = (1Gyr,0.01) in kinematics fit \\
$f_{\rm 5Gyr}$ & \nodata & 6 & 6 & \nodata & Fraction of SSP model (age,Z) = (5Gyr,0.02) in kinematics fit \\
$f_{\rm 15Gyr}$ & \nodata & 7 & 7 & \nodata & Fraction of SSP model (age,Z) = (15Gyr,0.04) in kinematics fit \\
$\chi ^2$     & 5 & 8 & 8 & \nodata & $chi 2$ of the kinematics fit \\
S/N           & \nodata & 9 & 9 & 11 & S/N per {\AA}nstrom in the rest frame \\
S/N$_{43}$    & 6 & \nodata & \nodata & \nodata & S/N per {\AA}nstrom in the rest frame for LCS grating \#43 observations \\
S/N$_{47}$    & 7 & \nodata & \nodata & \nodata & S/N per {\AA}nstrom in the rest frame for LCS grating \#47 observations \\
Emis.         & \nodata & \nodata & \nodata & 12 & Emission line flags. First digit refers to [\ion{O}{2}], second digit to H$\beta$, \\
              &         &         &         &    & 0 or blank=no emission, 1=emission, 7=line is outside wavelength coverage \\
\enddata
\end{deluxetable*}

%\clearpage

% =============== PERSEUS LCS ======
\startlongtable
\begin{splitdeluxetable*}{l rrrrrr rr rr rr rr rr B rr rr rr rr rr rr rr rr rr rr}
% this table  needs to be rotated
\rotate
\tablecaption{Abell 426 / Perseus: LCS Data \label{tab-perseuslcs} }
\tablewidth{0pt}
\tabletypesize{\scriptsize}
\tablenum{13}
\tablehead{
\colhead{ID} & \colhead{Redshift} & \colhead{$\log \sigma$} & \colhead{$\sigma _{\log \sigma}$} 
& \colhead{$\chi ^2$} &\colhead{S/N$_{43}$} & \colhead{S/N$_{47}$} 
& \colhead{H$\zeta _{\rm A}$} & \colhead{$\sigma _{\rm H\zeta}$}
& \colhead{CN3883}  & \colhead{$\sigma _{\rm CN3883}$}
& \colhead{CaHK}    & \colhead{$\sigma _{\rm CaHK}$}
& \colhead{D4000}   & \colhead{$\sigma _{\rm D4000}$}
& \colhead{H$\delta _{\rm A}$} & \colhead{$\sigma _{\rm H\delta}$}
& \colhead{CN$_2$}  & \colhead{$\sigma _{\rm CN_2}$}
& \colhead{G4300}   & \colhead{$\sigma _{\rm G4300}$}
& \colhead{H$\gamma _{\rm A}$} & \colhead{$\sigma _{\rm H\gamma}$}
& \colhead{Fe4383}  & \colhead{$\sigma _{\rm Fe4383}$}
& \colhead{C4668} & \colhead{$\sigma _{\rm C4668}$} 
& \colhead{H$\beta$} & \colhead{$\sigma _{\rm H\beta}$}
& \colhead{H$\beta _{\rm G}$} & \colhead{$\sigma _{\rm H\beta _{\rm G}}$}
& \colhead{Mg$b$} & \colhead{$\sigma _{{\rm Mg}b}$}
& \colhead{Fe5270} & \colhead{$\sigma _{\rm Fe5270}$}
& \colhead{Fe5335} & \colhead{$\sigma _{\rm Fe5335}$} \\
\colhead{(1)} & \colhead{(2)} & \colhead{(3)} & \colhead{(4)} & \colhead{(5)} & \colhead{(6)} & \colhead{(7)} & \colhead{(8)} & \colhead{(9)} & \colhead{(10)} &
\colhead{(11)} & \colhead{(12)} & \colhead{(13)} & \colhead{(14)} & \colhead{(15)} & \colhead{(16)} & \colhead{(17)} & \colhead{(18)} & \colhead{(19)} & \colhead{(20)} &
\colhead{(21)} & \colhead{(22)} & \colhead{(23)} & \colhead{(24)} & \colhead{(25)} & \colhead{(26)} & \colhead{(27)} & \colhead{(28)} & \colhead{(29)} & \colhead{(30)} &
\colhead{(31)} & \colhead{(32)} & \colhead{(33)} & \colhead{(34)} & \colhead{(35)} & \colhead{(36)} & \colhead{(37)}
}
\startdata
12074 & 0.01616 & 2.240 & 0.013 & 0.8 & \nodata & 23.0 & \nodata & \nodata & \nodata & \nodata & \nodata & \nodata & \nodata & \nodata & \nodata & \nodata & \nodata & \nodata & \nodata & \nodata & \nodata & \nodata & \nodata & \nodata & \nodata & \nodata & 1.69 & 0.29 & 1.91 & 0.24 & 4.07 & 0.24 & 3.13 & 0.23 & 3.02 & 0.22 \\
12098 & 0.02036 & 2.317 & 0.010 & 1.0 & 21.9 & 30.7 & 0.05 & 0.93 & 0.246 & 0.018 & 21.73 & 0.85 & 2.212 & 0.011 & -2.12 & 0.50 & 0.112 & 0.014 & 5.51 & 0.33 & -6.38 & 0.40 & 6.31 & 0.32 & 8.77 & 0.29 & 2.39 & 0.21 & 2.41 & 0.17 & 4.80 & 0.17 & 3.16 & 0.18 & 3.31 & 0.17 \\
12141 & 0.01395 & 2.249 & 0.018 & 1.0 & \nodata & 22.2 & \nodata & \nodata & \nodata & \nodata & \nodata & \nodata & \nodata & \nodata & \nodata & \nodata & \nodata & \nodata & \nodata & \nodata & \nodata & \nodata & \nodata & \nodata & \nodata & \nodata & 1.68 & 0.29 & 1.94 & 0.25 & 4.73 & 0.24 & 2.86 & 0.24 & 2.84 & 0.22 \\
12152 & 0.02095 & 2.290 & 0.011 & 1.2 & 24.8 & 29.5 & 1.66 & 0.73 & 0.297 & 0.015 & 22.72 & 0.71 & 2.140 & 0.009 & -2.77 & 0.43 & 0.112 & 0.012 & 5.70 & 0.29 & -6.22 & 0.35 & 4.98 & 0.29 & 7.85 & 0.26 & 2.25 & 0.21 & 2.39 & 0.17 & 4.49 & 0.19 & 3.06 & 0.18 & 2.83 & 0.17 \\
12157 & 0.01623 & 2.239 & 0.010 & 1.2 & 25.7 & 32.0 & -0.99 & 0.79 & 0.178 & 0.014 & 21.07 & 0.70 & 2.180 & 0.009 & -1.26 & 0.41 & 0.093 & 0.011 & 5.00 & 0.28 & -5.46 & 0.34 & 5.47 & 0.28 & 7.16 & 0.25 & 2.02 & 0.20 & 2.22 & 0.16 & 3.89 & 0.17 & 3.55 & 0.17 & 3.25 & 0.16 \\
12171 & 0.01881 & 2.346 & 0.012 & 1.2 & 27.5 & 28.6 & 1.51 & 0.46 & 0.149 & 0.009 & 15.08 & 0.55 & 1.554 & 0.006 & -0.06 & 0.34 & 0.089 & 0.010 & 3.70 & 0.26 & -4.03 & 0.30 & 3.75 & 0.27 & 7.51 & 0.26 & 0.89 & 0.22 & 1.00 & 0.18 & 4.12 & 0.19 & 2.55 & 0.19 & 2.88 & 0.18 \\
12176 & 0.01955 & 2.389 & 0.008 & 1.1 & \nodata & 28.0 & \nodata & \nodata & \nodata & \nodata & \nodata & \nodata & \nodata & \nodata & \nodata & \nodata & \nodata & \nodata & \nodata & \nodata & \nodata & \nodata & \nodata & \nodata & \nodata & \nodata & 1.38 & 0.24 & 2.04 & 0.19 & 4.80 & 0.19 & 2.63 & 0.19 & 2.83 & 0.18 \\
12193 & 0.02418 & 2.481 & 0.013 & 1.4 & 24.3 & 30.3 & -0.47 & 0.86 & 0.298 & 0.016 & 20.89 & 0.77 & 2.212 & 0.010 & -3.19 & 0.46 & 0.145 & 0.013 & 6.08 & 0.30 & -6.64 & 0.37 & 5.01 & 0.30 & 6.95 & 0.27 & 1.89 & 0.21 & 2.14 & 0.17 & 5.02 & 0.18 & 3.31 & 0.18 & 2.44 & 0.17 \\
12203 & 0.02035 & 2.196 & 0.010 & 1.0 & 23.5 & 28.2 & 1.27 & 0.74 & 0.252 & 0.015 & 22.24 & 0.75 & 2.030 & 0.009 & -1.76 & 0.45 & 0.123 & 0.013 & 5.17 & 0.31 & -5.59 & 0.37 & 5.52 & 0.30 & 7.68 & 0.27 & 1.78 & 0.23 & 1.99 & 0.18 & 4.56 & 0.19 & 2.93 & 0.19 & 2.92 & 0.18 \\
12208 & 0.01928 & 2.306 & 0.015 & 1.1 & 16.4 & 23.4 & 1.40 & 0.99 & 0.238 & 0.021 & 18.70 & 1.09 & 1.760 & 0.012 & -2.07 & 0.65 & 0.132 & 0.018 & 5.17 & 0.45 & -5.24 & 0.53 & 5.01 & 0.44 & 7.50 & 0.40 & 1.78 & 0.28 & 2.04 & 0.23 & 4.99 & 0.23 & 3.08 & 0.23 & 3.29 & 0.22 \\
\enddata
\tablecomments{Columns are explained in Table \ref{tab-columns}. \\ This table is available in its entirety in machine-readable form.}
\end{splitdeluxetable*}

% =========== A2029 A2142 GMOS-N =============
\startlongtable
\begin{splitdeluxetable*}{l rrrrrrrr rr rr rr rr rr B rr rr rr rr rr rr rr rr rr rr}
% this table  needs to be rotated
\rotate
\tablecaption{Abell 2029 and Abell 2142: GMOS-N Data \label{tab-gmos} }
\tablewidth{0pt}
\tabletypesize{\scriptsize}
\tablenum{14}
\tablehead{
\colhead{Cluster/ID} & \colhead{Redshift} & \colhead{$\log \sigma$} & \colhead{$\sigma _{\log \sigma}$} 
& \colhead{$f_{\rm 1Gyr}$} & \colhead{$f_{\rm 5Gyr}$} & \colhead{$f_{\rm 15Gyr}$} 
& \colhead{$\chi ^2$} &\colhead{S/N} 
& \colhead{H$\zeta _{\rm A}$} & \colhead{$\sigma _{\rm H\zeta}$}
& \colhead{CN3883}  & \colhead{$\sigma _{\rm CN3883}$}
& \colhead{CaHK}    & \colhead{$\sigma _{\rm CaHK}$}
& \colhead{D4000}   & \colhead{$\sigma _{\rm D4000}$}
& \colhead{H$\delta _{\rm A}$} & \colhead{$\sigma _{\rm H\delta}$}
& \colhead{CN$_2$}  & \colhead{$\sigma _{\rm CN_2}$}
& \colhead{G4300}   & \colhead{$\sigma _{\rm G4300}$}
& \colhead{H$\gamma _{\rm A}$} & \colhead{$\sigma _{\rm H\gamma}$}
& \colhead{Fe4383}  & \colhead{$\sigma _{\rm Fe4383}$}
& \colhead{C4668} & \colhead{$\sigma _{\rm C4668}$} 
& \colhead{H$\beta$} & \colhead{$\sigma _{\rm H\beta}$}
& \colhead{H$\beta _{\rm G}$} & \colhead{$\sigma _{\rm H\beta _{\rm G}}$}
& \colhead{Mg$b$} & \colhead{$\sigma _{{\rm Mg}b}$}
& \colhead{Fe5270} & \colhead{$\sigma _{\rm Fe5270}$}
& \colhead{Fe5335} & \colhead{$\sigma _{\rm Fe5335}$} \\
\colhead{(1)} & \colhead{(2)} & \colhead{(3)} & \colhead{(4)} & \colhead{(5)} & \colhead{(6)} & \colhead{(7)} & \colhead{(8)} & \colhead{(9)} & \colhead{(10)} &
\colhead{(11)} & \colhead{(12)} & \colhead{(13)} & \colhead{(14)} & \colhead{(15)} & \colhead{(16)} & \colhead{(17)} & \colhead{(18)} & \colhead{(19)} & \colhead{(20)} &
\colhead{(21)} & \colhead{(22)} & \colhead{(23)} & \colhead{(24)} & \colhead{(25)} & \colhead{(26)} & \colhead{(27)} & \colhead{(28)} & \colhead{(29)} & \colhead{(30)} &
\colhead{(31)} & \colhead{(32)} & \colhead{(33)} & \colhead{(34)} & \colhead{(35)} & \colhead{(36)} & \colhead{(37)} & \colhead{(38)} & \colhead{(39)}
}
\startdata
{\bf A2029:} \\
9 & 0.08073 & 1.878 & 0.041 & 0.16 & 0.84 & 0.00 & 0.7 & 20.4 & \nodata & \nodata & \nodata & \nodata & \nodata & \nodata & \nodata & \nodata & \nodata & \nodata & 0.182 & 0.036 & 6.60 & 0.83 & -6.82 & 0.94 & 3.00 & 1.05 & 5.70 & 0.78 & 2.15 & 0.29 & 2.23 & 0.19 & 4.17 & 0.33 & 2.16 & 0.30 & 1.96 & 0.32 \\
16 & 0.07482 & 2.129 & 0.021 & 0.32 & 0.23 & 0.46 & 1.0 & 37.1 & 0.70 & 0.97 & \nodata & \nodata & 24.12 & 1.56 & \nodata & \nodata & -1.56 & 0.63 & 0.113 & 0.019 & 5.16 & 0.44 & -5.40 & 0.48 & 4.53 & 0.55 & 6.69 & 0.43 & 1.60 & 0.16 & 1.89 & 0.10 & 4.55 & 0.15 & 2.43 & 0.15 & 2.79 & 0.16 \\
17 & 0.08048 & 2.033 & 0.030 & 0.15 & 0.82 & 0.03 & 1.1 & 30.9 & 0.74 & 1.14 & \nodata & \nodata & 21.96 & 1.93 & \nodata & \nodata & -1.52 & 0.75 & 0.036 & 0.022 & 5.26 & 0.53 & -4.27 & 0.56 & 3.76 & 0.66 & 5.26 & 0.52 & 1.77 & 0.19 & 1.90 & 0.12 & 4.27 & 0.18 & 2.95 & 0.17 & 2.37 & 0.19 \\
43 & 0.07643 & 2.033 & 0.036 & 0.01 & 0.81 & 0.18 & 0.6 & 16.8 & -0.71 & 1.91 & 0.129 & 0.065 & 18.74 & 3.35 & 1.722 & 0.036 & -1.68 & 1.36 & 0.010 & 0.000 & 6.70 & 0.99 & -6.66 & 1.15 & 4.70 & 1.26 & 1.02 & 1.02 & 1.85 & 0.36 & 2.20 & 0.23 & 3.00 & 0.33 & 2.47 & 0.32 & 2.28 & 0.36 \\
62 & 0.08236 & 1.881 & 0.033 & 0.29 & 0.39 & 0.33 & 0.7 & 21.5 & \nodata & \nodata & \nodata & \nodata & \nodata & \nodata & \nodata & \nodata & \nodata & \nodata & 0.150 & 0.034 & 4.20 & 0.82 & -3.83 & 0.83 & 3.90 & 0.97 & 7.68 & 0.74 & 1.39 & 0.28 & 1.95 & 0.18 & 4.20 & 0.26 & 3.23 & 0.25 & 2.90 & 0.28 \\
92 & 0.07363 & 1.961 & 0.026 & 0.39 & 0.07 & 0.54 & 2.0 & 55.7 & 1.39 & 0.58 & 0.240 & 0.023 & 23.47 & 0.98 & 2.208 & 0.016 & -1.92 & 0.41 & 0.079 & 0.012 & 5.23 & 0.28 & -5.21 & 0.31 & 4.40 & 0.35 & 5.35 & 0.29 & 1.90 & 0.11 & 2.15 & 0.07 & 3.90 & 0.10 & 2.85 & 0.10 & 2.48 & 0.11 \\
115 & 0.08586 & 1.773 & 0.035 & 0.34 & 0.00 & 0.66 & 0.6 & 23.2 & -0.25 & 1.80 & 0.093 & 0.066 & 19.10 & 2.92 & 2.337 & 0.047 & -1.80 & 1.07 & 0.060 & 0.030 & 5.25 & 0.72 & -5.11 & 0.77 & 4.34 & 0.88 & 5.06 & 0.70 & 2.04 & 0.25 & 2.31 & 0.16 & 3.41 & 0.23 & 2.20 & 0.24 & 3.32 & 0.26 \\
131 & 0.08266 & 2.132 & 0.039 & 0.00 & 0.90 & 0.10 & 1.3 & 30.9 & 0.22 & 1.16 & 0.207 & 0.044 & 25.33 & 1.88 & 2.151 & 0.030 & 0.20 & 0.76 & 0.058 & 0.023 & 6.82 & 0.53 & -5.62 & 0.59 & 4.01 & 0.66 & 6.04 & 0.53 & 1.77 & 0.19 & 1.85 & 0.12 & 4.98 & 0.17 & 2.66 & 0.17 & 2.67 & 0.19 \\
135 & 0.08121 & 1.988 & 0.038 & 0.21 & 0.00 & 0.79 & 0.9 & 22.5 & -0.21 & 1.62 & 0.297 & 0.060 & 20.64 & 2.73 & 2.051 & 0.039 & -2.36 & 1.12 & 0.143 & 0.032 & 5.73 & 0.75 & -5.67 & 0.82 & 3.31 & 0.94 & 6.64 & 0.72 & 1.53 & 0.26 & 1.89 & 0.17 & 4.45 & 0.25 & 2.87 & 0.24 & 2.56 & 0.27 \\
137 & 0.08446 & 2.083 & 0.030 & 0.41 & 0.10 & 0.49 & 1.0 & 33.7 & 2.82 & 0.80 & 0.244 & 0.034 & 22.86 & 1.54 & 1.981 & 0.022 & 0.45 & 0.63 & 0.026 & 0.020 & 3.62 & 0.50 & -4.10 & 0.50 & 5.35 & 0.57 & 6.44 & 0.48 & 1.59 & 0.17 & 1.72 & 0.11 & 4.13 & 0.16 & 2.98 & 0.16 & 2.44 & 0.18 \\
\enddata
\tablecomments{Columns are explained in Table \ref{tab-columns}. \\ This table is available in its entirety in machine-readable form.}
\end{splitdeluxetable*}

% ============= My SDSS measurements - all four clusters ===========
\startlongtable
\begin{splitdeluxetable*}{l rrrrrrrr rr rr rr rr rr B rr rr rr rr rr rr rr rr rr rr}
% this table  needs to be rotated
\rotate
\tablecaption{Perseus, Coma, A2029 and A2142: Measurements from SDSS Spectra \label{tab-sdss} }
\tablewidth{0pt}
\tabletypesize{\scriptsize}
\tablenum{15}
\tablehead{
\colhead{Cluster/ID} & \colhead{Redshift} & \colhead{$\log \sigma$} & \colhead{$\sigma _{\log \sigma}$} 
& \colhead{$f_{\rm 1Gyr}$} & \colhead{$f_{\rm 5Gyr}$} & \colhead{$f_{\rm 15Gyr}$} 
& \colhead{$\chi ^2$} &\colhead{S/N} 
& \colhead{H$\zeta _{\rm A}$} & \colhead{$\sigma _{\rm H\zeta}$}
& \colhead{CN3883}  & \colhead{$\sigma _{\rm CN3883}$}
& \colhead{CaHK}    & \colhead{$\sigma _{\rm CaHK}$}
& \colhead{D4000}   & \colhead{$\sigma _{\rm D4000}$}
& \colhead{H$\delta _{\rm A}$} & \colhead{$\sigma _{\rm H\delta}$}
& \colhead{CN$_2$}  & \colhead{$\sigma _{\rm CN_2}$}
& \colhead{G4300}   & \colhead{$\sigma _{\rm G4300}$}
& \colhead{H$\gamma _{\rm A}$} & \colhead{$\sigma _{\rm H\gamma}$}
& \colhead{Fe4383}  & \colhead{$\sigma _{\rm Fe4383}$}
& \colhead{C4668} & \colhead{$\sigma _{\rm C4668}$} 
& \colhead{H$\beta$} & \colhead{$\sigma _{\rm H\beta}$}
& \colhead{H$\beta _{\rm G}$} & \colhead{$\sigma _{\rm H\beta _{\rm G}}$}
& \colhead{Mg$b$} & \colhead{$\sigma _{{\rm Mg}b}$}
& \colhead{Fe5270} & \colhead{$\sigma _{\rm Fe5270}$}
& \colhead{Fe5335} & \colhead{$\sigma _{\rm Fe5335}$} \\
\colhead{(1)} & \colhead{(2)} & \colhead{(3)} & \colhead{(4)} & \colhead{(5)} & \colhead{(6)} & \colhead{(7)} & \colhead{(8)} & \colhead{(9)} & \colhead{(10)} &
\colhead{(11)} & \colhead{(12)} & \colhead{(13)} & \colhead{(14)} & \colhead{(15)} & \colhead{(16)} & \colhead{(17)} & \colhead{(18)} & \colhead{(19)} & \colhead{(20)} &
\colhead{(21)} & \colhead{(22)} & \colhead{(23)} & \colhead{(24)} & \colhead{(25)} & \colhead{(26)} & \colhead{(27)} & \colhead{(28)} & \colhead{(29)} & \colhead{(30)} &
\colhead{(31)} & \colhead{(32)} & \colhead{(33)} & \colhead{(34)} & \colhead{(35)} & \colhead{(36)} & \colhead{(37)} & \colhead{(38)} & \colhead{(39)}
}
\startdata
{\bf Perseus:} \\
12074 & 0.01621 & 2.069 & 0.017 & 0.10 & 0.62 & 0.28 & 3.0 & 137.2 & 0.95 & 0.10 & 0.203 & 0.004 & 19.60 & 0.18 & \nodata & \nodata & -1.57 & 0.08 & 0.103 & 0.002 & 4.48 & 0.06 & -5.41 & 0.07 & 4.71 & 0.09 & 7.77 & 0.10 & 1.42 & 0.04 & 1.62 & 0.03 & 4.13 & 0.04 & 2.99 & 0.05 & 2.72 & 0.06 \\
12097 & 0.01981 & 1.670 & 0.011 & 0.81 & 0.19 & 0.00 & 0.6 & 56.3 & 3.20 & 0.17 & 0.041 & 0.007 & 16.30 & 0.40 & 1.525 & 0.004 & 2.57 & 0.16 & -0.005 & 0.000 & 2.13 & 0.16 & -0.28 & 0.16 & 2.81 & 0.23 & 4.11 & 0.26 & 0.67 & 0.11 & 0.71 & 0.07 & 2.20 & 0.11 & 2.05 & 0.12 & 1.73 & 0.14 \\
12119 & 0.01546 & 2.064 & 0.018 & 0.08 & 0.81 & 0.10 & 1.2 & 83.5 & -1.12 & 0.17 & 0.023 & 0.006 & 5.95 & 0.32 & 1.968 & 0.004 & -0.92 & 0.12 & 0.052 & 0.004 & 2.71 & 0.11 & -2.74 & 0.11 & 2.62 & 0.15 & 4.28 & 0.18 & 1.03 & 0.07 & 1.24 & 0.05 & 2.78 & 0.07 & 1.79 & 0.08 & 2.15 & 0.10 \\
12132 & 0.01563 & 1.888 & 0.021 & 1.00 & 0.00 & 0.00 & 0.4 & 42.3 & 2.05 & 0.24 & -0.011 & 0.000 & 11.74 & 0.62 & \nodata & \nodata & 3.05 & 0.23 & -0.009 & 0.000 & 2.28 & 0.22 & -2.76 & 0.23 & 1.17 & 0.32 & 1.67 & 0.34 & -6.30 & 0.15 & -6.09 & 0.11 & 1.87 & 0.12 & 1.30 & 0.14 & 1.31 & 0.16 \\
12133 & 0.01783 & 2.098 & 0.014 & 0.48 & 0.21 & 0.31 & 1.1 & 88.3 & 1.96 & 0.15 & 0.126 & 0.007 & 18.38 & 0.33 & 1.971 & 0.004 & 0.42 & 0.13 & 0.026 & 0.004 & 3.95 & 0.11 & -3.16 & 0.12 & 3.68 & 0.16 & 5.61 & 0.17 & 1.35 & 0.07 & 1.56 & 0.04 & 3.36 & 0.06 & 2.46 & 0.07 & 2.36 & 0.08 \\
12160 & 0.01654 & 1.993 & 0.008 & 0.27 & 0.69 & 0.04 & 1.6 & 127.6 & 1.50 & 0.10 & 0.135 & 0.004 & 19.68 & 0.20 & 1.930 & 0.003 & -0.11 & 0.08 & 0.020 & 0.003 & 4.42 & 0.07 & -3.94 & 0.08 & 3.69 & 0.10 & 4.69 & 0.12 & 1.81 & 0.05 & 2.01 & 0.03 & 3.21 & 0.05 & 2.63 & 0.05 & 2.19 & 0.06 \\
12166 & 0.01346 & 1.873 & 0.022 & 1.00 & 0.00 & 0.00 & 0.6 & 56.4 & 4.42 & 0.16 & \nodata & \nodata & 15.04 & 0.41 & \nodata & \nodata & 4.49 & 0.16 & -0.012 & 0.000 & 1.60 & 0.16 & 1.11 & 0.15 & 0.96 & 0.23 & 0.01 & 0.27 & -1.60 & 0.11 & -1.46 & 0.08 & 2.04 & 0.11 & 1.47 & 0.12 & 1.55 & 0.14 \\
12171 & 0.01882 & 2.266 & 0.007 & 0.49 & 0.11 & 0.40 & 1.4 & 112.7 & 1.67 & 0.08 & 0.128 & 0.003 & 13.73 & 0.19 & 1.650 & 0.002 & 0.49 & 0.08 & 0.077 & 0.003 & 3.02 & 0.07 & -3.12 & 0.08 & 3.96 & 0.10 & 6.24 & 0.12 & 0.45 & 0.05 & 0.64 & 0.04 & 3.87 & 0.05 & 2.38 & 0.06 & 2.26 & 0.07 \\
12176 & 0.01971 & 2.319 & 0.011 & 0.00 & 0.72 & 0.28 & 2.3 & 91.9 & 0.59 & 0.13 & 0.294 & 0.005 & 22.28 & 0.23 & 2.293 & 0.004 & -1.69 & 0.10 & 0.128 & 0.003 & 5.24 & 0.08 & -5.72 & 0.10 & 4.68 & 0.13 & 7.97 & 0.15 & 1.51 & 0.06 & 1.76 & 0.04 & 4.60 & 0.06 & 2.82 & 0.07 & 2.51 & 0.09 \\
12185 & 0.02097 & 2.260 & 0.010 & 0.11 & 0.61 & 0.28 & 2.5 & 106.1 & 1.08 & 0.14 & 0.285 & 0.005 & 22.48 & 0.24 & 2.321 & 0.004 & -1.99 & 0.10 & 0.094 & 0.003 & 5.21 & 0.08 & -5.94 & 0.09 & 5.54 & 0.11 & 7.16 & 0.13 & 1.52 & 0.05 & 1.83 & 0.04 & 4.40 & 0.05 & 2.94 & 0.06 & 2.56 & 0.07 \\
\enddata
\tablecomments{Columns are explained in Table \ref{tab-columns}. \\ This table is available in its entirety in machine-readable form.}
\end{splitdeluxetable*}

% ============= Fully calibrated - all four clusters ===========
\startlongtable
\begin{splitdeluxetable*}{l rrrrrrrrrrr rr rr rr rr B rr rr rr rr rr rr rr rr rr rr rr}
% this table  needs to be rotated
\rotate
\tablecaption{Perseus, Coma, A2029 and A2142: Fully Calibrated Spectral Parameters\label{tab-all} }
\tablewidth{0pt}
\tabletypesize{\scriptsize}
\tablenum{16}
\tablehead{
\colhead{Cluster/ID} & \colhead{R.A.\ (J2000)} & \colhead{Dec (J2000)} & \colhead{Sample} 
& \colhead{$F$} & \colhead{$g'_{\rm rest}$} & \colhead{$(g'-r')_{\rm rest}$} 
& \colhead{Redshift} & \colhead{$\log \sigma$} & \colhead{$\sigma _{\log \sigma}$} &\colhead{S/N} & \colhead{Emis.}
& \colhead{H$\zeta _{\rm A}$} & \colhead{$\sigma _{\rm H\zeta}$}
& \colhead{CN3883}  & \colhead{$\sigma _{\rm CN3883}$}
& \colhead{CaHK}    & \colhead{$\sigma _{\rm CaHK}$}
& \colhead{D4000}   & \colhead{$\sigma _{\rm D4000}$}
& \colhead{H$\delta _{\rm A}$} & \colhead{$\sigma _{\rm H\delta}$}
& \colhead{CN$_2$}  & \colhead{$\sigma _{\rm CN_2}$}
& \colhead{G4300}   & \colhead{$\sigma _{\rm G4300}$}
& \colhead{H$\gamma _{\rm A}$} & \colhead{$\sigma _{\rm H\gamma}$}
& \colhead{Fe4383}  & \colhead{$\sigma _{\rm Fe4383}$}
& \colhead{C4668} & \colhead{$\sigma _{\rm C4668}$} 
& \colhead{H$\beta$} & \colhead{$\sigma _{\rm H\beta}$}
& \colhead{H$\beta _{\rm G}$} & \colhead{$\sigma _{\rm H\beta _{\rm G}}$}
& \colhead{Mg$b$} & \colhead{$\sigma _{{\rm Mg}b}$}
& \colhead{Fe5270} & \colhead{$\sigma _{\rm Fe5270}$}
& \colhead{Fe5335} & \colhead{$\sigma _{\rm Fe5335}$} \\
\colhead{(1)} & \colhead{(2)} & \colhead{(3)} & \colhead{(4)} & \colhead{(5)} & \colhead{(6)} & \colhead{(7)} & \colhead{(8)} & \colhead{(9)} & \colhead{(10)} &
\colhead{(11)} & \colhead{(12)} & \colhead{(13)} & \colhead{(14)} & \colhead{(15)} & \colhead{(16)} & \colhead{(17)} & \colhead{(18)} & \colhead{(19)} & \colhead{(20)} &
\colhead{(21)} & \colhead{(22)} & \colhead{(23)} & \colhead{(24)} & \colhead{(25)} & \colhead{(26)} & \colhead{(27)} & \colhead{(28)} & \colhead{(29)} & \colhead{(30)} &
\colhead{(31)} & \colhead{(32)} & \colhead{(33)} & \colhead{(34)} & \colhead{(35)} & \colhead{(36)} & \colhead{(37)} & \colhead{(38)} &
\colhead{(39)} & \colhead{(40)} & \colhead{(41)} & \colhead{(42)}
}
\startdata
{\bf Perseus:} \\
12074 & 48.69888 & 42.22270 & 1 & 0.54 & 13.63 & 0.894 & 0.01619 & 2.182 & 0.011 & 139.1 & 70 & 0.95 & 0.10 & 0.203 & 0.004 & 18.85 & 0.17 & \nodata & \nodata & -1.57 & 0.08 & 0.116 & 0.002 & 4.65 & 0.07 & -5.41 & 0.07 & 4.71 & 0.09 & 8.06 & 0.11 & 1.62 & 0.14 & 1.82 & 0.12 & 4.20 & 0.13 & 3.08 & 0.13 & 3.00 & 0.13 \\
12097 & 48.83565 & 41.61246 & 4 & 0.00 & 14.89 & 0.498 & 0.01981 & 1.726 & 0.011 & 56.3 & 11 & 3.20 & 0.17 & 0.041 & 0.007 & 15.67 & 0.39 & 1.525 & 0.004 & 2.57 & 0.16 & 0.008 & 0.000 & 2.21 & 0.17 & -0.28 & 0.16 & 2.81 & 0.23 & 4.26 & 0.27 & 0.73 & 0.12 & 0.75 & 0.08 & 2.31 & 0.11 & 2.08 & 0.12 & 1.89 & 0.15 \\
12098 & 48.83789 & 41.35541 & 1 & 0.26 & 13.29 & 0.791 & 0.02036 & 2.317 & 0.010 & 30.7 &  0 & \nodata & \nodata & 0.246 & 0.018 & 21.73 & 0.86 & 2.212 & 0.011 & -2.12 & 0.50 & 0.112 & 0.014 & 5.51 & 0.34 & -6.38 & 0.40 & 6.31 & 0.35 & 8.77 & 0.30 & 2.39 & 0.21 & 2.41 & 0.17 & 4.80 & 0.19 & 3.16 & 0.20 & 3.31 & 0.20 \\
12119 & 48.96732 & 41.85770 & 4 & 1.00 & 14.69 & 0.860 & 0.01546 & 2.120 & 0.018 & 83.5 & 70 & \nodata & \nodata & 0.023 & 0.006 & 5.72 & 0.31 & 1.968 & 0.004 & -0.92 & 0.12 & 0.065 & 0.004 & 2.81 & 0.11 & -2.74 & 0.11 & 2.62 & 0.15 & 4.44 & 0.18 & 1.12 & 0.08 & 1.32 & 0.05 & 2.92 & 0.08 & 1.82 & 0.09 & 2.35 & 0.10 \\
12132 & 49.00324 & 40.88569 & 4 & 0.03 & 13.80 & 0.432 & 0.01563 & 1.944 & 0.021 & 42.3 & 71 & 2.05 & 0.24 & -0.011 & 0.000 & 11.29 & 0.60 & \nodata & \nodata & 3.05 & 0.23 & 0.004 & 0.000 & 2.37 & 0.23 & -2.76 & 0.23 & 1.17 & 0.32 & 1.73 & 0.35 & \nodata & \nodata & \nodata & \nodata & 1.97 & 0.13 & 1.32 & 0.14 & 1.43 & 0.17 \\
12133 & 49.00412 & 42.07430 & 1 & 0.36 & 14.11 & 0.885 & 0.01783 & 2.154 & 0.014 & 88.3 & 70 & 1.96 & 0.15 & 0.126 & 0.007 & 17.67 & 0.32 & 1.971 & 0.004 & 0.42 & 0.13 & 0.039 & 0.004 & 4.11 & 0.12 & -3.16 & 0.12 & 3.68 & 0.16 & 5.82 & 0.18 & 1.47 & 0.07 & 1.66 & 0.05 & 3.52 & 0.07 & 2.49 & 0.07 & 2.58 & 0.09 \\
12141 & 49.02619 & 40.80464 & 1 & 0.87 & 13.49 & 0.807 & 0.01395 & 2.249 & 0.018 & 22.2 & 70 & \nodata & \nodata & \nodata & \nodata & \nodata & \nodata & \nodata & \nodata & \nodata & \nodata & \nodata & \nodata & \nodata & \nodata & \nodata & \nodata & \nodata & \nodata & \nodata & \nodata & 1.68 & 0.29 & 1.94 & 0.25 & 4.73 & 0.25 & 2.86 & 0.26 & 2.84 & 0.26 \\
12152 & 49.06599 & 41.18074 & 1 & 1.00 & 14.12 & 0.809 & 0.02095 & 2.290 & 0.011 & 29.5 &  0 & 1.66 & 0.73 & 0.297 & 0.015 & 22.72 & 0.71 & 2.140 & 0.009 & -2.77 & 0.43 & 0.112 & 0.012 & 5.70 & 0.30 & -6.22 & 0.35 & 4.98 & 0.31 & 7.85 & 0.27 & 2.25 & 0.21 & 2.39 & 0.18 & 4.49 & 0.20 & 3.06 & 0.20 & 2.83 & 0.21 \\
12157 & 49.10881 & 41.53037 & 1 & 0.67 & 14.09 & 0.815 & 0.01623 & 2.239 & 0.010 & 32.0 &  0 & \nodata & \nodata & 0.178 & 0.014 & 21.07 & 0.70 & 2.180 & 0.009 & -1.26 & 0.41 & 0.093 & 0.011 & 5.00 & 0.29 & -5.46 & 0.34 & 5.47 & 0.29 & 7.16 & 0.25 & 2.02 & 0.20 & 2.22 & 0.17 & 3.89 & 0.17 & 3.55 & 0.18 & 3.25 & 0.18 \\
12160 & 49.11466 & 41.62702 & 1 & 0.37 & 14.33 & 0.770 & 0.01654 & 2.049 & 0.008 & 127.6 & 70 & 1.50 & 0.10 & 0.135 & 0.004 & 18.92 & 0.19 & 1.930 & 0.003 & -0.11 & 0.08 & 0.033 & 0.003 & 4.60 & 0.07 & -3.94 & 0.08 & 3.69 & 0.10 & 4.86 & 0.12 & 1.96 & 0.05 & 2.15 & 0.03 & 3.37 & 0.05 & 2.66 & 0.06 & 2.40 & 0.07 \\
\enddata
\tablecomments{Columns are explained in Table \ref{tab-columns}. \\ This table is available in its entirety in machine-readable form.}
\end{splitdeluxetable*}

\startlongtable
\begin{splitdeluxetable*}{l rrr rr rr rr rr rr rr rr B rrrrrr rr}
% this table  needs to be rotated
\rotate
\tablecaption{Average Parameters, Ages, [M/H], [CN/Fe], and [Mg/Fe] \label{tab-average} }
\tablewidth{0pt}
\tablenum{18}
\tabletypesize{\scriptsize}
\tablehead{\colhead{Cluster} &
\colhead{$N$} & \colhead{$\log \sigma$} & \colhead{$\sigma _{\log \sigma}$} & 
\colhead{CN3883} & \colhead{$\sigma _{\rm CN3883}$} & 
\colhead{$({\rm H}\delta _{\rm A} + {\rm H}\gamma _{\rm A})'$} & \colhead{$\sigma _{({\rm H}\delta _{\rm A} + {\rm H}\gamma _{\rm A})'}$} & 
\colhead{log Fe4383} & \colhead{$\sigma _{\rm \log {\rm Fe4383}}$} & 
\colhead{log C4668} & \colhead{$\sigma _{\rm \log {\rm C4668}}$} & 
\colhead{$\log {\rm H}\beta_{\rm G}$} & \colhead{$\sigma _{\log {\rm H}\beta_{\rm G}}$} & 
\colhead{log Mg$b$} & \colhead{$\sigma _{\rm \log {\rm Mg}b}$} & 
\colhead{$\log \langle {\rm Fe} \rangle$} & \colhead{$\sigma _{\rm \log \langle {\rm Fe} \rangle}$} & 
\colhead{log age} & \colhead{$\sigma _{\log {\rm age}}$} & 
\colhead{[M/H]} & \colhead{$\sigma _{\rm [M/H]}$} & 
\colhead{[CN/Fe]} & \colhead{$\sigma _{\rm [CN/Fe]}$} & 
\colhead{[Mg/Fe]} & \colhead{$\sigma _{\rm [Mg/Fe]}$}  
}
\startdata
Perseus & 13 & 1.874 & 0.005 & 0.176 & 0.007 & -0.063 & 0.002 & 0.613 & 0.013 & 0.777 & 0.017 & 0.373 & 0.008 & 0.555 & 0.007 & 0.446 & 0.008 & 0.70 & 0.03 & 0.14 & 0.04 & -0.00 & 0.04 & 0.14 & 0.02 \\
Perseus & 7 & 1.975 & 0.007 & 0.191 & 0.007 & -0.074 & 0.003 & 0.633 & 0.013 & 0.793 & 0.016 & 0.332 & 0.012 & 0.579 & 0.008 & 0.454 & 0.008 & 0.80 & 0.04 & 0.14 & 0.04 & 0.04 & 0.04 & 0.16 & 0.02 \\
Perseus & 12 & 2.070 & 0.003 & 0.190 & 0.005 & -0.070 & 0.002 & 0.652 & 0.009 & 0.792 & 0.012 & 0.319 & 0.007 & 0.590 & 0.006 & 0.450 & 0.006 & 0.75 & 0.03 & 0.17 & 0.03 & 0.00 & 0.03 & 0.19 & 0.02 \\
Perseus & 8 & 2.130 & 0.004 & 0.201 & 0.007 & -0.077 & 0.003 & 0.617 & 0.016 & 0.800 & 0.016 & 0.331 & 0.014 & 0.599 & 0.009 & 0.426 & 0.010 & 0.84 & 0.05 & 0.14 & 0.04 & 0.12 & 0.04 & 0.26 & 0.02 \\
Perseus & 8 & 2.181 & 0.004 & 0.205 & 0.006 & -0.076 & 0.003 & 0.632 & 0.013 & 0.822 & 0.015 & 0.275 & 0.018 & 0.605 & 0.010 & 0.440 & 0.010 & 0.78 & 0.03 & 0.20 & 0.04 & 0.11 & 0.03 & 0.25 & 0.03 \\
Perseus & 10 & 2.237 & 0.004 & 0.216 & 0.007 & -0.093 & 0.004 & 0.678 & 0.012 & 0.844 & 0.013 & 0.311 & 0.020 & 0.649 & 0.009 & 0.464 & 0.010 & 0.90 & 0.04 & 0.22 & 0.04 & 0.08 & 0.03 & 0.27 & 0.02 \\
Perseus & 11 & 2.274 & 0.003 & 0.256 & 0.006 & -0.092 & 0.002 & 0.676 & 0.009 & 0.845 & 0.010 & 0.310 & 0.012 & 0.630 & 0.006 & 0.460 & 0.007 & 0.90 & 0.04 & 0.22 & 0.03 & 0.26 & 0.02 & 0.24 & 0.02 \\
Perseus & 19 & 2.322 & 0.002 & 0.236 & 0.006 & -0.093 & 0.003 & 0.669 & 0.011 & 0.859 & 0.011 & 0.278 & 0.015 & 0.659 & 0.006 & 0.457 & 0.007 & 0.88 & 0.03 & 0.25 & 0.03 & 0.19 & 0.03 & 0.30 & 0.01 \\
Perseus & 9 & 2.384 & 0.003 & 0.274 & 0.006 & -0.101 & 0.003 & 0.706 & 0.010 & 0.921 & 0.010 & 0.266 & 0.015 & 0.686 & 0.008 & 0.463 & 0.009 & 0.83 & 0.07 & 0.40 & 0.04 & 0.28 & 0.02 & 0.31 & 0.02 \\
Perseus & 8 & 2.477 & 0.003 & 0.293 & 0.011 & -0.101 & 0.006 & 0.723 & 0.017 & 0.933 & 0.015 & 0.259 & 0.023 & 0.707 & 0.009 & 0.478 & 0.009 & 0.75 & 0.09 & 0.47 & 0.06 & 0.30 & 0.03 & 0.32 & 0.02 \\
\enddata
\tablecomments{This table is available in its entirety in machine-readable form.}
\end{splitdeluxetable*}


\begin{thebibliography}{}

% * All 4 low z clusters
\bibitem[Abell(1958)]{abell:1958}
Abell, G.\ O. 1958, ApJS, 3, 211

% * contains all four clusters
\bibitem[Abell et al.(1989)]{abell:1989}
Abell, G.\ O., Corwin Jr., H.\ G., Olowin, R.\ P. 1989, ApJS, 70, 1

% * Cooling flow clusters ROSAT data
\bibitem[Allen(2000)]{allen:2000}
Allen, S.\ W. 2000, MNRAS, 315, 269

% * z=1.8 cluster
\bibitem[Andreon et al.(2014)]{andreon:2014}
Andreon, S., Newman, A.\ B., Trinchieri, G., et al. 2014, A\&A, 565, 120

% * GOGREEN survey paper
\bibitem[Balogh et al.(2017)]{balogh:2017}
Balogh, M.\ L., Gilbank, D.\ G., Muzzin, A., et al. 2017, MNRAS, 470, 4168

% * RXJ0142 spec analysis - use unpublished Perseus data
\bibitem[Barr et al.(2005)]{barr:2005}
Barr, J., Davies, R., J\o rgensen, I., Bergmann, M., \& Crampton, D. 2005, AJ, 130, 445

% * biweight sigma_cl
\bibitem[Beers et al.(1990)]{beers:1990}
Beers, T.\ C., Flynn, K., \& Gebhardt, K. 1990, AJ, 100, 32

% * KMOS FP for 19 galaxies uses my old Coma data as reference
\bibitem[Beifiori et al.(2017)]{beifiori:2017}
Beifiori, A., Mendel, J.\ T., Chan, J.\ C.\ C., et al. 2017, ApJ, 846, 120

% * MPA-JHU ref
\bibitem[Brinchmann et al.(2004)]{brinchmann:2004}
Brinchmann, J., Charlot, S., White, S.\ D.\ M., et al. 2004, MNRAS, 351, 1151

% * D4000 definition
\bibitem[Bruzual(1983)]{bruzual:1983}
Bruzual A., G. 1983, ApJ, 273, 105

% * NGC1275 star clusters
\bibitem[Canning et al.(2014)]{canning:2014}
Canning, R.\ E.\ A., Ryon, J.\ E., Gallagher, J.\ S., et al. 2014, MNRAS, 444, 336

% * z~1.6 stacked spectrum, uses Coma J1999 as reference
\bibitem[Cappellari et al.(2006)]{cappellari:2006}
Cappellari, M., Alighierei, S.\ de S., Cimatti, A., et al. 2009, ApJL, 704, L34

% * U-B and B-V CM relations redshift evolution
\bibitem[Cerulo et al.(2016)]{cerulo:2016}
Cerulo, P., Couch, W.\ J., Lidman, C., et al. 2016, MNRAS, 457, 2209

% K-corrections
\bibitem[Chilingarian et al.(2010)]{chilingarian:2010}
Chilingarian, I., Melchior, A.-L., \& Zolotukhin, I. 2010, MNRAS, 405, 1409 

% * redshift info RXJ1347 used for sample selection, low sigma_cluster
\bibitem[Cohen \& Kneib(2002)]{cohen:2002}
Cohen, J.\ G., \& Kneib, J.-P. 2002, ApJ, 573, 524

% * NGC1275 - with a good review in the start
\bibitem[Conselice et al.(2001)]{conselice:2001}
Conselice, C.\ J., Gallagher III, J.\ S., \& Wyse, R.\ F.\ G. 2001, ApJ, 122, 2281

% * SED models with individual elements varied  velocity dispersion versus age and abundance ratios
\bibitem[Conroy et al.(2014)]{conroy:2014}
Conroy, C., Graves, G.\ J., \& van Dokkum, P.\ G. 2014, ApJ, 780, 33

% * Coma cluster earliest reference
\bibitem[Curtis(1918)]{curtis:1918}
Curtis, H.\ D. 1918, Publications of Lick Observatory, 13, 9

% * CL1001 z=2.5 cluster
\bibitem[Daddi et al.(2017)]{daddi:2017}
Daddi, E., Jin, S., Strazzullo, V., et al. 2017, ApJL, 846, L31

\bibitem[Davidge \& Clark(1994)]{davidge:1994}
Davidge, T.\ J., \& Clark, C.\ C. 1994, AJ, 107, 946

% size evolution - 
\bibitem[Delaye et al.(2014)]{delaye:2014}
Delaye, L., Huertas-Company, H., Mei, S., et al. 2014, MNRAS, 441, 203

% * FP
\bibitem[Djorgovski \& Davis(1987)]{dd:1987}
Djorgovski, S., \& Davis, M. 1987, ApJ, 313, 59

\bibitem[Dressler(1980)]{dressler:1980}
Dressler, A. 1980, ApJ, 236, 351

% FP
\bibitem[Dressler et al.(1987)]{dressler:1987}
Dressler, A., Lynden-Bell, D., Burstein, D., et al. 1987, ApJ, 313, 42

% * dens-morph to z=0.5
\bibitem[Dressler et al.(1997)]{dressler:1997}
Dressler, A., Oemler Jr., A., Couch, W.\ J., et al. 1997, ApJ, 490, 577

% * A2142 SF and quenched galaxies infalling
\bibitem[Einasto et al.(2018)]{einasto:2018}
Einasto, M., Deshev, B., Lietzen, H., et al. 2018, A\&A, 610, A82

% * LF z=0.8-1.0 clusters  RXJ0152, RXJ1226, RXJ1415 - all GCP clusters
\bibitem[Ellis \& Jones(2004)]{ellis:2004}
Ellis, S.\ C., \& Jones, L.\ R. 2004, MNRAS, 348, 165

% * Xray data RXJ0152 subclusters, other higher z clusters
\bibitem[Ettori et al.(2004)]{ettori:2004}
Ettori, S., Tozzi, P., Borgani, S., \& Rostati, P. 2004, A\&A, 417, 13

% * LF z=1 to present DEEP2 COMBO-17 - descriptive figs of evolution scenarios
\bibitem[Faber et al.(2007)]{faber:2007}
Faber, S.\ M., Willmer, C.\ N.\ A., Wolf, C. et al.\ 2007, ApJ, 665, 265

% * Review article on cooling flow clusters
\bibitem[Fabian(1994)]{fabian:1994}
Fabian, A.\ C. 1994, ARA\&A, 32, 277

% * Millenium II simulations
\bibitem[Fakhouri et al.(2010)]{fakhouri:2010}
Fakhouri, O., Ma, C.-P., Boylan-Kolchin, M. 2010, MNRAS, 406, 2267

% * CM relations U-B vor SPARCS clusters
\bibitem[Foltz et al.(2015)]{foltz:2015}
Foltz, R., Rettura, A., Wilson, G., et al. 2015, ApJ, 812, 138

% * FP sort of including Perseus, states it is multiple overlapping planes
\bibitem[Fraix-Burnet et al.(2010)]{fraix:2010}
Fraix-Burnet, D., Dugue, M., Chattopadhyay, T., Chattopadhyay, A.\ K., \& Davoust, E., 2010, MNRAS, 407, 2207

\bibitem[Gebhardt et al.(2000)]{gebhardt:2000}
Gebhardt, K., Richstone, D., Kormendy, J., et al. 2000, AJ, 119, 1157

\bibitem[Gebhardt et.al.(2003)]{gebhardt:2003}
Gebhardt, K., Faber, S.\ M.; Koo, D.\ C., et al. 2003, ApJ, 597, 239

% * Perseus - no substructure
\bibitem[Girardi et al.(1997)]{girardi:1997}
Girardi, M., Esclera, E., Fadda, D., et al., 1997, ApJ, 482, 41

% * z=2.0 cluster
\bibitem[Gobat et al.(2013)]{gobat:2013}
Gobat, R., Strazzullo, V., Daddi, E., et al. 2013, ApJ, 776, 9

% Coma catalog
\bibitem[Godwin et al.(1983)]{godwin:1983}
Godwin, J.\ G., Metcalfe, N, \& Peach, J.\ V. 1983, MNRAS, 202, 113

% Def of HbetaG, use of MgFe
\bibitem[Gonz\'{a}lez(1993)]{gonzalez:1993}
Gonz\'{a}lez, J.\ J. 1993, PhD thesis, Univ.\ California, Santa Cruz

% vrad/sigma * Rcl/R500 from simulations
\bibitem[Haines et al.(2013)]{haines:2012}
Haines, C.\ P., Pereira, M.\ J., Sanderson, J.\ R., et al. 2012, ApJ, 754, 97

\bibitem[Haines et al.(2015)]{haines:2015}
Haines, C.\ P., Pereira, M.\ J., Smith, G.\ P., et al. 2015, ApJ, 806, 101

% age M/H alpha/Fe sigma - Coma included
\bibitem[Harrison et al.(2011)]{harrison:2011}
Harrison, C., D., Colless, M., Kuntschner, H., et al. 2011, MNRAS, 413, 1036

\bibitem[Hook et al.(2004)]{hook:2004}
Hook, I.\ M., J{\o}rgensen, I. Allington-Smith, J.\ R., et al. 2004, PASP, 116, 425

% mass-size and morphology
\bibitem[Huertas-Company et al.(2013)]{huertas:2013}
Huertas-Company, M., Shankar, F., Mei, S., et al. 2013, ApJ, 779, 29

% morphology transformations since z~3 CANDLES
\bibitem[Huertas-Company et al.(2016)]{huertas:2016}
Huertas-Company, M., Bernardi, M., P\'{e}rez-Gonz\'{a}lez, P.\ G. et al. 2016, MNRAS, 462, 4495

% * SDSS abundance ratios using Thomas models
\bibitem[Johansson et al.(2012)]{johansson:2012}
Johansson, J., Thomas, D., and Maraston, C. 2012, MNRAS, 421, 1908

% * Chandra data for RXJ0152
\bibitem[Jones et al.(2004)]{jones:2004}
Jones, L.\ R., Maughan, B.\ J., Ebeling, H., et al. 2004, 
in Clusters of Galaxies: Probes of Cosmological Structure and Galaxy Evolution, 
ed.\ J.\ S.\ Mulchaey, A.\ Dressler, \& A.\ Oemler (Pasadena, CA: Carnegie Observatories), 25

\bibitem[J{\o}rgensen(1999)]{IJ:1999}
J{\o}rgensen, I. 1999, MNRAS, 306, 607

\bibitem[J{\o}rgensen \& Chiboucas(2013)]{IJ:2013}
J{\o}rgensen, I., \& Chiboucas, K. 2013, AJ, 145, 77

% * Coma photometry
\bibitem[J{\o}rgensen \& Franx(1994)]{IJ:1994}
J{\o}rgensen, I., \&  Franx, M. 1994, ApJ, 433, 553

\bibitem[J{\o}rgensen et al.(1995a)]{IJ:1995a} 
J{\o}rgensen, I., Franx, M., \& Kj{\ae}rgaard, P. 1995a, MNRAS, 273, 1097

\bibitem[J{\o}rgensen et al.(1995b)]{IJ:1995b} 
J{\o}rgensen, I., Franx, M., \& Kj{\ae}rgaard, P. 1995b, MNRAS, 276, 1341

\bibitem[J{\o}rgensen et al.(1996)]{IJ:1996} 
J{\o}rgensen, I., Franx, M., \& Kj{\ae}rgaard, P. 1996, MNRAS, 280, 167

\bibitem[J{\o}rgensen et al.(2005)]{IJetal:2005}
J{\o}rgensen, I., Bergmann, M., Davies, R., et al. 2005, AJ, 129, 1249

\bibitem[J{\o}rgensen et al.(2006)]{IJetal:2006}
J{\o}rgensen, I., Chiboucas, K., Flint, K., et al. 2006, ApJL, 639, L9

\bibitem[J{\o}rgensen et al.(2007)]{IJetal:2007}
J{\o}rgensen, I., Chiboucas, K., Flint, K., et al. 2007, ApJL, 654, L179

\bibitem[J{\o}rgensen et al.(2014a)]{IJetal:2014}
J{\o}rgensen, I., Chiboucas, K., Toft, S., et al. 2014, AJ, 148, 117

\bibitem[J{\o}rgensen et al.(2017)]{IJetal:2017}
J{\o}rgensen, I., Chiboucas, K., Berkson, E., et al. 2017, AJ, 154, 251

% * Cluster catalog - photometry paper
\bibitem[J{\o}rgensen et al.(2018)]{IJetal:2018}
J{\o}rgensen, I., Chiboucas, K., Hibon, P., et al. 2018, ApJS, 235, 29

% * MPA-JHU ref
\bibitem[Kauffmann et al.(2003)]{kauffmann:2003}
Kauffmann, G., Heckman, T.\ M., White, S.\ D.\ M., et al. 2003, MNRAS, 341, 33

% * age metal alpha/fe sigma z=0.33 cluster
\bibitem[Kelson et al.(2006)]{kelson:2006}
Kelson, D.\ D., Illingworth, G.\ D., Franx, M., \& van Dokkum, P.\ G. 2006, ApJ, 653, 159

% * SPT-SZ z>1.25 clusters; passive galaxy composite spectrum
\bibitem[Khullar et al.(2018)]{khullar:2018}
Khullar, G., Bleem, L.\ E., Bayliss, M.\ B., et al. 2018, arXiv:1806.01962

% Def of hdga 
\bibitem[Kuntschner(2000)]{kuntschner:2000}
Kuntschner, H. 2000, MNRAS, 315, 184

% * CL0024 full spectrum fitting, age-metal-mass
\bibitem[Leethochawalit et al.(2018)]{leethochawalit:2018}
Leethochawalit, N., Kirby, E.\ N., Moran, S., et al. 2018, ApJ, 856, 15

% * RXJ1347 cluster velocity dispersion no data in paper
\bibitem[Lu et al.(2010)]{lu:2010}
Lu, T., Gilbank, D.\ G., Balogh, M.\ L., et al. 2010, MNRAS, 403, 1787

\bibitem[Maraston \& Str\"{o}mb\"{a}ck(2011)]{marastron:2011}
Maraston, C., Str\"{o}mb\"{a}ck, G.\ 2011, MNRAS, 418, 2785

% * A2142 substructure based on Chandra data
\bibitem[Markevitch et al.(2000)]{markevitch:2000}
Markevitch, M., Ponman, T.\ J., Nulsen, P.\ E.\ J., et al. 2000, ApJ 541, 542 

% * RXJ0152 original X-ray data analysis - binary cluster
\bibitem[Maughan et al.(2003)]{maughan:2003}
Maughan, B.\ J., Jones, L.\ R., Ebeling, H., et al. 2003, ApJ, 587, 589

% * ATLAS-3D line indices full spectral fitting
\bibitem[McDermid et al.(2015)]{mcdermid:2015}
McDermid, R.\ M., Alatalo, K., Blitz, L., et al. 2015, MNRAS, 448, 3484

% * possible substructure in Perseus, old X-ray data
\bibitem[Mohr et al.(1993)]{mohr:1993}
Mohr, J.\ J., Fabricant, D.\ G., \& Geller, M.\ J. 1993, ApJ, 413, 492

% CL0024 FP line indices
\bibitem[Moran et al.(2005)]{moran:2005}
Moran, S., Ellis, R.\ S., Treu, T., et al. 2005, ApJ, 634, 977

% * MS0451, CL0024 zform etc
\bibitem[Moran et al.(2007b)]{moran:2007}
Moran, S., Ellis, R.\ S., Treu, T., et al. 2007, ApJ, 671, 1503

% * SDSS DR12 lambda-vac - lambda-air
\bibitem[Morton(1991)]{morton:1991}
Morton, D.\ C. 1991, ApJS, 77, 119

% NOAO FP survey - index-sigma ages metal abundance ratios
\bibitem[Nelan et al.(2005)]{nelan:2005}
Nelan, J.\ E., Smith, R.\ J., Hudson, M.\ J., et al. 2005, ApJ, 632, 137

% * HzetaA def
\bibitem[Nantais et al.(2013)]{nantais:2013}
Nantais, J.\ B., Rettura, A., Lidman, C., et al. 2013, A\&A, 556, 112

% * z=1.8 cluster
\bibitem[Newman et al.(2014)]{newman:2014}
Newman, A.\ B., Ellis, R.\ S., Andreon, S., et al. 2014, ApJ, 788, 51

% * Coma weak lensing study, subhalos
\bibitem[Okabe et al.(2014)]{okabe:2014}
Okabe, N., Futamase, T., Kajisawa, M., \& Kuroshima, R. 2014, ApJ, 784, 90

% * A2142 substructure and velocity dispersion
\bibitem[Owers et al.(2011)]{owers:2011}
Owers, M.\ S., Nulsen, P.\ E.\ J., Couch, W.\ J. 2011, ApJ, 741, 122

% * RXJ1334 non-relaxed, A2029 strong-relaxes, A2142 non-relaxed
\bibitem[Parekh et al.(2015)]{parekh:2015}
Parekh, V., van der Heyden, K., Ferrari, C., Angus, G., \& Holwerda, B. 2015, A\&A, 575, A127

% PGC catalog, Perseus  HYPERLEDA
\bibitem[Paturel et al.(2003)]{paturel:2003}
Paturel, G., Petit, C., Prugniel, P., et al. 2003, A\&A, 412, 45

% * Perseus dwarf galaxies
\bibitem[Penny et al.(2009)]{penny:2009}
Penny, S.\ J., Conselice, C.\ J., de Rijcke, S., \& Held, E.\ V. 2009, MNRAS, 393, 1054

\bibitem[Piffaretti et al.(2011)]{piffaretti:2011}
Piffaretti, R., Arnaud, M., Pratt, G.\ W., Pointecouteau, \& Melin, J.-B. 2011, A\&A, 534, A109

% * Coma is not typical ...
\bibitem[Pimbblet et al.(2014)]{pimbblet:2014}
Pimbblet, K.\ A., Penny, S.\ J., \& Davies, R.\ L. 2014, MNRAS, 438, 3049

% * Cooling flows and SF Perseus, Coma, A2029
\bibitem[Rafferty et al.(2008)]{rafferty:2008}
Rafferty, D.\ A., McNamara, B.\ R., Nulsen, P.\ E.\ J. 2008, ApJ, 687, 899

% * EDiSC FP
\bibitem[Saglia et al.(2010)]{saglia:2010}
Saglia, R.\ P., S\'{a}nchez-Bl\'{a}zquez, P., Bender, R., et al. 2010, A\&A, 524, A6

\bibitem[Salpeter(1955)]{salpeter:1955}
Salpeter, E.\ E. 1955, ApJ, 121, 161

% * EDisCS line indices
\bibitem[S\'{a}nchez-Bl\'{a}zquez et al.(2009)]{sanchez:2009}
S\'{a}nchez-Bl\'{a}zquez, P., Jablonka, P., Noll, S., et al. 2009, A\&A, 499, 47

% * cited in Lynx W paper; uses JFK1995 phot
\bibitem[Saracco et al.(2014)]{saracco:2014}
Saracco, P., Casati, A., Gargiulo, G., et al. 2014, A\&A, 567, A94

% * Coma subhalo NGC4839 X-ray
\bibitem[Sasaki et al.(2016)]{sasaki:2016}
Sasaki, T., Matsushita, K., Sato, K., \& Okabe, N. 2016, PASJ, 68, 85

% Galactic extinction
\bibitem[Schlafly et al.(2011)]{schlafly:2011}
Schlafly, E.\ J., et al. 2011, ApJ, 737, 103

% * def of sersic profile
\bibitem[Sersic(1968)]{sersic:1968}
S\'{e}rsic, J.\ L. 1968, Atlas de Galaxias Australes (Cordoba: Observatorio Astronomico)

% NOAO FP sigma-line-env age-metal-abundance ratios
\bibitem[Smith et al.(2006)]{smith:2006}
Smith, R.\ J., Hudson, M.\ J., Lucey, J.\ R., et al. 2006, MNRAS, 369, 1419

% Coma dwarf galaxies, also lists Shapley super cluster age-metal-abundances
\bibitem[Smith et al.(2009a)]{smith:2009a}
Smith, R.\ J., Lucey, J.\ R., Hudson, M.\ J., et al. 2009a, MNRAS, 392, 1265

% Shapley super cluster age-metal-abundances
\bibitem[Smith et al.(2009b)]{smith:2009b}
Smith, R.\ J., Lucey, J.\ R., Hudson, M.\ J. 2009b, MNRAS, 400, 1690

% * Coma low mass galaxies and environment
\bibitem[Smith et al.(2012)]{smith:2012}
Smith, R.\ J., Lucey, J.\ R., Price, J., Hudson, M.\ J., \& Phillips, S.  2012, MNRAS, 419, 3167

% * A2029 and Coma velocity dispersions
\bibitem[Sohn et al.(2017)]{sohn:2017}
Sohn, J., Geller, M.\ J., Zahid, H.\ J., et al. 2017, ApJS, 229, 20

% * CL1426 z=1.75
\bibitem[Stanford et al.(2012)]{stanford:2012}
Stanford, S.\ A., Brodwin, M., Gonzalez, A.\ H., et al. 2012, ApJ, 753, 164

% * Older models used in Thomas et al 2005
\bibitem[Thomas et al.(2003)]{thomas:2003}
Thomas, D., Maraston, C., \& Bender, R. 2003, MNRAS, 339, 897 \\

% * zform vs mass
\bibitem[Thomas et al.(2005)]{thomas:2005}
Thomas, D., Maraston, C., Bender, R., \& de Oliviera, C.\ M. 2005, ApJ, 621, 673

% * SSP models used
\bibitem[Thomas et al.(2011)]{thomas:2011}
Thomas, D., Maraston, C., \& Johansson, J. 2011, MNRAS, 412, 2183 \\
(http://www.icg.port.ac.uk/\~{ }thomasd)

% * zform vs mass  steeper than 2005
\bibitem[Thomas et al.(2010)]{thomas:2010}
Thomas, D., Maraston, C., Schawinski, K., Sarzi, M., \& Silk, J. 2010, MNRAS, 404, 1775

% * SDSS velocity dispersions, emission line EWs
\bibitem[Thomas et al.(2013)]{thomas:2013}
Thomas, D., Steele, O., Maraston, C.,  et al. 2013, MNRAS, 431, 1383

% * Coma High S/N spectra age, metal, alpha/Fe
\bibitem[Trager et al.(2008)]{trager:2008}
Trager, S.\ C.,  Faber, S.\ M., \& Dressler, A. 2008, MNRAS, 386, 715

% * MPA-JHU reference
\bibitem[Tremonti et al.(2004)]{tremonti:2004}
Tremonti, C.\ A., Heckman, T.\ M., Kauffmann, G., et al. 2004, ApJ, 613, 898

% * A2029 and Coma SF galaxies
\bibitem[Tyler et al.(2013)]{tyler:2013}
Tyler, K.\ D., Rieke, G.\ H., \& Bai, L. 2003, ApJ, 773, 86

% * mass of cluster evolution
\bibitem[van den Bosch(2002)]{vandenbosch:2002}
van den Bosch, F.\ C.\ 2002, MNRAS, 331, 98

% * FP CL0024 uses Coma JFK1996
\bibitem[van Dokkum \& Franx(1996)]{vandokkum:1996}
van Dokkum, P.\ G., \& Franx, M. 1996, MNRAS, 281, 985

% FP to z~1 clusters and field
\bibitem[van Dokkum \& van der Marel(2007)]{vandokkum:2007}
van Dokkum, P.\ G., \& van der Marel, R.\ P. 2007, ApJ, 655, 30

% * MILES library and base models
\bibitem[Vazdekis et al.(2010)]{vazdekis:2010}
Vazdekis, A., S\'{a}nchez-Bl\'{a}zquez, P., Falc\'{o}n-Barroso, J., et al. 2010, MNRAS, 404, 1639 

% * Chandra relaxes clusters A2029, A2142
\bibitem[Vikhlinin et al.(2009)]{vikhlinin:2009}
Vikhlinin, A., Burenin, R.\ A., Ebeling, H., et al. 2009, ApJ, 692, 1033

% * Perseus sloshing of X-ray gas
\bibitem[Walker et al.(2017)]{walker:2017}
Walker, S.\ A., Hlavacek-Larrondo, J., Gendron-Marsolais, M., et al. 2011, MNRAS, 468, 2506

% * Coma X-ray substructure - first ref
\bibitem[White et al.(1993)]{white:1993}
White, S.\ D.\ M., Briel, U.\ G, \& Henry, J.\ P. 1993, MNRAS, 261, L8

% * Worthey original models age-metal degeneracy
\bibitem[Worthey(1994)]{worthey:1994}
Worthey, G. 1994, ApJS, 95, 107

% * Lick/IDS defs
\bibitem[Worthey et al.(1994)]{wortheyetal:1994}
Worthey, G., Faber, S.\ M., Gonz\'{a}lez, J.\ J., \& Burstein, D. 1994, ApJS, 94, 687

% * Lick/IDS high order Balmer defs
\bibitem[Worthey \& Ottaviani(1997)]{worthey:1997}
Worthey, G., \& Ottaviani, D.\ L. 1997, ApJS, 111, 377

% * uses JFK1996 as reference
\bibitem[Wuyts et al.(2004)]{wuyts:2004}
Wuyts, S., van Dokkum, P.\ G., Kelson, D.\ D., Franx, M., Illingworth, G.\ D. 2004, ApJ, 605, 677

\bibitem[Zabludoff et al.(1990)]{zabludoff:1990}
Zabludoff, A., Huchra, J.\ P., \& Geller, M.\ J. 1990, ApJS, 74, 1

% * FP A2218 uses Coma J1999 data
\bibitem[Ziegler et al.(2001)]{ziegler:2001}
Ziegler, B., Bower, P., Smail, I., et al. 2001, MNRAS, 325, 1571

% * Perseus earliest ref, Coma 
\bibitem[Zwicky(1942)]{zwicky:1942}
Zwicky, F., 1942, PASP, 54, 185

\end{thebibliography}
\end{document}